\documentclass[a4paper,11pt]{article}
\pdfoutput=1 

\usepackage{jheppub} 

\usepackage[T1]{fontenc} 
\usepackage[utf8]{inputenc}
\usepackage{amsmath,amssymb,amstext,amsfonts} 

\usepackage[dvipsnames]{xcolor}
\usepackage{graphicx}
\usepackage{hyperref}
\usepackage{cancel} 
\usepackage{float}
\usepackage{bbm}
\usepackage{mathtools}
\usepackage{comment}
\usepackage{dsfont}

\usepackage{lscape}
\usepackage{siunitx}
\usepackage{mathdots}
\usepackage{multirow}

\allowdisplaybreaks[4]
\usepackage{cleveref}

\makeatletter
\renewcommand*\env@matrix[1][\arraystretch]{%
  \edef\arraystretch{#1}%
  \hskip -\arraycolsep
  \let\@ifnextchar\new@ifnextchar
  \array{*\c@MaxMatrixCols c}}
\makeatother

\makeatletter
\g@addto@macro\bfseries{\boldmath}
\makeatother

\newcommand{\eps}{\epsilon}

\def\beq{\begin{equation}}
\def\eeq{\end{equation}}
\def\bsp#1\esp{\begin{split}#1\end{split}}

\newenvironment{gleichung*}{\begin{equation*}\begin{aligned}}{\end{aligned}\end{equation*}}

\usepackage{tikz}
    \usetikzlibrary{positioning}
    \usetikzlibrary{arrows}
    \usetikzlibrary{shapes}
    \usepackage{pgfplots}
    \usetikzlibrary{calc}
    \usetikzlibrary{decorations.markings}
     \usetikzlibrary{decorations.pathmorphing}
    \tikzset{snake it/.style={decorate, decoration=snake}}
\def\centerarc[#1](#2)(#3:#4:#5) 
    { \draw[#1] ($(#2)+({#5*cos(#3)},{#5*sin(#3)})$) arc (#3:#4:#5); }
    \usepackage{booktabs}

\definecolor{cnblue}{RGB}{7,82,154}

\DeclareMathOperator{\K}{K}

\DeclareMathOperator{\Sol}{Sol}
\DeclareMathOperator{\Sym}{Sym}
\DeclareMathOperator{\Gr}{Gr}
\DeclareMathOperator{\diag}{diag}
\newcommand{\rd}{\mathrm{d}}
\newcommand{\uz}{\underline{z}}
\newcommand{\ord}{\mathcal{O}}
\newcommand{\cA}{\mathcal{A}}
\newcommand{\cL}{\mathcal{L}}
\newcommand{\cD}{\mathcal{D}}

\title{Aspects of canonical differential equations for Calabi-Yau geometries and beyond}

\author[a]{Claude Duhr} 
\author[a]{Sara Maggio} 
\author[b]{Christoph Nega}
\author[c]{Benjamin Sauer}
\author[b]{Lorenzo Tancredi}
\author[b]{and Fabian J. Wagner} 

\affiliation[a]{Bethe Center for Theoretical Physics, Universit\"at Bonn, Da53115, Germany}
\affiliation[b]{Technical University of Munich, TUM School of Natural Sciences, Physics Department, James-Franck-Straße 1, 85748 Garching, Germany}
\affiliation[c]{Institut f\"ur Physik, Humboldt-Universit\"at zu Berlin,
10099 Berlin, Germany}

\emailAdd{cduhr@uni-bonn.de}
\emailAdd{smaggio@uni-bonn.de}
\emailAdd{c.nega@tum.de}
\emailAdd{benjamin.sauer@hu-berlin.de}
\emailAdd{lorenzo.tancredi@tum.de}
\emailAdd{fabianjohannes.wagner@tum.de}

\preprint{{\raggedleft
            BONN-TH-2025-11   \\
		    TUM-HEP 1559/25   \\      
            HU-EP-25/13-RTG   \\
}}

\abstract{We show how a method to construct canonical differential equations for multi-loop Feynman integrals recently introduced by some of the authors can be extended to cases where the associated geometry is of Calabi-Yau type and even beyond. This can be achieved by supplementing the method with information from the mixed Hodge structure of the underlying geometry. We apply these ideas to specific classes of integrals whose associated geometry is a one-parameter family of Calabi-Yau varieties, and we argue that the method can always be successfully applied to those cases. Moreover, we perform an in-depth study of the properties of the resulting canonical differential equations. In particular, we show that the resulting canonical basis is equivalent to the one obtained by an alternative method recently introduced in the literature. We apply our method to non-trivial and cutting-edge examples of Feynman integrals necessary for gravitational wave scattering, further showcasing its power and flexibility. }

\begin{document} 
\maketitle
\flushbottom

\section{Introduction}
\label{intro}

Feynman integrals are central quantities in Quantum Field Theory (QFT).
In the standard textbook approach, they are essential building blocks for computing scattering amplitudes, which, in turn, constitute the backbone for the calculation of most observables at particle colliders.
Through the idea of reverse unitarity~\cite{Anastasiou:2002yz}, cut Feynman integrals (in which a selected subset of propagators are substituted by Dirac delta functions or more complicated combinations of delta and theta functions) become the main ingredients for directly computing inclusive or differential distributions.
In addition, Feynman-related integrals also play a central role in many other branches of physics.
For example, the top-down construction of Effective Field Theories typically requires matching the Wilson coefficients of general operators to some more complicated QFT and involves computing Feynman integrals in special kinematical limits (typically through the method of asymptotic expansions~\cite{Beneke:1997zp, Smirnov:2002pj}). In a similar fashion, analogues of Feynman integrals also appear in the description of the scattering of compact objects in classical General Relativity in the so-called post-Minkowskian expansion~\cite{Goldberger:2004jt, Porto:2016pyg, Mogull:2020sak} and in the computation of cosmological correlators (see, e.g.,~ref.~\cite{Arkani-Hamed:2023kig}).
Despite the obvious differences between these integrals, it has recently become clear that the very same mathematics underlies their structure, which in turn means that similar techniques can be used to systematise their analytical and numerical evaluation.
At the core of all these computations are classes of special functions related to complex manifolds, for example, the Riemann sphere, elliptic curves, and their higher-genus or higher-dimensional generalisations, notably {Calabi-Yau} (CY) geometries.

An extremely powerful technique to study these integrals is the differential equations method~\cite{Kotikov:1990kg, Remiddi:1997ny, Gehrmann:1999as}, which in turn is based on the existence of \emph{integration-by-parts identities}~\cite{Tkachov:1981wb, Chetyrkin:1981qh} (IBPs) among Feynman integrals. IBPs allow one to reduce all Feynman integrals from a given family to a basis of so-called \emph{master integrals}. By construction, this basis is closed under differentiation with respect to the masses and the external kinematical invariants.
Systems of linear differential equations with rational functions as coefficients can, at least in principle, be derived completely algorithmically as long as dimensional regularisation~\cite{tHooft:1972tcz, Bollini:1972ui} is used consistently and all integrals are considered as functions of the dimensional regulator $\epsilon = (d-d_0)/2$, with $d_0$ an integer. What makes this method so powerful is arguably the fact that we are typically not interested in solving these equations for general values of $d$, but instead as a Laurent series in $\epsilon$. It has been realised that it is often possible to find special bases of Feynman integrals whose differential equations take a very evocative form, often referred to as a \emph{canonical form}~\cite{Henn:2013pwa}.\footnote{See also ref.~\cite{Kotikov:2010gf} for
first considerations in this direction.} The dependence on the dimensional regulator $\epsilon$ can then be factorised from the differential equation system, which also makes the analytic properties of their solutions close to $\epsilon=0$ completely manifest: Each order in the Laurent series can be expressed as Chen iterated integrals~\cite{ChenSymbol} over the differential forms that appear in the connection matrix of the differential equations.
If this is the case, provided that one has an understanding of these differential forms (in particular, of all relations among them, including those modulo exact forms) and of the corresponding boundary values, the equations can be solved straightforwardly in terms of independent sets of iterated integrals.

The term `\emph{canonical basis}' was originally introduced for Feynman integrals of polylogarithmic type (i.e.~defined on the Riemann sphere), taking inspiration from local integrals with unit leading singularities~\cite{Arkani-Hamed:2010pyv}, which had been introduced to compactly represent amplitudes in $\mathcal{N}=4$ super Yang-Mills theory. 
The corresponding integrals can be cast in the form of iterated integrals over dlog-forms, which renders their properties particularly transparent. A special subclass of these iterated integrals, when the dlog-forms involve only rational functions, go under the name of \emph{multiple polylogarithms} (MPLs)~\cite{Kummer, Remiddi:1999ew, Goncharov:1995, Goncharov:1998kja, Vollinga:2004sn, Duhr:2011zq}.
The underlying logarithmic structure of these integrals has inspired many semi-algorithmic techniques to find canonical bases based on the analysis of the residues of the corresponding integrands~\cite{Henn:2020lye, Wasser:2022kwg, Chen:2020uyk, Dlapa:2021qsl, Chen:2022lzr}, on the direct manipulation of the differential equations~\cite{Lee:2014ioa, Gehrmann:2014bfa, Meyer:2017joq, Dlapa:2020cwj, Lee:2020zfb} or on combinations thereof. 
While these procedures are powerful and have been the basis for solving many state-of-the-art problems, a general approach to finding canonical bases, even in the polylogarithmic case, is currently not available.

As is well known, starting at the two-loop order, Feynman integrals are characterised by new geometries, and dlog-forms are insufficient to span the space of possible solutions. In the last decade, substantial effort has been dedicated to understanding how the concept of canonical differential equations can be extended beyond polylogarithms, and much progress has been achieved, especially in the elliptic and one-parameter CY cases~\cite{Adams:2016xah, Adams:2018bsn, Pogel:2022ken, Frellesvig:2021hkr, Dlapa:2022wdu, Frellesvig:2023iwr, Gorges:2023zgv, Pogel:2022yat}.
There are various important differences between the polylogarithmic case and more general geometries. Arguably, the most relevant one in this context is the fact that the cohomology of the corresponding varieties cannot be spanned by algebraic differential forms without higher-order poles. For example, it is well known that the cohomology of a compact Riemann surface of genus $g\ge 1$ is spanned by differential forms of the first and second kind, which are respectively holomorphic everywhere or have higher-order poles with vanishing residue. In the case of a punctured Riemann surface, one also needs to add differential forms of the third kind, which have a non-vanishing residue. For higher-dimensional varieties, the situation becomes more complicated, and the classification into differentials of the first, second, and third kind gets replaced by the \emph{mixed Hodge structure} (MHS) carried by the cohomology group. At the same time, it is known that the cohomology group can be generated by forms with, at most, logarithmic singularities, albeit at the price of introducing non-algebraic differential forms. In this way, it becomes possible to define generalisations of polylogarithms to higher-genus curves, cf.,~e.g.,~refs.~\cite{Brown:2011wfj, Broedel:2014vla,Broedel:2017kkb, Enriquez_hyperelliptic,EnriquezZerbini, DHoker:2023vax,DHoker:2023khh, DHoker:2024ozn,Baune:2024biq, Baune:2024ber, DHoker:2025szl, DHoker:2025dhv}.

The fact that it is, in general, not possible to find a set of algebraic generators with, at most, logarithmic singularities for the cohomology of an algebraic variety constitutes a fundamental roadblock to the idea of determining canonical Feynman integrals from a sole analysis of the residues of their integrands and imposing the absence of double poles, as is often done for Feynman integrals that evaluate to MPLs. At the same time, the fact that it is possible to find a set of transcendental cohomology generators with logarithmic singularities gives hope that some of the properties can be extended in one way or another, in particular, the existence of a canonical basis, which satisfies a differential equation with at most simple poles.

The existence of forms with higher-order poles is also often equivalent to the statement that the problem involves the solution of a higher-order Picard-Fuchs differential operator. 
Considering the series expansions that define its solutions locally close to a point of so-called \emph{Maximal Unipotent Monodromy} (MUM), one can classify them according to the powers of logarithms with which they diverge. 
For an operator of degree $n+1$, one generally has a holomorphic solution and $n$ solutions, which diverge at the MUM-point with up to $n$ powers of logarithms.
Based on this, in ref.~\cite{Gorges:2023zgv}, a general algorithm was proposed to reorganise the master integrals according to their logarithmic behaviour and define a basis of functions with (locally) uniform transcendental weight. 
This has been achieved in practice by separating the matrix of homogeneous solutions (also called the \emph{Wronskian matrix}) into a \emph{semi-simple} and a \emph{unipotent} part, following ideas introduced in ref.~\cite{Broedel:2018qkq}.
Locally, the semi-simple part carries the (generalised) algebraic part of the solutions, and the unipotent part is the logarithmic (transcendental) one. 
The applicability of this algorithm has been explicitly demonstrated for many state-of-the-art problems containing integrals of elliptic type~\cite{Gorges:2023zgv, Becchetti:2025rrz, Becchetti:2025oyb} and for their generalisation to CY varieties~\cite{Duhr:2024bzt, Forner:2024ojj, Klemm:2024wtd, Driesse:2024feo} and higher-genus surfaces~\cite{Duhr:2024uid}. Despite various explicit examples of canonical forms beyond polylogarithms being available, a general theory of canonical bases, including questions related to their existence, their uniqueness, or their mathematical properties, is still lacking. For example, various methods have been proposed to construct canonical bases beyond polylogarithms, cf.,~e.g., refs.~\cite{Gorges:2023zgv, Frellesvig:2021hkr, Adams:2018yfj, Pogel:2024sdi, Pogel:2022yat, Pogel:2022vat, Pogel:2022ken}. It is currently not always clear how these different proposals are related (though some proposals seem to be more appropriate than others, cf.~ref.~\cite{Frellesvig:2023iwr} for a discussion).

The goal of this paper is to perform a systematic study of various aspects of canonical bases that can be constructed using the method of ref.~\cite{Gorges:2023zgv}. In the first part of this paper, we illustrate how integrand analyses can be leveraged to study integrals associated with more complicated geometries. This allows us to formulate a proposal for a roadmap of how the method of ref.~\cite{Gorges:2023zgv} can be extended and applied to those cases. Central to these ideas is the mixed Hodge structure underlying the geometry defined by the maximal cuts, as well as the separation of the period matrix into its semi-simple and unipotent parts. In the second part of this paper, we focus on a specific class of geometries that arise from one-parameter families of CY varieties. Our general roadmap allows us to argue that the method of ref.~\cite{Gorges:2023zgv} always succeeds in determining a canonical basis for these cases, and we can prove the equivalence of our canonical bases and one obtained by the method of refs.~\cite{Pogel:2022yat, Pogel:2022vat, Pogel:2022ken}. Moreover, having identified a large class of CY examples for which we can determine the canonical form, we perform a study of the properties of the resulting canonical basis. Finally, in the third part of our paper, we discuss several examples that go beyond the pure CY geometries considered so far. In particular, we show that while determining the mixed Hodge structure for a realistic example may become increasingly complicated, we can supplement our ideas with intuition from physics to determine the canonical bases also for non-trivial realistic examples.

Our paper is organized as follows: In section~\ref{sec:conventions}, we briefly review differential equations for Feynman integrals and introduce our notations and conventions. In section~\ref{main}, we perform in detail the integrand analysis for a two-loop sunrise integral, and we show how this information can be leveraged to identify a good starting basis to which the method of ref.~\cite{Gorges:2023zgv} can be applied. In section~\ref{sec:roadmap}, we propose our roadmap of how to extend the method of ref.~\cite{Gorges:2023zgv} to more complicated geometries. In section~\ref{sec:CY}, we briefly review CY varieties before we show in section~\ref{sec:general_eps_fac} how to construct the canonical bases for some classes of one-parameter families of CY varieties, and we discuss their properties. Finally, in section~\ref{sec:additional_examples}, we discuss additional examples, illustrating how the method can be applied to non-trivial examples beyond the pure CY cases discussed before. In section~\ref{conclusions}, we draw our conclusions and we include various appendices where we collect additional material omitted in the main text.


\section{Feynman integrals and their differential equations}
\label{sec:conventions}

Throughout this paper, we consider families of Feynman integrals defined by
\begin{equation}
\label{def:integralfamily}
I_{\nu_1,\dots,\nu_{n+m}}(\underline z;d) = \int \left(\prod_{j=1}^l\frac{\mathrm d^dk_j}{i\pi^{d/2}}\right)\frac{\prod_{j=1}^m N_j^{-\nu_{n+j}}}{\prod_{j=1}^n D_j^{\nu_j}}\, ,
\end{equation}
where $\nu_j \leq 0$ for all $j>n$. The set $\{D_1, D_2, \dots, D_n \}$ of propagators in the denominator is specified by the topology of the Feynman graph under consideration. The numerators $\{N_1, N_2, \dots, N_m \}$ are a minimal set of \emph{irreducible scalar products} (ISPs), i.e., scalar products involving loop momenta that cannot be expressed as linear combinations of the propagators. 
We work in dimensional regularisation in $d=d_0-2\eps$ dimensions, with $d_0$ an even integer, and we view the Feynman integrals as Laurent expansions in the dimensional regulator $\eps$. We collect all kinematic variables, i.e., all Lorentz invariants of independent external momenta and masses, into the vector $\underline z$.
Every integral belongs to a \emph{sector}, defined as the set of integrals from the family that share exactly the same propagators, i.e., $I_{\nu_1,\dots,\nu_{n+m}}$ and $I_{\nu'_1,\dots,\nu'_{n+m}}$ belong to the same sector if $\theta(\nu_i)  =\theta(\nu'_i)$, for all $1\le i\le n$, where $\theta$ denotes the Heaviside step function, $\theta(x)=1$ if $x>0$ and $\theta(x)=0$ otherwise. The sectors can be labelled by an $n$-tuple of $0$'s and $1$'s, which label negative and positive values of $\nu_i$, respectively. The \emph{corner-integral} associated to the sector labeled by $(\theta_1,\ldots,\theta_n)$, $\theta_i\in\{0,1\}$, is the integral $I_{\theta_1,\ldots,\theta_n,0,\ldots,0}$.

It is well known that not all Feynman integrals that differ only by the value of the $\nu_i$ are independent, but they are related by IBP relations~\cite{Tkachov:1981wb, Chetyrkin:1981qh}, which give rise to linear relations between integrals from the same family, but with different values of $\nu_i$. We can solve these linear relations and express all members of the family in terms of some basis integrals, called master integrals. The number of master integrals is always finite~\cite{Smirnov:2010hn, Bitoun:2017nre}. As a further consequence of the IBP relations, the vector of master integrals $I(\uz,\eps)$ satisfies a system of coupled first-order differential equations~\cite{Kotikov:1990kg, Remiddi:1997ny, Gehrmann:1999as}
\beq
\label{eq:DEgen}
\partial_{z_i}I(\uz,\eps) = A_i(\uz,\eps)I(\uz,\eps)\,,
\eeq
where $ A_i(\uz,\eps)$ is a matrix whose entries are rational functions in $\uz$ and $\eps$. 

We will not only consider Feynman integrals, but also their \emph{cuts}, obtained by putting a subset of propagators on-shell. There are various notions of cuts, differing by how the on-shell condition is implemented. For example, one may replace propagators by $\delta$ functions or use a residue prescription, which can, for example, be implemented by integrating along a contour that encircles some of the propagator poles (see ref.~\cite{Britto:2024mna} for a recent review). The precise definitions will be irrelevant for our purposes, because they all share some common properties. In particular, reverse-unitarity~\cite{Anastasiou:2002yz} implies that cut integrals satisfy the same IBP relations and differential equations as their uncut analogues, where we put to zero all integrals where a cut propagator is absent. We will mostly be interested in so-called \emph{maximal cuts}, where all propagators are put on-shell. The maximal cuts of the master integrals are solutions to the homogeneous part of the differential equations in that sector~\cite{Primo:2016ebd, Primo:2017ipr, Frellesvig:2017aai, Bosma:2017ens}. Note that in order to fully specify a maximal cut, we still need to specify the integration contour, and each independent integration contour furnishes an independent solution of the corresponding homogeneous equations~\cite{Bosma:2017ens, Primo:2017ipr}. The precise form of the integration contour for the maximal cut will often be irrelevant, and we suppress the dependence on the contour in the notation and simply denote a maximal cut of the master integral $I_k$ by $\textrm{Cut}(I_k)$. 

Closely connected to maximal cuts are leading singularities~\cite{Cachazo:2008vp}. The latter are typically
defined in an integer fixed number of dimensions $d=d_0$ by taking all residues of the integrand and then integrating the resulting differential form along some contour. As for the maximal cuts, one only needs to consider independent integration contours, which span the corresponding homology group.
There is then a clear similarity between the definition
of maximal cuts in a fixed (integer) number of dimensions
and that of leading singularities, with the former being, in general,
a subset of the latter. In fact, in contrast to
the maximal cut, leading singularities 
carry, in general, information about the subtopologies of the problem.
This distinction is important, especially if one refers to the concept of \emph{local integrals} with \emph{unit leading singularities}, introduced in the literature as a good basis of integrals in the polylogarithmic case~\cite{Arkani-Hamed:2010pyv, Henn:2013pwa, Henn:2020lye}. These integrals were introduced to optimally represent the transcendental part of scattering amplitudes, by rendering manifest the uniform transcendental structure of $\mathcal{N}=4$ super Yang-Mills. For this reason, they are typically selected by analysing the full set of leading singularities beyond the maximal cut.
Note that in cases where we can localise the integrand completely by taking residues, leading singularities evaluate to algebraic functions. For more complicated geometries, it may not be possible to localise the integral completely, and the leading singularities lead to transcendental integrals, cf.,~e.g.,~ref.~\cite{Primo:2017ipr, Bourjaily:2020hjv, Bourjaily:2021vyj}.
Just like in the case of maximal cuts, we will suppress the dependence of the leading singularity on the precise form of the integration contour, and we denote a leading singularity of the master integral $I_k$ by $\textrm{LS}(I_k)$.
In our treatment, we will often focus on the homogeneous
part of the differential equations, and we will therefore
always assume that we have taken a maximal cut.
In this case, since all contributions from subtopologies
are neglected, by construction, we have the relation
\beq
	\textrm{Cut}(I_k) =  \textrm{LS}(I_k) +\mathcal{O}(\eps)\,.
\eeq
This is also what is commonly done in the literature,
where leading singularities are often computed
assuming that all propagators have already been cut.

Returning to the differential equations satisfied by the master integrals~\eqref{eq:DEgen}, there is considerable freedom in how we choose a basis of master integrals. In particular, we may define a new basis via
\beq
J(\uz,\eps) = M(\uz,\eps)I(\uz,\eps)\,.
\eeq
In the new basis, the differential equation reads,
\beq
\partial_{z_i}J(\uz,\eps) = B_i(\uz,\eps)J(\uz,\eps)\,,
\eeq
where $B_i(\uz,\eps)$ is related to $A_i(\uz,\eps)$ by a gauge transformation,
\beq
B_i(\uz,\eps) = \left[M(\uz,\eps)A_i(\uz,\eps) + \partial_{z_i}M(\uz,\eps) \right] M(\uz,\eps)^{-1}\,.
\eeq

In general, the system of differential equations satisfied by the master integrals is highly coupled, and a judicious choice of basis may dramatically simplify the task of finding a solution. In ref.~\cite{Henn:2013pwa}, it was proposed that a convenient basis is the so-called \emph{canonical basis}, in which the dependence on the dimensional regulator factorises from the differential equation matrix,
\beq
\rd J(\uz,\eps) = \eps\,\mathcal{B}(\uz)J(\uz,\eps)\,,
\eeq
where we introduced the total differential $\rd = \sum_i\rd z_i\,\partial_{z_i}$.
The advantage of such a system in $\eps$-form is that the equations are decoupled order by order in $\eps$, which is sufficient to find the solution as a Laurent expansion in $\eps$ in terms of iterated integrals. However, the canonical form of the differential equations imposes further constraints on the form of the differential equations, which go beyond the factorisation of $\eps$. If the Feynman integrals evaluate to multiple polylogarithms, the canonical form is well understood: The matrix $\mathcal{B}(\uz)$ is a linear combination of dlog-forms, and the resulting iterated integrals are pure functions of uniform transcendental weight~\cite{Arkani-Hamed:2010pyv}. There are several methods to find a canonical form in the case of dlog-forms, many of which are based on studying the complete set of leading singularities of the integrals beyond the maximal cut.

In cases that go beyond MPLs and dlog-forms, there is no commonly accepted definition of a canonical basis, and various approaches and bases have been proposed.
In particular, there are different ways to define $\eps$-factorised bases, and experience over the past years has shown that some $\eps$-forms are better than others (cf., e.g., the discussion in ref.~\cite{Frellesvig:2023iwr}). Nevertheless, over the last couple of years, there has been a considerable number of examples of $\eps$-factorised differential equations beyond polylogarithms that deserve to be called in \emph{canonical form}. In particular, in ref.~\cite{Gorges:2023zgv}, a method for finding canonical bases for Feynman integrals beyond multiple polylogarithms has been introduced, based on earlier ideas in refs.~\cite{Brown:2015ylf, Broedel:2018qkq}.
One of the goals of this paper is to show how this method can be employed for types of geometries that had not been studied in detail in ref.~\cite{Gorges:2023zgv}. Before considering these more complicated examples, we describe in detail how the method of ref.~\cite{Gorges:2023zgv} works on a simple example, stressing especially the similarities and differences between the polylogarithmic and the elliptic cases.

\section{A motivational example}
\label{main}

In this section, we discuss a simple example of a Feynman integral that is related to a family of elliptic curves. It illustrates the features that will be encountered in later sections and highlights the importance of choosing a good starting basis of master integrals to derive a system of differential equations in canonical form. We stress that some ideas from this section have already been applied in the literature in one form or another, especially in ref.~\cite{Gorges:2023zgv} and subsequent works~\cite{Duhr:2024bzt, Forner:2024ojj, Duhr:2024uid, Becchetti:2025rrz, Becchetti:2025oyb}, but we feel that it is important to summarise them here in a coherent manner to motivate how they extend to more general geometries.

\begin{figure}[!h]
\centering
\begin{tikzpicture}
\coordinate (llinks) at (-2.5,0);
\coordinate (rrechts) at (2.5,0);
\coordinate (links) at (-1.5,0);
\coordinate (rechts) at (1.5,0);
\begin{scope}[very thick,decoration={
    markings,
    mark=at position 0.5 with {\arrow{>}}}
    ] 
\draw [-, thick,postaction={decorate}] (rechts) to [bend right=0]  (links);
\draw [-, thin,postaction={decorate}] (links) to [bend left=85]  (rechts);
\draw [-, thick,postaction={decorate}] (llinks) to [bend right=0]  (links);
\draw [-, thick,postaction={decorate}] (rechts) to [bend right=0]  (rrechts);
\end{scope}
\begin{scope}[very thick,decoration={
    markings,
    mark=at position 0.5 with {\arrow{<}}}
    ]
\draw [-, thick,postaction={decorate}] (links) to  [bend right=85] (rechts);
\end{scope}
\node (d1) at (0,1.1) [font=\scriptsize, text width=.2 cm]{$k_2$};
\node (d2) at (0,0.25) [font=\scriptsize, text width=.2 cm]{$k_1$};
\node (d3) at (0.75,-0.6) [font=\scriptsize, text width=3 cm]{$k_1-k_2-p$};
\node (p1) at (-2.0,.25) [font=\scriptsize, text width=1 cm]{$p$};
\node (p2) at (2.4,.25) [font=\scriptsize, text width=1 cm]{$p$};
\end{tikzpicture}
\caption{The two-loop sunrise graph with two different masses.}
\label{fig:sunrise}
\end{figure}
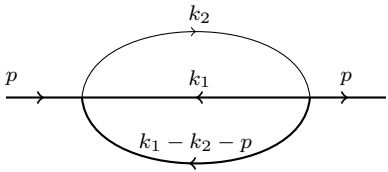

We consider the sunrise family of Feynman integrals (see~\cref{fig:sunrise}), 
defined by
\begin{equation}
\label{def:sunrise}
I_{\nu_1,\dots,\nu_{5}}(\uz;d) = \int \left(\prod_{j=1}^2\frac{\mathrm d^dk_j}{i\pi^{d/2}}\right)\frac{(k_1 \cdot p)^{-\nu_{4}}(k_2 \cdot p)^{-\nu_{5}}}{(k_1^2-m^2)^{\nu_1}(k_2^2-M^2)^{\nu_2}((k_1-k_2-p)^2-m^2)^{\nu_3}}\, ,
\end{equation}
where the set of kinematical variables is $\uz = \{\frac{m^2}{s},\frac{M^2}{s}\}$ and we have singled out $s$ as the only dimensionful scale. 
We will study this example in the two cases $M^2 \neq 0$ and $M^2 = 0$ (and we always assume $m^2\neq0$). 
It is well known (cf.~refs.~\cite{Sabry, Broadhurst:1987ei, Bauberger:1994by, Bauberger:1994hx, Laporta:2004rb, Bloch:2013tra, Adams:2013nia, Remiddi:2013joa, Adams:2014vja, Adams:2015gva, Adams:2015ydq, Remiddi:2016gno, Adams:2016xah, Remiddi:2017har, Hidding:2017jkk}) that whenever all three propagator masses are non-zero, 
this integral involves functions related to an elliptic curve, while 
the integral is of polylogarithmic type for $M^2=0$. As we will argue, a good choice of master integrals is intimately connected to the underlying geometry. This idea is also at the heart of many
leading-singularity-based methods for finding a canonical basis of master integrals that evaluate to polylogarithms (cf.,~e.g.,~refs.~\cite{Henn:2020lye, Wasser:2022kwg}). Extensions of the notion of leading singularities to elliptic geometries have been presented in refs.~\cite{Bourjaily:2020hjv, Bourjaily:2021vyj}. When applying this
approach to more complicated geometries, one encounters new features, which require an extension of
the leading-singularity analysis. One of the goals of this section is to identify from examples what these extensions are. 

As a starting point, we use the Baikov representation~\cite{Baikov:1996iu, Baikov:1996rk, Frellesvig:2017aai, Frellesvig:2024ymq} to compute maximal cuts and leading singularities. 
We work close to $d_0=2$ dimensions, which is the most natural (even) 
number of dimensions to analyse this integral.
The maximal cuts allow us to obtain information on the homogeneous
part of the differential equations satisfied by our integrals~\cite{Primo:2016ebd, Primo:2017ipr,Frellesvig:2017aai,Bosma:2017ens}. 
In the case of the sunrise family, this is the only non-trivial part, because all subtopologies are products of one-loop tadpole integrals.
Let us start by considering the corner-integral, $I_1 = I_{1,1,1,0,0}(\underline z;d)$.
Either from a loop-by-loop approach~\cite{Primo:2016ebd,Frellesvig:2017aai} or by parametrising both
loops at once and taking a further residue, we see that the leading singularities of $I_1$ in $d=2$ dimensions are given by a one-fold integral,
\begin{equation}\bsp
{\rm LS}\left( I_1 \right) &\,\propto \oint \frac{\mathrm dx_5}{\sqrt{M^2+s+2 x_5} \sqrt{M^2 s-x_5^2}
   \sqrt{4 m^2-M^2-s-2 x_5}}= \oint \frac{\mathrm dx_5}{\sqrt{P_4(x_5)}}\,,
   \label{eq:cutsunrise}
\esp\end{equation}
where the $\propto$ sign indicates that we are working modulo overall numerical normalisations.
We use the integral sign without specifying the contour since each independent integration contour provides an independent solution of the associated homogeneous equation~\cite{Bosma:2017ens, Primo:2017ipr} in two dimensions. 
In the equation above, we also defined the quartic polynomial 
\beq\bsp
P_4(x_5) &= (M^2+s+2 x_5)(M^2 s-x_5^2)(4 m^2-M^2-s-2 x_5)\,, \\ 
&= 4(x_5-a_1)(x_5-a_2)(x_5-a_3)(x_5-a_4)\,, \label{eq:P4}
\esp\eeq
where the four roots are given by
\begin{align}
    a_1 = - \sqrt{M^2 s}\,, \quad a_2 = - \frac{M^2+s}{2}
    \,, \quad a_3 = \frac{4m^2-M^2-s}{2}\,, \quad a_4 = + \sqrt{M^2 s}\,.
\label{eq:rootsquar}
\end{align}
For arbitrary non-zero values of $M^2$,
the denominator is the square root of a quartic polynomial, which defines an elliptic curve:
\begin{align}
\mathcal E:\quad y^2 = P_4(x_5) =  (M^2+s+2 x_5)(M^2 s-x_5^2)(4 m^2-M^2-s-2 x_5)   \,. \label{eq:Esunset}
\end{align}
The integrand in eq.~\eqref{eq:cutsunrise} has no poles and four branch points at the four solutions of the equation $P_4(x_5)=0$. Thus, as long as the masses are non-zero, the geometry associated to the sunrise integral is the family of elliptic curves defined by eq.~\eqref{eq:Esunset}. If $M^2=0$, however, the elliptic curve may degenerate. As we will now discuss, a good choice of master integrals takes into account the geometry. We therefore discuss the two cases $M^2 =0$ and $M^2 \neq 0$ separately.

\subsection{The polylogarithmic case: $M^2 = 0$}
From eq.~\eqref{eq:Esunset}, we see that for $M^2=0$ two of the branch points coincide, and the elliptic curve degenerates into two copies of the Riemann sphere connected by the branch cut associated with the
remaining square root. The final geometry is then topologically 
equivalent to a single Riemann sphere. While it is well-known how to find a system of differential equations in canonical form for this case, we discuss it in some detail because it allows us to highlight the differences to the elliptic case later on.

\begin{figure}[h]
    \centering
    \includegraphics[width=0.8\textwidth]{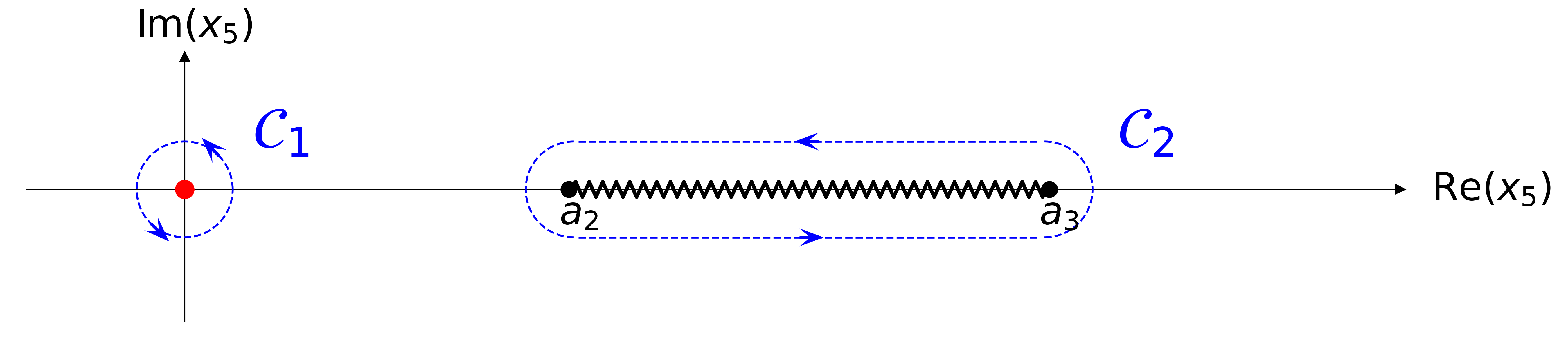}
    \caption{The contours for the two-mass sunrise integral.}
    \label{fig:contours}
\end{figure}

By solving the associated IBPs, we see that there are two independent master integrals,
which we can choose as
\beq
I_1 = I_{1,1,1,0,0}(\uz; d) \textrm{~~~and~~~} I_2 = I_{1,1,1,0,-1}(\uz; d)\,.
\eeq
Assuming $m^2>0$ and $s<0$ for definiteness,
we see from eq.~\eqref{eq:cutsunrise} that the leading singularities of the first integral in $d=2$ take the form
\begin{equation}
{\rm LS}\left( I_1 \right)\Big|_{M^2=0} \propto \oint \frac{\mathrm dx_5}{x_5\, \sqrt{s+2 x_5} 
   \sqrt{4 m^2-s-2 x_5}}\,.
   \label{eq:cutsunrise2}
\end{equation}
The integrand of eq.~\eqref{eq:cutsunrise2} has a single pole at $x_5=0$, and a branch cut between the two roots $x_{5,1} = -s/2$ and $x_{5,2} = (4 m^2-s)/2$. Correspondingly, we can consider two independent integration contours: $\mathcal C_1$ is a small circle around the single pole, and $\mathcal C_2$ encircles the branch cut (see~\cref{fig:contours}). 
Since there is no pole at infinity, these two cycles cannot be independent. 
One easily finds that, modulo overall prefactors and with $s\to s+i0$,
\begin{equation}
\begin{aligned}
  \oint_{\mathcal C_1}\frac{\mathrm dx_5}{x_5\, \sqrt{s+2 x_5} 
   \sqrt{4 m^2-s-2 x_5}}&\propto
   \oint_{\mathcal C_2}\frac{\mathrm dx_5}{x_5\, \sqrt{s+2 x_5} 
   \sqrt{4 m^2-s-2 x_5}} 
   \propto\frac{2\pi}{\sqrt{s(s-4m^2)}}\,.
\end{aligned}
\end{equation}
We see that we can completely localise the integrand of $I_1$ by taking residues. Alternatively, we could also rationalise the square root. The result is a rational integrand with two poles, whose residues are equal and opposite due to the Global Residue Theorem.

Let us repeat the same analysis for $I_2$. To obtain its integrand, we just need to
multiply the integrand of $I_1$ by $(k \cdot p) = x_5$. This gives
\begin{equation}
{\rm LS}\left( I_2 \right)\Big|_{M^2=0} \propto \oint \frac{\mathrm dx_5\,}{ \sqrt{s+2 x_5} 
   \sqrt{4 m^2-s-2 x_5}}\,.
   \label{eq:cutsunrise2a}
\end{equation}
The integrand now has a single pole at $x_5 = \infty$ in addition to the branch cut. Again, the residue at the simple pole is related to the integral over $\mathcal{C}_2$, and we find
\begin{align}
\oint_{\mathcal{C}_2} \frac{\mathrm dx_5}{ \sqrt{s+2 x_5} 
   \sqrt{4 m^2-s-2 x_5}} \propto i\pi
   \,. 
\end{align}

Let us interpret these results. In $d=2$ dimensions, the integrands of the maximal cuts of $I_1$ and $I_2$ only have simple poles and are associated with two different algebraic leading singularities. In other words, our integrand analysis on the maximal cut provides a justification for why we have chosen $I_1$ and $I_2$ as a basis of master integrals, rather than a non-trivial linear combination of them.
Being algebraic, the leading singularities satisfy two first-order linear differential equations with rational coefficients. This means that their homogeneous differential equations are expected to 
decouple in the limit $d=2$. Finally, after appropriate normalisation by their leading singularities, both integrals will have a constant leading singularity on the maximal cut.
By direct calculation, one can in fact easily see that 
\begin{equation}
    J_1 = \sqrt{s(s-4m^2)} I_1 \,, \qquad J_2 = I_2\,,
\end{equation}
satisfy the following differential equation in $s$,
\begin{align} \label{eq:caneqmassless}
   \partial_s\! \begin{pmatrix}
       \textrm{Cut}(J_1) \\  \textrm{Cut}(J_2)
   \end{pmatrix}  =
(d-2)   
   \left(
\begin{array}{cc}
 \frac{2 (s-m^2)}{s (s-4 m^2)} & \frac{3}{\sqrt{s(s-4m^2)}} \\
 -\frac{1}{2 \sqrt{s(s-4m^2)}} & -\frac{1}{2 s} \\
\end{array}
\right)\begin{pmatrix}
        \textrm{Cut}(J_1) \\  \textrm{Cut}(J_2)
   \end{pmatrix} \,.
\end{align} 
Clearly, the homogeneous differential equation is in $\epsilon$-factorised form for $d=2-2\eps$. Moreover, the entries of the differential equation matrix can easily be identified as dlog-forms. This, in turn, is the hallmark of the canonical differential equation in the polylogarithmic case! It implies that the coefficient of $\eps^k$ in the Laurent expansion around $d=2$ will only involve iterated integrals of dlog-forms of length $k$. The latter are \emph{pure functions of weight $k$} in the sense of ref.~\cite{Arkani-Hamed:2010pyv}, and this notion is closely connected (and in this case, in fact, identical) to the notion of \emph{transcendental weight} introduced in the physics literature~\cite{Kotikov:2010gf}.

\subsection{The elliptic case: $M^2 \neq 0$}
\label{subsec:ellipticsunrise}
Let us now see how the previous analysis changes if $M^2 \neq 0$. At variance with the previous case, solving IBPs now exposes three independent master integrals, which fulfil a set of three coupled differential equations.
A possible choice is 
\beq\bsp \label{eq:elliptic_sunrise_basis}
I_1 = I_{1,1,1,0,0}( \uz; d)\,, \qquad I_2 = I_{1,1,1,0,-1}( \uz; d)\,,\qquad
I_3 = I_{1,1,1,0,-2}( \uz; d)\,.
\esp\eeq
Note that $I_3$ becomes linearly dependent for $M^2=0$.
As before, let us study their leading singularities in $d=2$ dimensions. We can write
\begin{equation}
{\rm LS}\left( I_r \right) \propto \oint \mathrm dx_5 \frac{x_5^{r-1}}{\sqrt{P_4(x_5)}}\,,  \qquad r=1,2,3\,,
\end{equation}
with $P_4$ defined in eq.~\eqref{eq:P4}. At this point, there is no reason why we should pick the basis in eq.~\eqref{eq:elliptic_sunrise_basis}, and we could, in fact, have chosen any other basis. In particular, we could have picked any linear combination of the three integrals in eq.~\eqref{eq:elliptic_sunrise_basis}. In the polylogarithmic case $M^2=0$,  based on our integrand analysis, we could identify a distinguished basis, where each basis integrand has only simple poles and  
is associated with a different leading singularity. Let us see how the situation changes in the elliptic case and how an integrand analysis can still help us to identify a distinguished basis.

\subsubsection{Integrand analysis on the maximal cut}
For $r=1$ we reproduce eq.~\eqref{eq:cutsunrise}. In fact, the integrand in eq.~\eqref{eq:cutsunrise} defines a holomorphic differential form on the elliptic curve $\mathcal{E}$ in eq.~\eqref{eq:Esunset}, and this holomorphic differential is unique up to normalisation. We see that the geometry immediately furnishes a distinguished master integral, namely the one that for $\eps=0$ corresponds to the holomorphic differential on the elliptic curve. This motivates our choice of $I_1$. Due to the existence of four distinct branch points, there are two independent choices of cycles, $\mathcal{C}_1$ and $\mathcal{C}_2$, on $\mathcal{E}$,\footnote{A standard choice is the contour encircling the branch cut between $a_1$ and $a_2$ and the contour encircling the branch points $a_2$ and $a_3$.} so that we can compute \emph{two different} leading singularities for $I_1$ in $d=2$ dimensions:
\begin{equation}
\pi_0 = \oint_{\mathcal{C}_1}\frac{\mathrm d x_5}{\sqrt{P_4(x_5)}}\,, \qquad
\pi_1 = \oint_{\mathcal{C}_2} \frac{\mathrm d x_5}{\sqrt{P_4(x_5)}}\,.  \label{eq:persun}
\end{equation}
These integrals may be evaluated in terms of the complete elliptic integral of the first kind,
\beq
\K(\lambda) = \int_0^1\frac{\rd x}{\sqrt{(1-x^2)(1-\lambda \, x^2)}}\,.
\eeq
We stress that, unlike for $M^2=0$ where the integrals over the two cycles were equal up to normalisation, these two integrals define distinct functions in the elliptic case. In fact, $\pi_0$ and $\pi_1$ define two independent periods of the elliptic curve $\mathcal{E}$, or equivalently, they are two independent solutions
of the second-order Picard-Fuchs differential operator attached to this family of elliptic curves (see below).

Let us now consider ${\rm LS}\left( I_2 \right)$. 
With the transformation $x_5 \to 1/y_5$, we get
\begin{align}
{\rm LS}\left( I_2 \right) &\propto \oint \frac{ \mathrm dy_5}{y_5\, \sqrt{4(1-y_5a_1)(1-y_5a_2)(1-y_5a_3)(1-y_5a_4)}} 
\nonumber \\
&\sim \oint \mathrm dy_5 \left[ \frac{ 1}{2y_5 } - \frac{s+M^2-2m^2}{4} + \mathcal{O}(y_5)\right] \,.
\label{eq:cutI2}
\end{align}
This exposes a single pole at $x_5 = \infty$ (or $y_5 = 0$\,). We may correspondingly consider a new independent contour $\mathcal{C}_3$, which encircles the pole at $x_5=\infty$. When evaluated on the contours $\mathcal{C}_1$ or $\mathcal{C}_2$, we obtain an elliptic integral of the third kind, which fulfils a first-order
inhomogeneous differential equation with elliptic integrals of first and second kind in the inhomogeneous term. As the residue at $x_5=\infty$ is already constant, this
analysis singles out a second master integral.

Finally, consider the leading singularities of the third master integral. 
With the same change of variables as for $I_2$, we find 
\begin{align}
{\rm LS}\left( I_3 \right) \propto \oint  \mathrm dy_5 &\left[ \frac{1}{ 2y_5^2} - \frac{s+M^2-2m^2}{4 y_5}  \,
+ \mathcal{O}(y_5^0)\right] \,. \label{eq:cutI3}
\end{align}
The integrand has both a double and a simple pole at $x_5 = \infty$. 
The simple pole is proportional to the first symmetric polynomial
of the four roots $S_1 = a_1+a_2+a_3+a_4 = s+M^2-2m^2$, and so we may add the appropriate multiple of $I_2$ to cancel the simple pole. The resulting integral can then be evaluated solely in terms of complete elliptic integrals of the first and second kind and, therefore, provides a third candidate master integral.

The previous discussion leads us to consider the following new basis of master integrals:
\begin{align}
{I}_1^{\textrm{new}} = I_1\,, \qquad {I}_2^{\textrm{new}} = I_3 + \frac{S_1}{2} \, I_2 \,, \qquad
     I_3^{\textrm{new}} = I_2 \,.
\end{align}
This basis provides a direct correspondence between our basis and abelian differentials of the first, second and third kind on $\mathcal{E}$.
Indeed, we find that in this basis, ${I}_3^{\textrm{new}}$ decouples from the differential equations for the first two integrals on the maximal cut at $\eps=0$.
The latter, instead,  fulfil a coupled system of 
two differential equations, which is a standard result of the theory of elliptic integrals.

Let us try to make a connection with what is usually done with elliptic Feynman integrals (and beyond). Typically,  instead of choosing ${I}_2^{\textrm{new}}$ as described above, one directly chooses a derivative of $I_1$ as the second master integral.
We can now see why this is a convenient choice from the point of view of an integrand analysis. If the first master integral has been chosen appropriately, at least on the maximal cut at $\eps=0$, its derivative can be expressed as a linear combination of first- and second-kind integrals, without any contamination from third-kind integrals. As we will see later on, this is a very general feature of geometrical origin and can be traced back to Griffiths transversality. Importantly, when chosen in this way, the master integrals that correspond to 
integrals of the third kind naturally decouple from those of the first and second kind.

Let us summarise our findings. The fact that the maximal cut is related to the family of elliptic curves in eq.~\eqref{eq:Esunset} provides a natural way to choose a distinguished basis of master integrals by relating them to abelian differentials of the first, second and third kind on $\mathcal{E}$. While in our example we have explicitly constructed the linear combination that corresponds to the differential of the second kind, we have argued that we could also just have picked the derivative of ${I}_1$. Once the appropriate differential forms of the first and third kind have all been identified, this avoids having to perform a further integral analysis to identify the form of the second kind. From now on we will work with this special choice, and we choose as the basis of master integrals
\beq
\widetilde{I}_1 = I_1 =  I_{1,1,1,0,0}(\underline z; d)\,, \qquad \widetilde{I}_2 = \partial_z\widetilde{I}_1\,,\qquad
\widetilde{I}_3 =I_2= I_{1,1,1,0,-1}(\underline z; d)\,.
\eeq
We stress an important point: While our integrals depend on two dimensionless kinematic ratios, we can pick either of them to define the differential of the second kind (or even any linear combination of derivatives). To make this freedom explicit, we therefore indicate the derivative acting on $\widetilde{I}_1$ as $\partial_z$, where $z$  can be any kinematic variable. 
Finally, while we were guided by the underlying geometry to identify our distinguished basis, this basis is not unique. Indeed, while the holomorphic differential on an elliptic curve is unique up to normalisation, a differential of the second kind is only defined up to normalisation and adding a multiple of the holomorphic differential. Similarly, a differential of the third kind is only defined up to normalization\footnote{In our case, the normalization of $\widetilde{I}_3$ is partially fixed by requiring the residue to be constant.} and adding a linear combination of the differentials of the first and second kind. Another way of stating this is as follows: Let $V_{3} = \langle \widetilde{I}_1, \widetilde{I}_2,\widetilde{I}_3\rangle$ be the vector space generated by the three master integrals. If we define the subspaces
\beq\label{eq:V1V2_def}
V_1 = \langle \widetilde{I}_1\rangle\,,\qquad V_{2} = \langle \widetilde{I}_1, \widetilde{I}_2\rangle\,,
\eeq
then this defines a \emph{filtration} of $V_3$
\beq\label{eq:filtration_ell}
0\subset V_1 \subset V_2 \subset V_3\,.
\eeq
Different distinguished bases will then always be such that they respect this filtration. 

\subsubsection{Canonical differential equations in the elliptic case}

Let us now consider the system of differential equations satisfied by the basis of master integrals singled out in the previous section. While this is a two-variable problem, for the following discussion, we will focus on a single-variable slice and study the differential equations in a variable that we refer to generically as $z$. We stress that the integrand analysis in the previous section does not single out any special kinematical variable. For this reason, as long as the periods of the geometry depend on all kinematical invariants in the problem, we expect that performing the remaining part of our construction in one variable will also fix the non-trivial part of the rotation necessary to get to a full canonical basis in all remaining variables, modulo simple extra rotations that only depend on the remaining variables. We will see that this is indeed the case in the problem considered here.
We can act with $\partial_z$ on the vector of master integrals. Focussing on the homogeneous part, we obtain
\begin{align}\label{eq:DEeps3x3a}
    \partial_z\!\begin{pmatrix} \mathrm{Cut}(\widetilde{I_1}) \\ \mathrm{Cut}(\widetilde{I_2}) \\ \mathrm{Cut}(\widetilde{I_3}) \end{pmatrix}
    = \left[ A(\underline z) + \epsilon \, B(\underline z) + \epsilon^2 \, C(\underline z) \right]
    \!\begin{pmatrix} \mathrm{Cut}(\widetilde{I_1}) \\ \mathrm{Cut}(\widetilde{I_2}) \\ \mathrm{Cut}(\widetilde{I_3}) \end{pmatrix} \, , 
\end{align}
with
\begin{equation}\bsp
\label{eq:DEeps3x3}
    A(\underline z) = \begin{pmatrix}
    0 & 1 & 0 \\
    a_{21}(\underline z) & a_{22}(\underline z) & 0 \\
    a_{31}(\underline z) & a_{32}(\underline z) & 0
    \end{pmatrix}  , \
    B(\underline z) =  \begin{pmatrix}
    0 & 0 & 0 \\
    b_{21}(\underline z) & b_{22}(\underline z) & b_{23}(\underline z)  \\
    b_{31}(\underline z) & 0 & b_{33}(\underline z)
    \end{pmatrix} , \
    C(\underline z) = \begin{pmatrix}
    0 & 0 & 0 \\
    c_{21}(\underline z) & 0 & c_{23}(\underline z)  \\
    0 & 0 & 0
    \end{pmatrix} \, .
\esp\end{equation}
For instance, for the particular choice $z=\frac{m^2}{s}$, 
and putting $s=1$ for ease of typing, we have

\begin{alignat}{3}
    a_{21}(\underline z) =& \ \frac{2 (1-6 m^2+M^2)}{m^2 
    R} \, , 
    \quad && 
    a_{21}(\underline z) = -\frac{48 m^4-16 m^2
   (M^2+1)+(M^2-1)^2 }{m^2 R} \, , 
   \nonumber \\
    a_{31}(\underline z) =& \ 1 \, ,
    \quad && 
    a_{32}(\underline z) = \ \frac{1}{2} (4 m^2-M^2-1) \, , 
    \nonumber \\
    b_{21}(\underline z) =& \ \frac{6 M^2-56 m^2 + 6}{m^2 R} \, ,
    \quad &&
    b_{22}(\underline z) = \ -\frac{80 m^4-24 m^2
   (M^2+1)+(M^2-1)^2}{m^2 R} \, , \nonumber \\
    b_{23}(\underline z) =& \ -\frac{12}{m^2 R} \, , 
    \quad &&
    b_{31}(\underline z) = \ 2 \, , 
    \qquad  
    b_{33}(\underline z) = \ 0 \, , \nonumber \\
    c_{21}(\underline z) =& \ \frac{4 (1-16 m^2+M^2)}{m^2 R} \, ,
    \quad &&
    c_{23}(\underline z) = \ -\frac{24}{m^2 R} \, .
    \label{eq:sunriseentriesexplicit}
\end{alignat}
with $R = 16 m^4-8 m^2 \left(M^2+1\right)+\left(M^2-1\right)^2$.
We stress that the peculiar structure of the higher-order terms in $\epsilon$ follows purely from a direct computation and we will come back to this point in section~\ref{sec:general_eps_fac}. 
As a first important feature, we notice that the matrix $A(\underline z)$ has a third column of zeros, which is a direct consequence of having chosen for $\widetilde{I}_3$ a candidate whose maximal cut has a single pole at infinity with residue normalised to a constant, as discussed above. Or, in other words, our choice of basis respects the filtration in eq.~\eqref{eq:filtration_ell}.

We now discuss how (and why)
the algorithm of ref.~\cite{Gorges:2023zgv} leads to a system of $\eps$-factorised differential equations that can be called \emph{canonical}.
However, in doing so, we will keep the discussion general for any system of the form~\eqref{eq:DEeps3x3a} and not resort to the particular functional form of~\cref{eq:DEeps3x3} for our example of the sunrise to showcase how the same strategy may be applied to similar problems.

As explained above, $\mathrm{Cut}(\widetilde{I_3})$ decouples for $\eps=0$. We, therefore, separate the discussion into two and first concentrate on the subsystem formed by formed by $\big( \mathrm{Cut}(\widetilde{I_1}), \mathrm{Cut}(\widetilde{I_2})\big)^T$:
\begin{align}\label{eq:DEQ_sunrise_cut}
    \partial_z \begin{pmatrix} \mathrm{Cut}(\widetilde{I_1}) \\\mathrm{Cut}(\widetilde{I_2}) \end{pmatrix}
    = \left[ \widehat{A}(\underline{z}) + \mathcal{O}(\epsilon)
    \right]
    \begin{pmatrix} \mathrm{Cut}(\widetilde{I_1})\\ \mathrm{Cut}(\widetilde{I_2})  \end{pmatrix}\, , 
    \textrm{~~with~~}
    \widehat{A}(\underline z) = \begin{pmatrix}
    0 & 1  \\
    a_{21}(\underline z) & a_{22}(\underline z) 
    \end{pmatrix}\,.
\end{align}
The functions $a_{21}(\underline{z})$ and $a_{22}(\underline{z})$ are directly related to the Picard-Fuchs equation in $z$ of the underlying elliptic geometry,
\begin{equation}
    \left[ \partial_z^2 -a_{22}(\underline z) \partial_z  -a_{11}(\underline z) \right]\pi_i = 0 \, , \qquad i=0,1 \, .\label{eq:sunrise_PF}
\end{equation}
In the polylogarithmic case, the hallmark of a system in the canonical form is the appearance of dlog-forms, which is tightly connected to the concepts of pure functions and transcendental weight. We thus expect that these concepts also play an important role in identifying a canonical form for our elliptic example. As a starting point, we therefore focus on a somewhat simpler problem, namely how to interpret the transcendental weight for $\eps=0$. 

\paragraph{The splitting of the Wronskian and the transcendental weight.}
Up to this point, one could, at least in principle, have defined the periods $\pi_i$ globally 
through the integral representation for the periods in eq.~\eqref{eq:persun}.
In the following, it will be more useful to consider the local behaviour of these solutions. 
To simplify the exposition, and in view of the application of our method to one-parameter cases in later sections, we consider here the dependence of the integral on one single variable, but we stress that all these considerations can easily be extended to a multi-parameter problem.
With $z$ any such kinematic variables,~\cref{eq:sunrise_PF} is just an ordinary second-order differential equation in $z$.
The Frobenius method then guarantees that close to a
regular singular point $z_0$, the solutions of a 
Fuchsian linear differential equation
with rational coefficients 
can always be found in terms of generalised power series close to $z=z_0$, i.e., linear combinations of power series multiplied by non-integer powers of $z-z_0$ and/or logarithms. It follows from a theorem by Landman~\cite{landman} that the periods of a family of $n$-dimensional varieties can diverge at most like the $n^{\textrm{th}}$ power of a logarithm. 
Note that this does not imply that all powers of logarithms arise at every singular point. A regular singular point where periods diverge like the $n^{\textrm{th}}$ power of a logarithm is called a 
point of Maximal Unipotent Monodromy (MUM). For our case $n=1$, this is indeed the case for all regular singular points. Without loss of generality, we assume that  $z=z_0=0$ is a MUM-point, and so there is a basis of solutions of the Picard-Fuchs equation~\eqref{eq:sunrise_PF} of the form
\beq
 \label{eq:ForbeniusElliptic}
 \varpi_0(z) = 1 + \sum_{j=1}^\infty c_j z^j \textrm{~~~and~~~}  \varpi_1(z) = \varpi_0(z) \log(z) + \sum_{j=1}^\infty d_j z^j \,,
\eeq
where the $c_j$ and $d_j$ are complex numbers (typically, they will be rational numbers for the cases we are considering). This basis of solutions is commonly referred to as a Frobenius basis. The two periods $\pi_0$ and $\pi_1$ are then linear combinations (with complex coefficients) of $ \varpi_0$ and $\varpi_1$, and we may choose a fundamental solution matrix (or Wronskian) $W$ of the subsystem at $\eps=0$ formed by $\big( \mathrm{LS}(\widetilde{I_1}), \mathrm{LS}(\widetilde{I_2})\big)^T$ as
\begin{align} \label{eq:Wrdef}
    W = \begin{pmatrix} \varpi_0 & \varpi_1 \\ \partial_z \varpi_0 & \partial_z \varpi_1\end{pmatrix}\,, \qquad 
    \mbox{with} \qquad
    \partial_z W = \begin{pmatrix}
    0 & 1  \\
    a_{21}(\underline z) & a_{22}(\underline z)
    \end{pmatrix}  W\,.
    \end{align}
Let us now discuss how we can assign a notion of transcendental weight to the entries of $W$.
For geometries beyond those that give rise to dlog-forms, there is no commonly accepted definition of the notion of transcendental weight. Here we rely on the extension of the notions of pure functions and transcendental weight to elliptic geometries introduced in ref.~\cite{Broedel:2018qkq}. In a nutshell, ref.~\cite{Broedel:2018qkq} starts from the observation that \emph{locally close to the MUM-point}, $\varpi_0$ is a power series, so it is natural to assign transcendental weight 0 to it. The same reasoning applies to $\partial_z\varpi_0$ as the derivative of a power series (which is itself a power series). 
Further, according to the definition in ref.~\cite{Broedel:2018qkq}, the ratio
\beq\label{eq:tau_def}
\tau(z) = \frac{\varpi_1(z)}{\varpi_0(z)} = \log(z) + \ord(z)\,,
\eeq
is a pure function of transcendental weight 1 (close to the MUM-point at $z=0$),\footnote{In ref.~\cite{Broedel:2018qkq}, $\tau(z)$ is assigned weight 0, because it is conventionally defined as $\tau(z) = \frac{\varpi_1(z)}{2\pi i \varpi_0(z)} = \frac{\log(z)}{2\pi i} + \ord(z)$.} because it behaves locally as $\log(z)$ and satisfies a unipotent differential equation. As a consequence, the solution $\varpi_1$ is not a pure function because, close to the MUM-point, it is the product of $\varpi_0$ and $\tau(z)$. Moreover, $\varpi_0$ is not pure as it does not satisfy a unipotent differential equation.
We stress that this assignment of the transcendental weight and the associated notions of pure functions are only valid locally, and they change if we choose a different MUM-point.

At this point, we have to address two issues. First, while the entries of the first column of $W$ both have the same weight 0, this is not the case for the first row, where $\varpi_1 = \varpi_0\tau$ entails a mixture of weights with $\tau$ having transcendental weight $1$. Second, the assignment of a transcendental weight to $\partial_z\varpi_1$ is subtle, because the latter is not independent of the other three entries of $W$. Indeed, it is well-known that the determinant $\Delta = \det W$ of the Wronskian is always an algebraic function, and thus has transcendental weight 0. We thus see that the second column contains a mixture of transcendental weights. Since we expect systems in canonical form to be closely related to pure functions, which in turn should have uniform transcendental weight, we must find a way to disentangle this mixture of transcendental weights and expose the pure part. 

The solution advocated in refs.~\cite{Broedel:2018qkq, Gorges:2023zgv} is to split the Wronskian matrix into its \textit{semi-simple} and its \textit{unipotent} parts.
Note that this splitting is not unique, but we can make it unique by requiring the unipotent part to be an upper unitriangular matrix, i.e., an upper triangular matrix with 1's on the diagonal.
For the case at hand, this splitting takes the form
\begin{equation}
\label{eq:sesiunipotsplitting}
  \underbrace{  \begin{pmatrix}
        \varpi_0 & \varpi_1  \\
        \partial_z \varpi_0 & \partial_z\varpi_1
    \end{pmatrix}}_{W} =   
      \underbrace{  \begin{pmatrix}
        \varpi_0 & 0 \\
        \partial_z \varpi_0 & \frac{\Delta}{ \varpi_0}
    \end{pmatrix}}_{W^\text{ss}} 
    \underbrace{\begin{pmatrix}
        1 & \tau \\
        0 & 1
    \end{pmatrix} }_{W^\text{u}} \,.
\end{equation}
By inspection, it is easy to see that $W^\text{u}$ satisfies the unipotent differential equation
\begin{equation}
    \partial_z W^\text{u} = \begin{pmatrix}
        0 & \partial_z\tau \\
        0 & 0
    \end{pmatrix} W^\text{u} = 
    \begin{pmatrix}
        0 & \frac{\Delta}{\varpi_0^2} \\
        0 & 0
    \end{pmatrix} W^\text{u}\,.
    \label{eq:diffeqWu}
\end{equation}

As a result of the splitting in eq.~\eqref{eq:sesiunipotsplitting}, all entries of  $W^\text{ss}$ and  $W^\text{u}$ have a well-defined transcendental weight (which, we reiterate, is only defined locally at the MUM-point). In particular, all entries of $W^\text{ss}$ have weight 0, and they can be interpreted as the generalisation of the algebraic leading singularities in the polylogarithmic case. 
The unipotent part $W^\text{u} = \big(W^\text{ss}\big)^{-1}W$  is obtained by normalising (in some sense) the entries of $W$ to have unit leading singularities. Its entries thus play the role of pure functions in the polylogarithmic case. 
We emphasise here that $\rd\tau$ can indeed take the role of a generalised dlog-form. In particular, one can verify that
in the limit $M^2\rightarrow 0$, we have
\begin{align}
    \rd\tau \xrightarrow{M\rightarrow 0}{} \rd\!\log\left( \frac{s-\sqrt{s(s-4m^2)}}{s+\sqrt{s(s-4m^2)}} \right) \, .
\end{align}

\paragraph{Realignment of the transcendental weight and the $\eps$-expansion.}
After this discussion of how the splitting of the period matrix into its semi-simple and unipotent parts allows one to assign transcendental weights for $\eps=0$ (locally to a MUM-point), let us now return to the full system in eq.~\eqref{eq:DEQ_sunrise_cut}. In order to achieve a consistent assignment of the transcendental weights, we need to multiply the vector of master integrals $(\widetilde{I}_1,\widetilde{I}_2)^T$ by $\big(W^{\textrm{ss}}\big)^{-1}$.
At this point, however, we need to address another issue, which did not appear during the discussion at $\eps=0$. From the polylogarithmic case, we would expect that the coefficient of $\eps^k$ only involves pure functions of uniform transcendental weight $k$. However, while all the entries of $W^{\textrm{u}}$ are pure functions, it is easy to see that the entries of the second column have different weights. This can easily be remedied by a simple rescaling by powers of $\eps$ to ensure that there is the expected correlation between the order in the $\eps$-expansion and the transcendental weight. We then define the new basis
\beq\label{eq:DWss_ell}
\begin{pmatrix}
\widetilde{J}_1 \\ \widetilde{J}_2\end{pmatrix} = 
D_1(\eps)\big(W^{\textrm{ss}}\big)^{-1}\begin{pmatrix}
\widetilde{I}_1 \\ \widetilde{I}_2\end{pmatrix}\,,
\eeq
where we defined the diagonal matrix
\beq\label{eq:D_n_def}
D_n(\eps) = \diag(\eps^n,\eps^{n-1},\ldots,\eps,1)\,.
\eeq

\paragraph{$\eps$-factorisation of the full $3\times 3$ system.}

After this extensive discussion of the $2\times2$ subsystem, let us return to the full $3\times3$ system in eq.~\eqref{eq:DEeps3x3a} and illustrate how the presence of the third master integral affects the discussion compared to the $2\times2$ case.

Following the discussion above,
as a first step, we should extend eq.~\eqref{eq:DWss_ell} to accommodate the third master integral. As was argued above, the entries of $W^\text{ss}$ can be seen as the generalisation of the algebraic leading singularities in the polylogarithmic case. For $\widetilde{I}_3=I_2$, this means that it should be normalised by its algebraic leading singularity $L_3(\underline{z})$, which corresponds to the residue at the additional simple pole at $x_5=\infty$ in eq.~\eqref{eq:cutI2}. 
In this example, since the integral already has a constant residue, this step is trivial, and we can choose $L_3(\underline{z})=1$. 
Concerning the $\eps$-rescaling, recall that the integrand of $\mathrm{LS}(\widetilde{I_3})$ corresponds to the integrand of $\mathrm{LS}(\widetilde{I_1})$ up to the additional simple pole. Consequently, we expect that the $\eps$-expansion of the first and third integral should be kept aligned. Based on these considerations, we can now define the new basis 
\begin{equation}
\label{eq:DWss_ell3}
    \begin{pmatrix}
    \widetilde{J}_1 \\ \widetilde{J}_2 \\ \widetilde{J}_3\end{pmatrix} = 
    \begin{pmatrix}
        D_1(\eps) & \begin{matrix} 0 \\ 0 \end{matrix} \\
        \begin{matrix} 0 \ \ & \ \ 0 \end{matrix} & \eps
    \end{pmatrix}
    \begin{pmatrix}
        \big(W^{\textrm{ss}}\big)^{-1} & \begin{matrix} 0 \\ 0 \end{matrix} \\
        \begin{matrix} 0 \ \ & \ \ 0 \end{matrix} & L_3^{-1} 
    \end{pmatrix}
    \begin{pmatrix}
    \widetilde{I}_1 \\ \widetilde{I}_2 \\ \widetilde{I}_3 \end{pmatrix}\,.
\end{equation}
After this transformation, the system of differential equations takes the form,
\begin{align}\label{eq:deqJell3}
    \partial_z  \begin{pmatrix}
        \textrm{Cut}(\widetilde{J}_1) \\ \textrm{Cut}(\widetilde{J}_2) \\ \textrm{Cut}(\widetilde{J}_3)
    \end{pmatrix} &= 
    \left[
    \widetilde{A}(\underline{z})+\eps\,\widetilde{B}(\underline{z})
    \right]
    \begin{pmatrix}
       \textrm{Cut}( \widetilde{J}_1) \\ \textrm{Cut}(\widetilde{J}_2) \\ \textrm{Cut}(\widetilde{J}_3)
    \end{pmatrix}\,,
\end{align}
with
\begin{align}
\widetilde{A}(\underline{z}) &= \begin{pmatrix}
        0 &  0 & 0 \\
        \frac{\varpi_0^2}{\Delta} b_{21}(\underline z) + \frac{\varpi_0 \, \partial_z \varpi_0}{\Delta} b_{22}(\underline z) & 0 & \frac{\varpi_0}{\Delta} b_{23}(\underline z) \\
        \varpi_0 \, a_{31}(\underline z) + \partial_z \varpi_0 \, a_{32}(\underline z) & 0 & 0
    \end{pmatrix} \, , \\
    \widetilde{B}(\underline{z}) &=  \begin{pmatrix}
        0 &  \frac{\Delta}{\varpi_0^2} & 0 \\
        \frac{\varpi_0^2}{\Delta} c_{21}(\underline z) & b_{22}(\underline z) & \frac{\varpi_0}{\Delta} c_{23}(\underline z) \\
        \varpi_0 \, b_{31}(\underline z) & \frac{\Delta}{\varpi_0} a_{32}(\underline z) & b_{33}(\underline z)
    \end{pmatrix}\,. \label{eq:Btilde_3x3}
\end{align}
The system is still not in $\eps$-factorised form, but we note that $\widetilde{A}(\underline{z})$ is nilpotent. In fact, by swapping the position of the second and third integral in this basis, this contribution becomes lower-triangular,
\begin{align}\label{eq:deqJell3swapped}
    \partial_z  \begin{pmatrix}
        \textrm{Cut}(\widetilde{J}_1) \\ \textrm{Cut}(\widetilde{J}_3) \\ \textrm{Cut}(\widetilde{J}_2)
    \end{pmatrix} &= 
    \left[
    \begin{pmatrix}
        0 &  0 & 0 \\
        \varpi_0 \, a_{31}(\underline z) + \partial_z \varpi_0 \, a_{32}(\underline z) & 0 & 0 \\
        \frac{\varpi_0^2}{\Delta} b_{21}(\underline z) + \frac{\varpi_0 \, \partial_z \varpi_0}{\Delta} b_{22}(\underline z) & \frac{\varpi_0}{\Delta} b_{23}(\underline z) & 0 \\
    \end{pmatrix}
    +\mathcal{O}(\eps)
    \right]
    \begin{pmatrix}
       \textrm{Cut}( \widetilde{J}_1) \\ \textrm{Cut}(\widetilde{J}_3) \\ \textrm{Cut}(\widetilde{J}_2)
    \end{pmatrix}\,.
\end{align}
The fact that $\widetilde{A}(\underline{z})$ is nilpotent implies that its contribution in~\cref{eq:deqJell3} can be removed by a rotation of the form
\begin{equation}
 \label{eq:G_rot_ell3}
\begin{pmatrix}
J_1 \\ J_2 \\ J_3 \end{pmatrix} = G(\underline{z})\begin{pmatrix}
\widetilde{J}_1 \\ \widetilde{J}_2 \\ \widetilde{J}_3 \end{pmatrix}\,,\qquad G(\underline{z}) = \begin{pmatrix} 1&0&0\\ g_1(\underline{z}) &1 & g_2(\underline{z}) \\ g_3(\underline{z}) &0&1\end{pmatrix}\,.
\end{equation}
The peculiar structure of the matrix $G(\underline z)$ can also be explained as follows: After multiplying by the inverse of the semi-simple part, the normalisation of the integrals is fixed. This explains the 1's on the diagonal. Moreover, $G(\underline z)$ is a unipotent matrix, meaning that $G(\underline z) - \mathds{1}$ is a nilpotent matrix (which can again easily be seen by swapping $\widetilde{J}_2$ and $\widetilde{J}_3$). As a consequence, this rotation preserves the filtration from the MHS at the leading order in $\eps$ and the alignment of the transcendental weights in the $\eps$-expansion. The former is reflected in the $(1,3)$ entry being zero in agreement with the $V_1 \subset V_3$ part of the filtration in~\cref{eq:filtration_ell}, while the latter corresponds to the vanishing of the $(1,2)$ and $(3,2)$ entries, as the first and third integral are rescaled with a relative factor of $\eps$ with respect to the second one.

The remaining functions $g_i(\underline{z}), \, i=1,2,3 \, , $ are determined by demanding that the system of differential equations for this new basis is $\eps$-factorised,
\begin{align}\label{eq:deqJell_final_3}
    \partial_z  \begin{pmatrix}
        \textrm{Cut}(J_1) \\ \textrm{Cut}(J_2) \\ \textrm{Cut}(J_3)
    \end{pmatrix} &= 
\eps\,\mathcal{B}(\underline{z})
    \begin{pmatrix}
        \textrm{Cut}(J_1) \\ \textrm{Cut}(J_2) \\ \textrm{Cut}(J_3)
    \end{pmatrix}\, ,
\end{align}
with
\beq \label{eq:deqB_final_3}
\mathcal{B}(\underline{z}) = G(\underline{z})\widetilde{B}(\underline{z})G(\underline{z})^{-1}\,.
\eeq
Concretely, the functions $g_i(\underline{z})$ are obtained by solving the first order differential equations
\begin{align}
    \partial_z g_1(\underline{z}) &= - \frac{\varpi_0^2}{\Delta} b_{21}(\underline z) - \frac{\varpi_0 \, \partial_z \varpi_0}{\Delta} b_{22}(\underline z) - g_2(\underline{z}) \left(\varpi_0 a_{31}(\underline z) +\partial_z \varpi_0 \, a_{32}(\underline z) \right) \, , \\
    \partial_z g_2(\underline{z}) &= - \frac{\varpi_0}{\Delta} b_{23}(\underline z) \, , \\
    \partial_z g_3(\underline{z}) &= - \varpi_0 a_{31}(\underline z) -\partial_z \varpi_0 \, a_{32}(\underline z) \, . \label{eq:g3diffeq}
\end{align}
The solution to the differential equation~\eqref{eq:deqJell_final_3} will be expressed order by order in $\eps$ in terms of iterated integrals over the kernels defined by the entries of $\mathcal{B}(\underline{z})$ given by~\cref{eq:deqB_final_3}. Inspecting $\widetilde{B}(\underline{z})$ in~\cref{eq:Btilde_3x3}, we see that $\mathcal{B}(\underline{z})$ does not depend on $\partial_z \varpi_0$ explicitly, while the transformation to the canonical basis does, see~\cref{eq:DWss_ell3,eq:sesiunipotsplitting}.

Let us briefly discuss the differential equation for $g_3$. From the matrix $A$ in eq.~\eqref{eq:DEeps3x3}, as well as from the relations $\mathrm{Cut}(\widetilde{I_1}) = \varpi_0 + \mathcal{O}(\eps)$ and $\mathrm{Cut}(\widetilde{I_2}) = \partial_z \varpi_0 + \mathcal{O}(\eps)$ for a particular choice of contour, we can see that 
\begin{equation}
    \partial_z \mathrm{Cut}(\widetilde{I_3}) = a_{31}(\underline{z}) \, \varpi_0 + a_{32}(\underline{z}) \, \partial_z \varpi_0 + \mathcal{O}(\eps) \, .
\end{equation}
Comparing to the differential equation~\eqref{eq:g3diffeq} for $g_3(\underline{z})$, and using the fact that $\mathrm{Cut}(\widetilde{I_3})=\mathrm{LS}(\widetilde{I_3})+\ord(\eps)$, we obtain (up to a numerical constant):
\begin{equation}
    \mathrm{LS}(\widetilde{I_3}) = - g_3(\underline{z})\,.
    \end{equation}
Hence, we may resort to the leading singularity of $\widetilde{I}_3=I_2$ at $\eps=0$ to obtain another integral representation for $g_3(\underline{z})$. In fact, it was to be expected that not only the solutions at $\eps=0$ for $\mathrm{LS}(\widetilde{I_1})$ and $\mathrm{LS}(\widetilde{I_2})$ but also for $\mathrm{LS}(\widetilde{I_3})$ are needed to completely decouple the system of differential equations at $\eps=0$.

To conclude the discussion of the homogenous system, let us come back to the fact that the master integrals actually depend on a second dimensionless kinematic variable in addition to the one we have singled out and denoted by $z$. In fact, it turns out that the one-variable analysis performed above is sufficient to find a basis such that the differential equation with respect to any of the kinematic variables is in canonical form. We have also verified explicitly that, independently of the concrete choice for $z$, we always land on the same basis.

Finally, let us briefly mention how to include the contributions from subtopologies. If we assume that the subtopologies have already been transformed into canonical form, we may add to $(J_1, J_2, J_3)$ linear combinations of master integrals from subtopologies, and adjust the coefficients of the linear combination so that the system for $(J_1, J_2, J_3)$ is also in canonical form. We mention that this transformation also preserves a filtration on the space of Feynman integrals, albeit in this case, the trivial filtration induced by the propagators, or equivalently by the sectors. We highlight here that this bottom-up way of building
a canonical $\epsilon$-factorised basis is rather standard and is often employed also in the polylogarithmic case when it is not possible to find a canonical basis resorting only to an integrand analysis, see for example ref.~\cite{Gehrmann:2014bfa}.

\paragraph{Transformation behaviour under analytic continuation.}

The construction of the basis satisfying canonical differential equations relied heavily on the ability to understand the transcendental weight structure of the (transcendental) leading singularities. However, we stress here again that this analysis was only valid \emph{locally}, close to the MUM-point $\underline z= \underline 0$. While in the example, we effectively worked on a one-parameter slice and analysed the differential equation close to the point $z=0$ (i.e.,~for~\cref{eq:sunriseentriesexplicit}, this corresponds to $\frac{m^2}{s}=0$), here we re-introduced the multi-parameter notation $\uz=\underline 0$ to make clear that the MUM-point must be in general specified by fixing all variables in a multi-parameter case. If one wishes to go to another singular point $\underline z_0 \neq \underline 0$, it is possible to repeat the same construction, as all singular points are MUM-points in the elliptic case~\cite{Bonisch:2021yfw}. However, in order to understand the connection between the resulting bases, since the series expansions defining the periods in the Frobenius basis (see~\cref{eq:ForbeniusElliptic}) only have a finite radius of convergence, we need to be able to do the analytic continuation from one singular point to another, in particular of the holomorphic solution and the semi-simple part of the Wronskian. 

Let $\varpi_0^{[z_0]}$ and $\varpi_1^{[z_0]}$ denote the Frobenius basis of periods close to $\underline z=\underline z_0$, i.e.~the first one is a power series solution and the second is a solution logarithmically diverging at $ \underline z= \underline z_0$. From the fact that both the periods $(\varpi_0,\varpi_1)^T$ at the original MUM-point $\underline z= \underline0$ and the second basis $(\varpi_0^{[z_0]},\varpi_1^{[z_0]})^T$ at the point $ \underline z= \underline z_0$ satisfy the same second-order differential equation (see~\cref{eq:sunrise_PF}), it follows that it must be possible to write one basis as a linear combination with purely numerical coefficient of the other. In other words, we have
\begin{align}
    \begin{pmatrix} \varpi_0 \\ \varpi_1 \end{pmatrix} = 
    \begin{pmatrix} d & c \\ b & a \end{pmatrix} \begin{pmatrix} \varpi_0^{[z_0]} \\ \varpi_1^{[z_0]} \end{pmatrix} \quad \text{and}\quad 
    W = W^{[z_0]} \begin{pmatrix} d & c \\ b & a \end{pmatrix}^T\,,
    \quad \text{with}\quad \begin{pmatrix} d & c \\ b & a \end{pmatrix}\in \mathrm{SL}(2,\mathbb C) \, .
\label{eq:trafoperiodsMUM}
\end{align}
Here, $W$ is the Wronskian matrix at the original MUM-point $\underline 0$ and $W^{[z_0]}$ is the Wronskian at the singular point $ \underline z_0$ constructed in the same way as $W$. From this transformation behaviour under analytic continuation, it is not hard to see that
\begin{align}
    \begin{pmatrix} 1 & \tau \\ 0 & 1 \end{pmatrix} = 
    \begin{pmatrix} 1 & \frac{a\tau^{[z_0]}+b}{c \tau^{[z_0]}+d} \\ 0 & 1 \end{pmatrix} \,,\quad\text{with}\quad
    \tau^{[z_0]} = \frac{\varpi_1^{[z_0]}}{\varpi_0^{[z_0]}} \, .
\end{align}
Moreover, we find for the semi-simple part~\cite{Duhr:2019rrs}
\begin{align}\label{eq:SS_modular}
    \begin{pmatrix} \varpi_0 & 0 \\ \partial_z \varpi_0 & \frac{\Delta}{\varpi_0} \end{pmatrix} =
    \begin{pmatrix} \varpi_0^{[z_0]} & 0 \\ \partial_z \varpi_0^{[z_0]} & \frac{\Delta}{\varpi_0^{[z_0]}} \end{pmatrix} \begin{pmatrix} c \tau^{[z_0]}+d  & 0 \\ c & \frac{1}{c \tau^{[z_0]}+d}\end{pmatrix} \, ,
\end{align}
where $\Delta$ is the determinant of the Wronskian.
Note that the semi-simple parts of the Wronskian at two singular points are connected by a matrix that only depends on the semi-simple quantity $\tau^{[z_0]}$.
Furthermore, let us also note that
\begin{align}\label{eq:SS_transform}
    \begin{pmatrix} c \tau^{[z_0]}+d  & 0 \\ c & \frac{1}{c \tau^{[z_0]}+d}\end{pmatrix} 
    \begin{pmatrix} 0 & 1 \\ -1 & 0 \end{pmatrix}
    \begin{pmatrix} c \tau^{[z_0]}+d  & 0 \\ c & \frac{1}{c \tau^{[z_0]}+d}\end{pmatrix}^T =
    \begin{pmatrix} 0 & 1 \\ -1 & 0 \end{pmatrix} \, ,
\end{align}
i.e., the analytic continuation leaves the intersection pairing invariant, and it preserves, in particular, the Legendre relation for complete elliptic integrals.
In~\cref{sec:commentsana}, we will argue that these are the necessary transformation behaviours such that we can obtain a canonical differential equation also at another singular point $ \underline z =  \underline z_0$, by just replacing the solutions $(\varpi_0,\varpi_1)^T$ by their counterparts at the new singular point. Since, in the elliptic case, all singular points are MUM-points, this is also exactly what we would have found if we had repeated the above construction at the new singular point. In other words, constructing the canonical basis at a MUM-point and working out the analytic continuation of the corresponding periods allows us to write down the canonical basis and canonical differential equations at any other singular point immediately.

\section{A roadmap to more complicated geometries}
\label{sec:roadmap}

The example from the previous section illustrates the method of ref.~\cite{Gorges:2023zgv} in the elliptic case. In the following, we summarise the main steps, and we present a proposal of how these steps have a natural generalisation to more complicated geometries. Also, in this section, we largely use the multi-dimensional notation $\uz$ and assume that we start our construction from a MUM-point, which for definiteness, we fix at $\uz = \underline 0$.

\subsection{Towards geometries beyond elliptic curves}

Let us consider a family of Feynman integrals. We assume that we have determined a set of master integrals and the associated system of differential equations. We also assume that the master integrals were chosen in a way that respects the natural order on the sectors, so that the system of differential equations is in block-triangular form. Our goal is to apply the method of ref.~\cite{Gorges:2023zgv} to bring the system into an $\eps$-factorised form. We proceed by induction in the sectors, from the lowest to highest. Let us assume that we have managed to bring all the sectors below a given sector into $\eps$-factorised form. We now describe the steps one needs to perform to bring also this sector into $\eps$-factorised form.

\paragraph{Identifying the geometries associated with the maximal cuts.}

We start by making an initial choice for the basis integrals, in particular for the dimension in which they are evaluated. Let us assume that we want to evaluate the master integrals in $d=d_0-2\eps$, where $d_0$ is an even integer. Note that it is not necessary to evaluate all integrals with the same value of $d_0$, because we can use dimensional-shift relations to relate integrals in different dimensions to each other~\cite{Tarasov:1996br, Lee:2009dh}. A good choice of $d_0$ for a given sector can often be obtained by analysing the integrand of a parametric representation, e.g., the Baikov representation. For instance, for the sunrise integral studied in the previous section, an analysis of the Baikov parametrisation reveals that for $d_0=2$, the integrand naturally leads us to associate a family of elliptic curves to it. In cases where this approach does not yield enough information, it can be helpful to study the factorisation of the associated Picard-Fuchs operator into irreducible components (cf.,~e.g.,~ref.~\cite{Adams:2017tga}), 
although this can become very challenging in the multivariate case. We note that, very generally, maximal cuts of Feynman integrals in integer dimensions compute periods of toric varieties~\cite{Vanhove:2018mto}. 
We also note that it may be possible to identify more than one geometry from a given leading singularity. An example of this is the family of ice cone integrals with equal propagator masses, where one obtains two different families of CY varieties from the same maximal cut~\cite{Duhr:2022dxb}.\footnote{While an $\eps$-form for the differential equations of the two-loop ice cone integrals was derived in ref.~\cite{Gorges:2023zgv}, we discuss the three-loop case in~\cref{reficecone}.}

\paragraph{Identifying a distinguished basis compatible with the mixed Hodge structure.} 

Let us focus on one of the geometries identified in the previous step, and let us assume that it describes an $n$-dimensional manifold. Each geometry on the maximal cut gives rise to a set of (independent) differential $n$-forms, which generate (a subspace of) the $n^{\textrm{th}}$ cohomology group $H^n(X,\mathbb{Q})$. A deep result in algebraic geometry~\cite{PMIHES_1971__40__5_0, PMIHES_1974__44__5_0} asserts that the cohomology of an algebraic variety always carries a mixed Hodge structure (MHS). We provide a very short review of MHS in appendix~\ref{app:mhs}. Here it suffices to say that the $n^{\textrm{th}}$ cohomology group $H^n(X,\mathbb{Q})$ of an algebraic variety is always equipped with two filtrations. There is an increasing filtration called the \emph{weight filtration},
\beq
0=W_{-1}\subseteq W_0 \subseteq W_1\subseteq \ldots W_{2n} = H^n(X,\mathbb{Q})\,,
\eeq 
and a decreasing filtration on the complexification called the \emph{Hodge filtration}
\beq\label{eq:Hodge_filtration_DEQ}
0\subseteq\ldots\subseteq F^p \subseteq F^{p-1} \subseteq \ldots \subseteq F^1\subseteq F^0 = H^n(X,\mathbb{C})\,.
\eeq 
The graded pieces $W_k/W_{k-1}$ naturally carry a pure Hodge structure of weight $k$ induced by the Hodge filtration. For the definition of a pure Hodge structure, we refer to section~\ref{eq:CY_ops} and appendix~\ref{app:mhs}. We may think of pure Hodge structures as the MHS carried by the cohomology of projective smooth varieties (meaning, they are non-singular and can be described as the zero-set of some homogeneous polynomials), in which case the MHS is concentrated in weight $n$: $0=W_{n-1} \subset W_n = H^n(X,\mathbb{Q})$.

The weight-graded pieces tell us, in some sense, how the cohomology of $X$ contains pieces coming from `simpler' varieties.
As we argue now, the MHS captures precisely what was needed in the previous example to find our distinguished basis. In particular, the weight filtration captures (among other things) the fact that some forms may have simple poles (but also that the underlying geometry may not be smooth but singular). 
For instance, in the example of the previous section, we expect that the basis of master integrals `knows' about the cohomology $H^1(\mathcal{E},\mathbb{Q})$ of the elliptic curve defined in eq.~\eqref{eq:Esunset}. This elliptic curve is smooth and projective (it is non-singular for non-zero propagator masses, and it is obviously defined as the zero-set of a polynomial), and hence $H^1(\mathcal{E},\mathbb{Q})$ carries a pure Hodge structure of weight 1. This defines the $W_1$ part of the weight filtration, which can itself be identified with the space $V_2$ in eq.~\eqref{eq:V1V2_def}. The Hodge filtration on the pure Hodge structure on $H^1(\mathcal{E},\mathbb{C})$ provides the identification of the subspace $V_1$ of differentials of the first kind in eq.~\eqref{eq:V1V2_def}. We know that our complete space of master integrals $V_3$ contains an additional basis element $\widetilde{I}_3$, which is only defined up to adding a linear combination of $\widetilde{I}_1$ and $\widetilde{I}_2$. The complete space $V_3$ can be identified with the $W_2$ part of the weight filtration. Working modulo linear combinations of $\widetilde{I}_1$ and $\widetilde{I}_2$ naturally leads us to consider the quotient $W_2/W_1$, which is one-dimensional and captures the existence of the master integral $\widetilde{I}_3$. We thus see that the MHS encodes the structure of our distinguished basis obtained from the leading singularity analysis. While in the present case, the same result could have been obtained by simply classifying the abelian differentials into first, second, and third kinds, such a classification may no longer be possible in higher dimensions, and these considerations get replaced by the MHS.

We, therefore, propose that the distinguished basis of master integrals is obtained by choosing a basis that is compatible with the MHS obtained from the leading singularities. Note that this is also consistent with the idea that our basis should contain derivatives.
Indeed, the Hodge filtration satisfies Griffiths transversality,
\beq\label{eq:Griffiths_DEQ}
\partial_{z}F^k\subseteq F^{k-1}\,,\qquad 0\le k\le n\,,
\eeq
where $z$ is some kinematic variable on which the integral depends.
$F^n$ will always consist of purely holomorphic $n$-forms (i.e., without any poles). Then,~\cref{eq:Griffiths_DEQ} implies that we can construct from the latter new master integrals compatible with the Hodge filtration by successive differentiation.

\paragraph{Splitting of the period matrix and normalization of the differential forms.} 

At this point, we have identified a basis of differential forms compatible with the MHS of the geometry attached to the maximal cut. 
In particular, we have separated off the differential forms with simple poles, and the basis of forms without poles is chosen to be compatible with the Hodge filtration. These differential forms are in one-to-one correspondence with the master integrals on the maximal cut in $d=d_0$ dimensions. 
Since we have chosen our basis such that it aligns with the MHS, we can look at the weight-graded pieces $W_k/W_{k-1}$, which carry a pure Hodge structure of weight $k$ (in essence, these correspond to the blocks on the diagonal for this basis choice). To the pure Hodge structure for each weight-graded piece, we can associate a period matrix (the corresponding block on the diagonal).
This period matrix can be separated into a semi-simple and a unipotent part according to ref.~\cite{Broedel:2018qkq}, and a general formula of how this splitting can be achieved is given in ref.~\cite{Broedel:2019kmn}.
We then define a new basis by multiplying by the inverse of the semi-simple part of the period matrix in $d=d_0$ dimensions for each weight-graded piece. 
In this new basis, the normalisation of the master integrals is also fixed. In particular, this step involves normalising all residues to be constant. 

We note that, in order to have a clean separation into the semi-simple and unipotent parts, one needs to have very good control over all relations satisfied by the entries of the period matrix, including quadratic relations. 
For example, in order to achieve the splitting in eq.~\eqref{eq:sesiunipotsplitting} in the elliptic case, we needed to use the fact that the determinant of the period matrix is a rational function, which is equivalent to the Legendre relation among complete elliptic integrals. 
As a consequence of the twisted Riemann bilinear relations from twisted cohomology~\cite{Cho_Matsumoto_1995}, maximal cuts of Feynman integrals in dimensional regularisation will always satisfy quadratic relations, both for $\eps=0$ and $\eps\neq 0$~\cite{Lee:2018jsw, Bonisch:2021yfw, Duhr:2024rxe}. 

Let us comment on the structure of the differential equations at this step. The original differential equation matrix only involves rational functions. After performing a gauge transformation with the inverse of the semi-simple part of the period matrix, the entries of the differential equations matrix will be drawn from an algebra of functions $\cA^{\textrm{ss}}$ generated by rational functions and the entries of the semi-simple part of the period matrix (and their derivatives). 
We see that after this step, we can associate in a natural way $\cA^{\textrm{ss}}$ to each maximal cut. 
In our example, we see from eq.~\eqref{eq:sesiunipotsplitting} that this algebra of functions is generated by $\varpi_0$ and $\tfrac{1}{\varpi_0}$, and their derivatives. 
In the one-parameter case, i.e., for the equal-mass sunrise graph, after passing to the variable $\tau$ defined in eq.~\eqref{eq:tau_def}, this algebra can be identified with (a subalgebra of) the algebra of quasi-modular forms for some congruence subgroup of $\mathrm{SL}(2,\mathbb{Z})$ (which is the monodromy group of the periods).

\paragraph{Realignment of the transcendental weight and the $\eps$-expansion.} 

We now assume that the system of differential equations on the maximal cut has a MUM-point.
We have a natural assignment of transcendental weights close to the MUM-point. Entries of the semi-simple part are assigned transcendental weight 0. The entries of the unipotent part will typically diverge like powers of logarithms. In particular, we rescale all master integrals in such a way that the coefficients of $\eps^k$ diverge like the $k^{\textrm{th}}$ power of a logarithm. In the sunrise example, this rescaling is illustrated in eq.~\eqref{eq:DWss_ell}. We note that typically, the rescaling is correlated to the Hodge filtration of a given weight-graded piece, and all master integrals that lie in $F^k/F^{k+1}$ are rescaled by the same power of $\eps$.

\paragraph{$\eps$-factorisation of the maximal cuts.}  

After the previous steps, we have identified a normalised basis on the maximal cut that respects the MHS of the associated geometries for $\eps=0$ and whose transcendental weights are aligned with the leading term in the $\eps$-expansion. 
The resulting system of differential equations will, however, typically not yet be in $\eps$-form (see, for instance~\cref{eq:deqJell3} in the sunrise example). 

This is achieved by a rotation by a matrix $G(\uz)$, similar to the rotation in eq.~\eqref{eq:G_rot_ell3}. The matrix $G(\uz)$ is not completely arbitrary. First of all, the matrix $G(\uz)$ must respect the normalisation of the integrals, so the elements on the diagonal must be equal to 1. Moreover, it must also respect both the filtrations from the MHS at leading order in $\eps$ and the alignment of the transcendental weights and the $\eps$-expansion. In the example from the previous section, this implies that the matrix $G(\underline{z})$ is unipotent, cf.~\cref{eq:G_rot_ell3}. Mixings violating the filtrations from the MHS may, however, still occur at higher orders in $\eps$.

The entries of this rotation matrix are then fixed by the requirement that the new system of differential equations is $\eps$-factorised. This leads to a system of first-order differential equations for the matrix entries. We emphasise that this system may or may not admit solutions inside the algebra $\cA^{\textrm{ss}}$. In particular, the solution may require the introduction of new classes of functions, which can be expressed as (iterated) integrals of functions from $\cA^{\textrm{ss}}$. 

At this point, we have to make an important comment: a priori, it may not be possible to find a rotation that brings the new system into an $\eps$-factorised form. For all known examples that arise from Feynman integrals, it was possible to find such a rotation and we observed this rotation to be unipotent. In section~\ref{sec:general_eps_fac}, we will argue that, at least for some classes of Feynman integrals associated to one-parameter families of CY varieties, this is indeed always possible. It would be interesting to have a more solid understanding of if and why this step is always possible for Feynman integrals and, if so, what the reason for this is.

\paragraph{$\eps$-factorisation of the subtopologies.}

Once the differential equation on the maximal cut, i.e., the homogeneous part, has been transformed into $\eps$-factorised form, we still need to achieve $\eps$-factorisation for the inhomogeneous part. 
The procedure is similar in spirit to the previous step. We can shift the basis obtained at the previous step by a linear combination that respects the natural order of the sectors, and the coefficients of the linear combination are again determined by requiring that the new system is in $\eps$-form. 
This, again, leads to a system of first-order linear differential equations to be satisfied by the coefficients of the linear combination. Since different sectors typically give rise to different geometries, it is more difficult to interpret these new functions as associated to iterated integrals over some functions drawn from a specific algebra of functions. 

Extending the integrand and leading singularity analysis performed in the previous steps by relaxing as many of the cut constraints as feasible, can help drastically simplify this step in practice (see also the discussion in~\cref{sec:conventions}). If such an analysis is not possible, we note that the solutions to these first-order differential equations might exhibit a non-trivial, albeit rational $\eps$-dependence.

\subsection{Comments on analytic continuation}
\label{sec:commentsana}

So far, we have explained how to construct an $\epsilon$-factorised differential equation close to a MUM-point. 
The existence of the MUM-point allowed us to locally identify the transcendental weights and it also provides a distinguished holomorphic period $\varpi_0(\underline{z}) = 1+\mathcal{O}({z_k})$.\footnote{More generally if $h^{n,0}>1$, for example for curves of genus $h^{1,0}=g\ge2$, we can identify a matrix of holomorphic periods. We normalise them such that this matrix is the identity at the MUM-point, cf.~ref.~\cite{Duhr:2024uid}. Since in the remainder of this paper, we will discuss CY varieties with $h^{n,0}=1$, we restrict the discussion to this case.} The power series that defines $\varpi_0(\underline{z})$ typically only has a finite radius of convergence, and it needs to be properly analytically continued to the whole parameter space. In the following, we sketch how the construction of the canonical basis is affected by the analytic continuation. 

Let us assume that we have constructed a canonical basis at a MUM-point, which, without loss of generality, we take to be at $\underline{z}=\underline0$. Our goal is to discuss how to analytically continue the canonical basis to some other point $\underline{z}_0\neq \underline 0$. For simplicity, for now, we assume $\underline{z}_0$ to also be a MUM-point, and we will come back to the situation where it is not at the end. Note that we could just have applied the method of ref.~\cite{Gorges:2023zgv} directly to the choice of MUM-point $\underline{z}=\underline{z}_0$. We now argue that analytic continuation from the MUM-point $\underline{z}=\underline0$ to the MUM-point $\underline{z}=\underline{z}_0$ leads to exactly the same result as if we had applied the method directly to the MUM-point $\underline{z}=\underline{z}_0$.

We start by discussing the analytic continuation of the periods and the Wronskian. 
Let $W$ and $\varpi_0$ denote the Wronskian and the normalised holomorphic period at the MUM-point $\underline{z}=0$. Since $\underline{z}_0$ is also a MUM-point, we can also identify a distinguished normalised holomorphic solution $\varpi_0^{[z_0]} = 1+\mathcal{O}({z_k}-{z}_{0,k})$ close to $\underline{z}=\underline{z}_0$, and $W^{[z_0]}$ is the Wronskian matrix of a basis of local solutions in a neighbourhood of that point.

At this point, we note that $\varpi_0^{[z_0]}$ is, in general, not the analytic continuation of $\varpi_0$, but it is a linear combination of all the independent solutions at $\uz= \underline0$. More generally, since both $W$ and $W^{[z_0]}$ are fundamental solution matrices of the Gauss-Manin connection, there is a constant matrix $M^{[z_0]}$ such that
\beq\label{eq:W_cont}
W^{[z_0]}(\underline{z}) = W(\underline{z})M^{[z_0]}\,.
\eeq
The matrices $M^{[z_0]}$ are not arbitrary, but they must preserve the quadratic relations coming from the monodromy-invariant intersection pairing in (co)homology. More precisely, there is a constant matrix $\Sigma$ and a matrix of rational functions $Z(\underline{z})$ such that
\beq\label{eq:Hodge-Riemann-bilinear}
W^{[z_0]}(\underline{z})\Sigma W^{[z_0]}(\underline{z})^T=W(\underline{z})\Sigma W(\underline{z})^T = Z(\underline{z})\,,
\eeq
which implies
\beq
M^{[z_0]}\Sigma M^{[z_0]T}=\Sigma\,.
\eeq
Note that the matrix $\Sigma$ is always either symmetric or antisymmetric.

It will be important to understand how the semi-simple and unipotent parts $W^{\textrm{ss}}$ and $W^{\textrm{u}}$ from~\cref{eq:sesiunipotsplitting} behave under analytic continuation. The only possibility compatible with~\cref{eq:W_cont} is
\begin{equation}
\begin{aligned}
W^{\textrm{ss},[z_0]}(\underline{z}) &\,= W^{\textrm{ss}}(\underline{z})R^{[z_0]}(\underline{z})\,,\\
\label{eq:Wu_cont}W^{\textrm{u},[z_0]}(\underline{z}) &\,= R^{[z_0]}(\underline{z})^{-1}W^{\textrm{u}}(\underline{z})M^{[z_0]}\,.
\end{aligned}
\end{equation}
Unlike $M^{[z_0]}$, the matrix $R^{[z_0]}(\underline{z})$ is not required to be constant. From~\cref{eq:Wu_cont}, we see that it is a ratio of two matrices whose entries are built from unipotent quantities, and so we expect $R^{[z_0]}(\underline{z})$ to be a function of unipotent quantities. 
In the elliptic case, we have worked this out in~\cref{eq:SS_modular} in the previous section, and we will present more examples in sections~\ref{sec:anal_cont_n2} and~\ref{sec:anal_cont_n3}.
We will also need to understand what the quadratic relations in~\cref{eq:Hodge-Riemann-bilinear} become once we decompose the Wronskian into its semi-simple and unipotent parts. For all cases we studied, we have the relations
\beq\bsp
W^{\textrm{u},[z_0]}(\underline{z})\Sigma W^{\textrm{u},[z_0]}(\underline{z})^T &\,= W^{\textrm{u}}(\underline{z})\Sigma W^{\textrm{u}}(\underline{z})^T =\Sigma\,,\\
\label{eq:Hodge-Riemann-bilinear-ss}
W^{\textrm{ss},[z_0]}(\underline{z})\Sigma W^{\textrm{ss},[z_0]}(\underline{z})^T&\,=W^{\textrm{ss}}(\underline{z})\Sigma W^{\textrm{ss}}(\underline{z})^T = Z(\underline{z})\,.
\esp\eeq
As a consequence, we get
\beq
R^{[z_0]}(\underline{z})\Sigma R^{[z_0]}(\underline{z})^T=\Sigma\,.
\label{eq:randsigma}
\eeq
We have seen an instance of these relations in~\cref{eq:SS_transform}, and we will present additional ones in section~\ref{sec:CY}.

Let us interpret these results. We can choose a Wronskian sucht that $W^{\textrm{ss},[z_0]}$ only differs from $W^{\textrm{ss}}$ by the matrix $R^{[z_0]}$, which is entirely built from unipotent quantities, and they satisfy the quadratic relations in~\cref{eq:Hodge-Riemann-bilinear-ss}. This implies that $W$ and $W^{\textrm{ss}}$ satisfy exactly the same relations, and so $W^{\textrm{ss},[z_0]}$ can be obtained from $W^{\textrm{ss}}$ by simply replacing the local solutions around $\underline{z}=0$, e.g., $\varpi_0$, by their counterparts around $\underline{z}=\underline{z}_0$, e.g., $\varpi_0^{[z_0]}$. This has an important consequence for the transformation to the canonical basis: The rotation that brings the system into canonical form close to $\underline{z}=\underline{z}_0$ is obtained by simply replacing the local solutions around $\underline{z}=0$ by their counterparts around $\underline{z}=\underline{z}_0$. In particular, the integral representations for the new functions introduced with the last rotation have the same functional form in both canonical bases, up to replacing the periods in the integrand with their appropriate counterparts. Note that this result is not unexpected, because it is exactly the result we would have obtained by applying the method of ref.~\cite{Gorges:2023zgv} directly to the MUM-point $\underline{z}=\underline{z}_0$.

So far, we have assumed that $\underline{z}_0$ is a MUM-point. Let us now discuss what changes when we analytically continue the canonical basis to a point that is not MUM (possibly even a regular point). Upon inspection, we see that all the arguments go through exactly as in the case of a MUM-point. The only place where we used the MUM property was when we singled out the distinguished solutions $\varpi_0$ and $\varpi_0^{[z_0]}$, which are assigned transcendental weight 0. If $z_0$ is not a MUM-point, then there may not be a unique distinguished holomorphic solution, but possibly many other ones. For reasons that will be explained in section~\ref{sec:roadmap_summary}, we pick a holomorphic solution $\varpi_0^{[z_0]}$ normalised by $\varpi_0^{[z_0]}(\underline{z}) = 1+\mathcal{O}({z_k}-{z}_{0,k})$. Once such a normalised period has been chosen, all the steps for the analytic continuation described earlier will go through. In particular, also in this case $W^{\textrm{ss},[z_0]}$ and the new functions preserve their functional form, up to replacing the periods by their appropriate counterparts at $\underline{z}=\underline{z}_0$. We note, however, that there may not be a unique choice for the normalised holomorphic period. Consequently, {at a point that is not a MUM-point, the canonical form obtained by our method may not be unique}. Conversely, if $\underline{z}_0$ is a MUM-point, we expect the canonical form to be unique (up to constant rotations) because there is a distinguished normalised holomorphic solution. This explains why the existence of a MUM-point plays such a central role in the method of ref.~\cite{Gorges:2023zgv}, but we now also see that it is by no means essential: We can construct a canonical basis locally around any point, and the canonical bases constructed locally around two different points are related by analytic continuation.

\subsection{Summary and discussion}
\label{sec:roadmap_summary}
The previous subsections summarise our proposal for the main steps that one needs to perform in order to transform a system into canonical form using the method of ref.~\cite{Gorges:2023zgv}. The method of ref.~\cite{Gorges:2023zgv} was originally mainly tested in elliptic cases and simple generalisations thereof (as the symmetric square of a one-parameter K3 surface, which appears in the three-loop banana graph).
Starting from these results, here we propose a general roadmap for its natural extension to more complicated geometries. The main new ingredient is the realisation that the MHS provided by the geometries associated with the maximal cuts defines a natural filtration by which to choose a distinguished initial basis of master integrals. In particular, the MHS naturally suggests to always separate integrals with and without simple poles in the choice of basis, putting on firmer mathematical grounds the first step of the procedure originally suggested in ref.~\cite{Gorges:2023zgv}.
The feasibility and power of this procedure in many cases of increasing complexity have been shown in detail in refs.~\cite{Gorges:2023zgv, Duhr:2024rxe, Duhr:2024bzt, Forner:2024ojj, Klemm:2024wtd, Driesse:2024feo,Becchetti:2025oyb}. 

We note that our proposal is the natural extension of existing approaches to find a canonical form for differential equations. Indeed, by comparing to the elliptic example discussed in section~\ref{main}, we have already explained how the idea of choosing a starting basis aligned with the MHS captures the essence of the integrand analysis in those cases. Moreover, our proposal also captures differential equations in dlog form, where a good basis choice is associated with the independent leading singularities. Finally, quite recently, its generalisation to the hyperelliptic case has been worked out~\cite{Duhr:2024uid}, and also, in that case, one can identify a direct connection between a good basis choice and the MHS underlying the hyperelliptic geometries. In the remainder of this paper, we explain how these ideas can be applied to more complicated classes of geometries, in particular CY varieties.

Let us make a comment at this point: a cornerstone of the method of ref.~\cite{Gorges:2023zgv} is the splitting of the period matrix associated to the maximal cut into a semi-simple and a unipotent part according to the prescriptions from refs.~\cite{Broedel:2018qkq, Broedel:2019kmn}. This splitting was introduced in ref.~\cite{Broedel:2018qkq} based on empirical observations made for Feynman integrals associated to elliptic curves~\cite{Broedel:2019hyg} (and K3 surfaces that are symmetric squares of elliptic curves~\cite{Broedel:2019kmn}), as a proposal of how to generalise the concepts of `pure functions' and `transcendental weights', which have been very successful in understanding Feynman integrals and scattering amplitudes in QFT. However, ref.~\cite{Broedel:2018qkq} did not provide any underlying mathematical principle if and how this splitting generalises to more complicated geometries, like higher-dimensional CY varieties or Riemann surfaces of higher genus. 
Given the fundamental role that the splitting plays for the method of ref.~\cite{Gorges:2023zgv}, one may thus wonder if this is not a weak point of the method. We now argue that the very same splitting has appeared in the mathematical literature in a context that is very close to the context we are considering here. 

First, it is well known that in the context of periods for CY varieties, one can find a basis in which the Gauss-Manin connection is nilpotent, which implies that in this gauge, the period matrix is unipotent. The semi-simple part then arises simply as the gauge transformation from the original period matrix to the gauge in which it is unipotent. We will review this procedure in some detail in the next section for certain classes of one-parameter families. For a discussion of the multi-parameter case, we refer to ref.~\cite{Ducker:2025wfl}.

Second, we note that the splitting has also appeared in ref.~\cite{Movasati2013} in the context of classifying triplets $(H, F^{\bullet},\omega)$, where $(H, F^{\bullet})$ is a pure Hodge structure and $\omega$ is a basis of $H$ compatible with the Hodge filtration $F^{\bullet}$ on $H$. We note that this is, in essence, the same setup that we consider here because our distinguished basis of master integrals is indeed required to be compatible with the (mixed) Hodge structure. Hence, there appears to be a mathematical connection between choosing a basis compatible with the Hodge structure and the splitting of the period matrix into semi-simple and unipotent parts. Reference~\cite{Movasati2013} studies in detail the cases of elliptic and hyperelliptic curves, as well as one-parameter families of CY three-folds, confirming our expectation that the proposed splitting has an appropriate interpretation also beyond elliptic curves. Finally, ref.~\cite{Movasati2013} argues that the entries of the semi-simple part of the period matrix provide natural candidates to attach an \emph{algebra of quasi-modular forms} to large classes of Hodge structures. This algebra is naturally identified with our algebra $\cA^{\textrm{ss}}$. It then makes sense to qualify $\cA^{\textrm{ss}}$ as the algebra of quasi-modular forms associated to a given maximal cut, thereby nicely generalising the observations from the elliptic case.

Finally, let us conclude by commenting on the difference between \emph{$\eps$-factorised} and \emph{canonical} differential equations. Currently, there is no commonly accepted definition of canonical differential equations beyond those that only involve dlog-forms, except that it must be $\eps$-factorised. However, the factorisation of $\eps$ alone is insufficient to define a canonical basis (cf.,~e.g.,~ref.~\cite{Frellesvig:2023iwr} for a comparison of different $\eps$-factorised systems). 
The differential equations produced by the method of ref.~\cite{Gorges:2023zgv} are by construction $\eps$-factorised. In the following, we discuss additional properties that we observe for the $\eps$-factorised systems obtained via the method of ref.~\cite{Gorges:2023zgv}. We stress that we have no firm proof that these properties always hold, but we have observed them for the cases we have studied. In addition, these properties are trivially satisfied by canonical systems in dlog-form (assuming, without loss of generality, that the arguments of the dlog forms are chosen to be algebraically independent), thereby giving confidence that the $\eps$-factorised systems obtained via the method of ref.~\cite{Gorges:2023zgv} are the appropriate generalisation of canonical differential equations beyond dlog-forms.
We observe the following properties for the systems of differential equations obtained by our method:
\begin{enumerate}
\item They are $\eps$-factorised.
\item They only have logarithmic singularities.
\item The differential forms that span the differential equation matrix define independent cohomology classes, i.e., they are linearly independent up to total derivatives (see below). 
\end{enumerate}
As already mentioned, these properties are satisfied by all canonical differential equations in dlog-form, and we will provide examples from cases where the underlying geometry is CY. We, therefore, believe that these properties provide the appropriate definition of a system of differential equations in canonical form.  Note that systems satisfying the first and third properties were dubbed in \emph{C-form} in ref.~\cite{Duhr:2024xsy}.

Let us make some comments about these properties. First, we mention that the property of having at most simple poles hinges on having performed the appropriate analytic continuation of the canonical basis away from the MUM-point, as described in the previous subsection. In all our examples, we find that the initial canonical basis produced by our method has only at most simple poles at the MUM-point. If this canonical basis is analytically continued to another singular point, say $\underline{z}_0$, then there may be more than one choice for the distinguished holomorphic period $\varpi_0^{[z_0]}$ close to $\underline{z}_0$. We were always able to choose the distinguished period such that it is normalised $\varpi_0^{[z_0]}(\underline{z}_0)=1+\mathcal{O}({z_k}-\underline{z}_{0,k})$ and such that the canonical differential equation only has simple poles at $\underline{z}=\underline{z}_0$. We note that this requirement may still not fix $\varpi_0^{[z_0]}$ uniquely. Hence, at a point that is not a MUM-point, the canonical form may not be unique (not even up to constant rotations). If the canonical form is in dlog form, or if the singular point is a MUM-point, then the canonical form is unique (up to constant rotations).

Finally, let us comment on the third property, namely the linear independence of the differential forms. Assume that after applying our method, we arrive at an $\eps$-factorised differential equation of the form
\beq\label{eq:sample_DEQ}
\rd J(\underline{z},\eps) = \eps\,\mathcal{B}(\underline{z})\,J(\underline{z},\eps)\,,
\eeq
and $\mathcal{B}(\underline{z})$ is an $N\times N$ matrix of differential one-forms ($N$ is the number of master integrals) with at most logarithmic singularities (provided that the canonical basis is correctly analytically continued). These differential forms are not arbitrary, but by construction they take the form $\sum_i\rd z_i\,f_i(\underline{z})$, where $f_i$ are function from the algebra $\cA_{\textrm{ext}}^{\textrm{ss}}$, obtained by adjoining to the algebra $\cA^{\textrm{ss}}$ the new functions needed to rotate the system into canonical form. 

Consider the complex vector space $\mathbb{A}$ generated by the entries of $\mathcal{B}$. In other words, $\mathbb{A}$ is a vector space whose elements are one-forms of the form $\sum_{i,j=1}^Nc_{ij}\mathcal{B}_{ij}$, $c_{ij}\in\mathbb{C}$. We do not require the one-forms $\mathcal{B}_{ij}$ to be linearly independent over $\mathbb{C}$, and we obviously have $\dim d \equiv \dim\mathbb{A}\le N^2$. We may pick a basis $\omega_1,\ldots,\omega_d$ of $\mathbb{A}$, and expand $\mathcal{B}$ in that basis,
\beq\label{eq:mat_basis_dec}
\mathcal{B} = \sum_{i=1}^dB_i\,\omega_i\,,
\eeq
where the $B_i$ are constant $N\times N$ matrices. The one-forms $\omega_1,\ldots,\omega_d$ are obviously linearly independent over $\mathbb{C}$. However, this does not imply that the iterated integrals that appear order by order in the $\eps$-expansion of $J(\underline{z},\eps)$ are linearly independent. For the iterated integrals to be linearly independent, the differential forms $\omega_i$ must be linearly independent up to total derivatives from $ \mathcal{A}_{\textrm{ext}}$, by which we mean that the equation
\beq\label{eq:independence_general}
\sum_{i=1}^d c_i\,\omega_i = \rd f\,,\qquad c_i\in\mathbb{C}\,,\quad f \in \mathcal{A}_{\textrm{ext}}^{\textrm{ss}}\,,
\eeq
only admits the trivial solution $c_i=0$, $1\le i\le d$ (which implies that $f$ must be a constant)~\cite{deneufchatel:hal-00558773,Duhr:2024xsy}.
Clearly, this latter criterion is stronger than just ordinary linear independence. If the $\omega_i$ are linearly independent dlog forms of algebraic arguments, then they are also independent up to total derivatives (of algebraic functions). Looking at our explicit examples, we find that the method of ref.~\cite{Gorges:2023zgv} always leads to a basis $\omega_i$ that is linearly independent up to total derivatives from $ \mathcal{A}_{\textrm{ext}}^{\textrm{ss}}$.


\section{One-parameter families of Calabi-Yau varieties}
\label{sec:CY}

In the previous section, we proposed a roadmap for constructing canonical differential equations for Feynman integrals in dimensional regularisation. In later sections, we will apply these ideas to cases where the underlying geometry is a one-parameter family of CY varieties. In this section, we, therefore, review some basic material on CY varieties, in particular, their periods and the Hodge structure on their middle-dimensional cohomology (middle cohomology, in short). Since the splitting of the period matrix is an essential ingredient in the construction of the canonical differential equation, we discuss this splitting in detail.

\subsection{Review of CY varieties and their periods}
\label{eq:CY_ops}
A CY $n$-fold is an $n$-dimensional K\"ahler variety $X$ with vanishing first Chern class. Its middle cohomology carries a pure Hodge structure of weight $n$, and we have a direct sum decomposition
\beq
H^n(X,\mathbb{C}) = \bigoplus_{p+q=n}H^{p,q}(X)\,,
\eeq
where $H^{p,q}(X)$ is generated by $(p,q)$-forms. The Hodge filtration is defined by
\beq\label{eq:CY_Hodge}
F^p = \bigoplus_{k=p}^{n} H^{k,n-k}(X)\,,\qquad 0\subset F^n\subseteq F^{n-1}\subseteq\ldots\subseteq F^0 = H^n(X,\mathbb{C})\,.
\eeq
The Hodge numbers are $h^{p,q} = \dim H^{p,q}(X)$ and $b_n = \sum_{p+q=n}h^{p,q}$. The vanishing of the first Chern class implies that $h^{n,0}=1$, and so, up to normalisation, there is a unique $(n,0)$ cohomology class, which we can represent by the $(n,0)$-form $\Omega$ on $X$. If we choose a basis  of the middle homology group $H_n(X,\mathbb{Z})$, say $\mathcal{C}_1,\ldots,\mathcal{C}_{b_n}$, we can form the vector of periods,
\beq
\pi = \big(\pi_0,\ldots,\pi_{b_n-1}\big) = \Big(\int_{\mathcal{C}_1}\Omega,\ldots,\int_{\mathcal{C}_{b_n}}\Omega\Big)\,.
\eeq

We now consider a family of CY $n$-folds depending on $m$ moduli $\underline{z}$, and the periods are then (multi-valued) functions of the moduli. One can show that $m=h^{n-1,1}$ (except for $n=2$, where $m$ equals the number of transcendental cycles). From now on, we focus on a special class of one-parameter families, namely those where the Hodge numbers satisfy 
\beq\label{eq:Hodge_numbers_CY_ops}
h^{n,0} = h^{n-1,1} = \ldots = h^{1,n-1}=h^{0,n} = 1\,, \qquad b_n=n+1\,.
\eeq
Note that a one-parameter family which does not fall into this class only appears for $n\ge4$ (see, e.g., refs.~\cite{Gerhardus:2016iot, Duhr:2023eld,Ducker:2025wfl}). 
Since there is only a single modulus, we will write $z$ instead of $\underline{z}$ from now on. 
The periods are then annihilated by an $(n+1)^\textrm{th}$-order ordinary differential operator called the \emph{Picard-Fuchs operator} of the family,

\beq
\label{eq:CYop}
 \mathcal L_0^{(n+1)}\, \pi_i = 0\,, \quad \text{for } i=0,1,\hdots,n \,,
 \eeq
 with
\begin{equation}
    \mathcal L_0^{(n+1)} = \partial_z^{n+1} + p_n(z) \partial_z^n + \hdots + p_0(z) \,,
\label{eq:pfcy}
\end{equation}
and $p_i$, $i=0,\ldots,n$, are rational functions. Note that the Picard-Fuchs operator is Fuchsian. We recall that a linear differential operator is called Fuchsian if it only has regular singularities, which implies that $p_{n+1-i}(z)$ has poles of order at most $i$.  We also make the further assumption that the Picard-Fuchs operator has a MUM-point, which we may assume without loss of generality to be at $z=0$. 
Our class of families of CY varieties seems to be quite restrictive, but it captures all known examples of one-parameter families of CY varieties relevant to Feynman integrals, in particular the equal-mass banana and ice cone integrals~\cite{Broedel:2019kmn, Bonisch:2021yfw, Pogel:2022vat, Duhr:2022dxb}, Yangian-invariant fishnet integrals in 2 dimensions~\cite{Duhr:2022pch, Duhr:2023eld} and integrals that appear in the computation of black-hole scattering~\cite{Klemm:2024wtd, Driesse:2024feo, Frellesvig:2024rea, Frellesvig:2023bbf, Dlapa:2021npj, Bern:2021dqo}. Moreover, the relevant Picard-Fuchs operators are known as \emph{CY operators} in the mathematics literature, and their properties are very well studied (cf., e.g., refs.~\cite{Almkvist2, BognerCY, BognerThesis, CYoperators, Almkvist:2023yja}). In the following, we review some of these properties.

Since we assumed the Picard-Fuchs operator to have a MUM-point at $z=0$, it is convenient to choose as the basis for the solution space a \emph{Frobenius basis} $\varpi_i$ for $i=0,1,\hdots,n$ (see also the discussion before~\cref{eq:ForbeniusElliptic}). Then, locally close to the origin, we can write
\begin{equation}
    \varpi_i = z^r\sum_{j=0}^i\frac{1}{j!}\log^j(z) \, S_{i-j}(z)  \, ,
\label{eq:frob}
\end{equation}
where $r$ is the \emph{local exponent} or \emph{indicial} of our solutions and $S_i(z) = \sum_{j=0}^\infty \sigma_{i,j}z^j$ are local holomorphic series. We normalise them by $\sigma_{i,0} = \delta_{i0}$. The Frobenius basis $\varpi_i$, $i=0,1,\hdots, n$, exhibits the full logarithmic tower up to $\log^n(z)$. The periods $\pi_i$ can be written as linear combinations with complex coefficients of the Frobenius basis elements. 
The Wronskian in the Frobenius basis takes the form
\begin{equation}\label{eq:CY_Wronskian}
    W_n(z) = \begin{pmatrix}
                            \varpi_0 & \varpi_1 & \hdots & \varpi_{n} \\
                            \partial_z\varpi_0 & \partial_z\varpi_1 & \hdots & \partial_z\varpi_{n} \\ 
                            \vdots & \vdots & \vdots & \vdots \\
                            \partial_z^{n}\varpi_0 & \partial_z^{n}\varpi_1 & \hdots & \partial_z^{n}\varpi_{n}
                       \end{pmatrix} \, .
\end{equation}
The middle cohomology of a family of CY $n$-folds comes equipped with a monodromy-invariant pairing. In the Frobenius basis, its intersection matrix can be chosen to be
\beq
\Sigma_n = (-1)^n \Sigma_n^T=\begin{pmatrix}0 & \ldots & 0&1\\0 & \ldots & -1&0 \\ \vdots&\iddots&\vdots&\\(-1)^n & \ldots & 0&0\end{pmatrix} \, .
\eeq
The periods and their derivatives satisfy the quadratic relations (see~\cref{eq:Hodge-Riemann-bilinear})
\beq\label{eq:qual_Rel_wronskian}
W_n(z)\Sigma_nW_n(z)^T = Z_n(z)\,,
\eeq
where $Z_n(z)$ is a matrix of rational functions, and Griffiths transversality implies $Z_{n,ij}(z)=0$ if $i<j$. The elements on the anti-diagonal are given by the so-called \emph{$n$-point coupling},
\beq
Z_{n,i,n-i+1}(z) = (-1)^{i+1}C_n(z)\,, \textrm{~~~with~~~}  \frac{\partial_z C_n(z)}{C_n(z)} = -\frac2{n+1}\, p_n(z) \,.
\label{eq:defnptcoupling}
\eeq
Let us make these relations explicit up to $n=3$. In the elliptic case, i.e., $n=1$, $Z_1$ is completely fixed by the one-point coupling $C_1(z)$, 
\begin{equation}
    Z_1 = C_1(z)\begin{pmatrix}
        0 & 1 \\
        -1 & 0
    \end{pmatrix} \, .
\end{equation}
Note that~\cref{eq:qual_Rel_wronskian} implies $\det{W_1(z)}=C_1(z)$. For the K3 case, $n=2$, we find
\begin{equation}
    Z_2 = C_2(z)\begin{pmatrix}
        0 & 0 & 1 \\
        0 & -1 & \frac{1}{3}p_2(z) \\
        1 & \frac{1}{3}p_2(z) & \frac{1}{9}p_2(z)^2+\frac{1}{3}\partial_z p_2(z)-p_1(z)
    \end{pmatrix}\, ,
\end{equation}
while for a CY three-fold 
\begin{equation}
    Z_3 = C_3(z)\begin{pmatrix}
        0 & 0 & 0 & 1 \\
        0 & 0 & -1 & \frac{1}{2}p_3(z) \\
        0 & 1 & 0 & \frac{1}{2} p_3(z)+\partial_z p_3(z)-2p_2(z)  \\
        -1 & \frac{1}{2}p_3(z) & \frac{1}{2} p_3(z)+\partial_z p_3(z)-2p_2(z) & 0
    \end{pmatrix} \, .
\end{equation}
The entries below the anti-diagonal are determined by requiring that the relations obtained by differentiating~\cref{eq:qual_Rel_wronskian} are consistent with~\cref{eq:qual_Rel_wronskian} itself and the Picard-Fuchs equation \eqref{eq:pfcy}.

The existence of a monodromy-invariant paring, together with the quadratic relations from Griffith transversality~\cite{MR717607}, is equivalent to the differential operator being \emph{essentially self-adjoint}~\cite{BognerThesis}. The adjoint operator~\cite{Bonisch:2021yfw, Almkvist:2023yja} is defined as
\begin{equation}
    \mathcal L_0^{*(n+1)} = \sum_{i=0}^{n+1} \left( -\partial_z  \right)^i p_i(z) \, .
\label{eq:adjop}
\end{equation}
The differential operator $\mathcal L_0^{(n+1)}$ is called essentially self-adjoint if there is an algebraic function $A(z)$ such that
\begin{equation}
    \mathcal L_0^{(n+1)} A(z) = A(z) \mathcal L_0^{*(n+1)} \, ,
\label{eq:essselfadj}
\end{equation}
The function $A(z)$ is proportional to the $n$-point coupling $C_n(z)$. For $n\ge 2$, eq.~\eqref{eq:essselfadj} puts constraints on the form of the rational functions $p_i(z)$. For example, for $n=2$ and $n=3$ respectively, we find the constraints: 
\beq\bsp
    \partial_z^2 p_2(z) &= -\frac{4}{9}  p_2(z)^3-2 p_2(z) \partial_z p_2(z)+2 p_2(z) p_1(z)+3 \partial_z p_1(z)-6 p_0(z) \, , \\
    \partial_z^2 p_3(z) &= -\frac{1}{4} p_3(z)^3-\frac{3}{2} p_3(z) \partial_z p_3(z)+p_3(z) p_2(z)+2 \partial_z p_2(z)-2 p_1(z) \, .
\label{eq:concoeffsadj}
\esp\eeq
From the first two solutions, we can construct
\begin{equation}
    t(z) = \frac{\varpi_1(z)}{\varpi_0(z)} = \log(z) + \ord(z)\,, \quad q(z) = e^{t(z)} = z + \ord(z^2) \, ,
\label{eq:canvar}
\end{equation}
which generalizes the $\tau$ parameter of an elliptic curve from eq.~\eqref{eq:tau_def} and defines, locally around the MUM-point, a canonical variable on the moduli space of our family.  The inverse of this map is also known as the \emph{mirror map}
\begin{equation}
    z(q) = q + \ord(q^2) \, .
\label{eq:mirrormap}
\end{equation}
The \emph{structure series} $\alpha_i(z)$ for $i=0,1,\hdots, n$ and the corresponding \emph{$Y$-invariants} 
(see,~e.g., refs.~\cite{BognerCY, BognerThesis}), are defined recursively as follows: $\alpha_i = u_{i,i}(z)^{-1}$ with
\begin{equation}
    u_{i,j}(z) = \theta_z \left( \frac{u_{i-1,j}(z)}{u_{i-1,i-1}(z)} \right) \quad\text{and}\quad u_{0,j}(z) = \varpi_j(z) \, ,
\label{eq:defu}
\end{equation}
where we used the logarithmic derivative $\theta_z=z\partial_z$.
In particular, we find $\alpha_0(z) = 1/\varpi_0(z)$ and $\alpha_1(z)=(\theta_z t(z))^{-1}$. The $Y$-invariants are ratios of the structure series, 
\begin{equation}
    Y_i(z) = \frac{\alpha_1(z)}{\alpha_{i+1}(z)}\, , \quad\text{for } i=1,2, \hdots, n-2 \, .
\end{equation}
Notice that we have $Y_0=1$ as well as the symmetry properties $\alpha_i = c\, \alpha_{n-i+1}$ for some $c\in\mathbb C$. This implies 
\beq\label{eq:Y_self-duality}
Y_i(z) = Y_{n-i-1}(z)\,.
\eeq
We can also write down explicit expressions of the structure series and the Yukawa coupling in terms of minors of Wronskian. If we define the minors\footnote{Note that we use the logarithmic derivative $\theta_z$ and not the ordinary derivative $\partial_z$ to define the minors.}
\begin{equation}
    M\{i_1,i_2,\hdots,i_r \} = \text{det}\begin{pmatrix}
                            \varpi_{i_1} & \varpi_{i_2} & \hdots & \varpi_{i_r} \\
                            \theta_z\varpi_{i_1} & \theta_z\varpi_{i_2} & \hdots & \theta_z\varpi_{i_r} \\ 
                            \vdots & \vdots & \vdots & \vdots \\
                            \theta_z^{r-1}\varpi_{i_1} & \theta_z^{r-1}\varpi_{i_2} & \hdots & \theta_z^{r-1}\varpi_{i_r}
                       \end{pmatrix} \, ,
\end{equation}
we have~\cite{Duhr:2022dxb}
\beq
\alpha_i(z) = \frac{M[i]^2}{M[i-1]M[i+1]}\,,\qquad \textrm{with}\qquad M[i] = M\{0,1\ldots,i-1\}\,.
\eeq
The structure series $\alpha_i$ and the $Y$-invariants are local analytic series in $z$ around the MUM-point $z=0$. By changing variables from $z$ to $q$ using the mirror map, we can equivalently see these functions (and also $\varpi_0(z)$) as holomorphic functions in $q$. In the following, if $f(z)$ is holomorphic in a neighbourhood of $z=0$, we use the notation $f(q)$ to denote the function $f(z(q))$. We can use the $Y$-invariants to write the CY operator $\mathcal L_0^{(n+1)}$ in a factorised form
\begin{equation}
\begin{aligned}
    \mathcal L_0^{(n+1)} &= \beta(z)\theta_z \alpha_n \theta_z \alpha_{n-1} \theta_z \hdots \theta_z \alpha_0 \\
    &= C \beta(q) \left(\alpha_1(q) \right)^{-1}\left[\theta_q^2 \frac{1}{Y_1(q)} \theta_q\frac{1}{Y_2(q)}\theta_q \hdots \theta_q\frac{1}{Y_2(q)}\theta_q \frac{1}{Y_1(q)}\theta_q^2   \right] \frac{1}{\varpi_0(q)} \, ,
\end{aligned}
\label{eq:normalform}
\end{equation}
also known as the \emph{local normal form}. Here, $\beta(z)$ and $C\in\mathbb C$ are necessary for the proper normalisation of the operator.

Let us conclude by making some comments about the arithmetic properties of the functions that we just introduced. The mirror symmetry conjecture implies that CY varieties typically come in pairs, and the coefficients that appear in the $q$-expansion of the holomorphic period $\varpi_0(q)$, the mirror map $z(q)$ and the Yukawa couplings $Y_i(q)$ on one side are expected to be related to topological properties of the mirror. In particular, it follows that the coefficients of these $q$-expansions are expected to be integer numbers.

\subsection{The semi-simple and unipotent parts of the Wronskian}
\label{sec:SS_U}
We now discuss the splitting of the Wronskian in eq.~\eqref{eq:CY_Wronskian} into its semi-simple and unipotent parts,
\beq\label{eq:W_Wss_Wu}
W_n(z) = W_n^{\textrm{ss}}(z)W_n^{\textrm{u}}(z)\,.
\eeq
Using the general formulas from ref.~\cite{Broedel:2019kmn}, it is possible to express the entries of $W_n^{\textrm{ss}}(z)$ and $W_n^{\textrm{u}}(z)$ in terms of minors of the Wronskian, e.g.,
\begin{equation}
    W^\text{u}_n(z) = \begin{pmatrix}
1 & \frac{M{\{1\}}}{M{\{0\}}} & \frac{M{\{2\}}}{M{\{0\}}} & \hdots & & &  \\
0& 1 & \frac{M{\{0,2\}}}{M{\{0,1\}}} & \frac{M{\{0,3\}}}{M{\{0,1\}}} & \hdots & &  \\
& 0& 1 & \frac{M{\{0,1,3\}}}{M{\{0,1,2\}}} & \frac{M{\{0,1,4\}}}{M{\{0,1,2\}}} & \hdots &  \\
& \ldots& 0& 1 & \frac{M{\{0,1,2,4\}}}{M{\{0,1,2,3\}}} & \frac{M{\{0,1,2,5\}}}{M{\{0,1,2,3\}}} & \hdots  \\
& & \ldots& 0& 1 & \ddots &   \\
& & & \ldots& 0& 1 & \frac{M{\{0,1,2,\hdots,n-2,n\}}}{M{\{0,1,2,\hdots,n-1\}}}  \\
& & & & \ldots& 0& 1  \\
    \end{pmatrix} \, .
\end{equation}

The expression in terms of minors is very general. In the present case of one-parameter CY varieties, it is possible to find an easier expression.\footnote{For an extension to families of CY varieties depending on more parameters, see ref.~\cite{Ducker:2025wfl}.} First, by direct computation, we can see that $ W^\text{u}_n(z)$ satisfies the differential equation
\beq
\theta_q W^\text{u}_n(q) = N_n(q) \, W^\text{u}_n(q)\,,
\eeq
where $N_n(q)$ is the nilpotent matrix
\begin{equation}\label{eq:N_n(q)}
    N_n(q) = \begin{pmatrix}
        0 & 1 & 0 &\ldots & & &  \\
        & 0 & Y_1(q) & 0 &\ldots & &  \\
        & \ldots& 0 & Y_2(q) & 0 &\ldots &  \\
        & & \ldots& 0 & \ddots & 0 &\ldots  \\
        & & & \ldots& 0 & Y_1(q) & 0  \\
        & & & & \ldots& 0 & 1  \\
        & & & & & \ldots& 0  \\
    \end{pmatrix} \, .
\end{equation}
This differential equation can easily be solved in terms of iterated integrals~\cite{Duhr:2022dxb},
\beq
\frac{\varpi_i(q)}{\varpi_0(q)} = I(Y_0,Y_1,\ldots,Y_{i-1};q)\,,\qquad i=1,\ldots,n\,,
\eeq
where introduced the following notation for iterated integrals
\beq
I(f_1,\ldots,f_p;q) = \int_0^{q}\frac{\rd q'}{q'}I(f_2,\ldots,f_p;q')\,,
\eeq
and the recursion starts with $I(;q)=1$. Note that the integrals will typically diverge at $q=0$, and we need to interpret the lower integration boundary as a tangential base point, cf.~ref.~\cite{Brown:mmv}. When written in terms of the iterated integrals, the unipotent part of the period matrix takes the simple form
\begin{equation}
    W^\text{u}_n(q) = \begin{pmatrix}
1 &I(1;q) & I(1,Y_1;q) &I(1,Y_1,Y_2;q)& \hdots &I(1,Y_1\ldots,Y_1;q) &  I(1,Y_1\ldots,1;q)\\
0& 1 & I(Y_1;q) &I(Y_1,Y_2;q) & \hdots & I(Y_1\ldots,Y_1;q) & I(Y_1\ldots,1;q) \\
0& 0&1 &I(Y_2;q) & \hdots &I(Y_2\ldots,Y_1;q) & I(Y_2\ldots,1;q) \\
& \vdots& &\ddots& &  \vdots &\vdots   \\
0& 0& 0&0& \ldots&  1 & I(1;q)  \\
0& 0& 0& 0& \ldots& 0& 1  \\
    \end{pmatrix} \, .
\end{equation}
Using elementary properties of iterated integrals, it is easy to check that this matrix satisfies the quadratic identity (see~\cref{eq:Hodge-Riemann-bilinear-ss})~\cite{Duhr:2022dxb},
\beq\label{eq:quad_rel_Wu}
W^\text{u}_n(q)\Sigma_nW^\text{u}_n(q)^T = \Sigma_n\,.
\eeq
Inserting eqs.~\eqref{eq:W_Wss_Wu} and~\eqref{eq:quad_rel_Wu} into eq.~\eqref{eq:qual_Rel_wronskian}, we obtain a quadratic relation satisfied by the semi-simple part
\beq\label{eq:qual_rel_ss}
W_n^{\textrm{ss}}(z)\Sigma_nW_n^{\textrm{ss}}(z)^T = Z_n(z)\,.
\eeq
We see that the quadratic relation~\eqref{eq:qual_Rel_wronskian} splits into quadratic relations for $W_n^{\textrm{ss}}(z)$ and $W_n^{\textrm{u}}(q)$ separately. Remarkably, the rational matrix $Z_n(z)$ only enters the quadratic relations for the semi-simple part. In fact, we find that the entries of $W_n^{\textrm{ss}}(z)$ take values in the field $\cA^{\textrm{ss}}$ (cf. section~\ref{sec:roadmap}), which is generated by algebraic functions in the mirror map $z(q)$, the holomorphic period $\varpi_0(q)$ and the Yukawa couplings $Y_i(q)$, as well as their derivatives. Note that since these functions have integral $q$-expansions, and since the integrality of the coefficients is preserved by differentiation, all elements from $\cA^{\textrm{ss}}$ have integral $q$-expansions.
As an illustration, we present the explicit expressions for $W_n^{\textrm{ss}}(q)$ up to $n=3$: 
\begin{align}
\label{eq:SeSiExplicit}
     W^\text{ss}_1(q) &= \begin{pmatrix}
        \varpi_0 & 0 \\
        \varpi_0' & \frac{C_1(z)}{\varpi_0}
    \end{pmatrix} \, , \\
 \nonumber    W^\text{ss}_2(q) &= \begin{pmatrix}
        \varpi_0 & 0 & 0 \\
        \varpi_0' & {\scriptstyle\sqrt{C_2(z)}} & 0 \\
        \frac{\varpi_0'^2}{2\varpi_0}+\frac{2p_2(z)^2+3 p_2'(z)-9p_1(z)}{18}\varpi_0 - \frac{p_2(z)}{3}\varpi_0' \ \ & \left(\frac{\varpi_0'}{\varpi_0} - \frac{p_2(z)}{3} \right){\scriptstyle \sqrt{C_2(z)}} \ \ & \frac{C_2(z)}{\varpi_0}
    \end{pmatrix} \, , \\
\nonumber     W^\text{ss}_3(q) &= \begin{pmatrix}
        \varpi_0 & 0 & 0 & 0 \\
        \varpi_0' & \frac{\varpi_0}{z\alpha_1} & 0 & 0 \\
        \varpi_0'' & \frac{2\varpi_0'}{z\alpha_1}-\left( \frac{1}{z^2\alpha_1} + \frac{\alpha_1'}{z\alpha_1^2} \right)\varpi_0 & \frac{zC_3(z)\alpha_1}{\varpi_0} & 0 \\
        \varpi_0''' & \frac{2\varpi_0'^2}{z\varpi_0\alpha_1} 
        - \frac{\varpi_0''}{z\alpha_1} 
        + v_1(z)\varpi_0-v_2(z)\varpi_0' \ & 
        \left(\frac{z\varpi_0'\alpha_1}{\varpi_0^2}
        -\frac{zp_3(z)\alpha_1}{2\varpi_0}\right) {\scriptstyle C_3(z)}\ & \frac{{\scriptstyle C_3(z)}}{\varpi_0}
    \end{pmatrix} \, , 
    \end{align}
with
\beq\bsp
     v_1(z) &\,= \frac{p_3(z)\alpha_1'}{2z\alpha_1^2} + \frac{2p_3(z)+zp_3(z)^2+2zp_3'(z)-4zp_2(z)}{4z^2\alpha_1} \, , \\
     v_2(z) &\,= \frac{\alpha_1'}{z\alpha_1^2} + \frac{1+zp_3(z)}{z^2\alpha_1} \, .
     \esp\eeq
In deriving these formulas, it was particularly important to use the subsequent identities that follow from the quadratic relations in eq.~\eqref{eq:qual_rel_ss} for $n=2$ and $n=3$, respectively,
\begin{align}
 \nonumber  \varpi_0'' &\,= \frac{\varpi_0'^2}{2\varpi_0}+\frac{2p_2(z)^2+3 p_2'(z)-9p_1(z)}{18}\varpi_0 - \frac{p_2(z)}{3}\varpi_0' \qquad &&\text{for} \ n=2 \, , \\
 \label{eq:quadexplicitly}   \alpha_1'' &\,= (zp_3(z)-2)\frac{\varpi_0'\alpha_1}{z\varpi_0}-2\frac{\varpi_0'\alpha_1'}{\varpi_0}+4\frac{\varpi_0''\alpha_1}{\varpi_0}-2\frac{\varpi_0'^2\alpha_1}{\varpi_0^2}+2\frac{\alpha_1'^2}{\alpha_1^2}  && \\
        &\quad -\frac{2zp_3(z)+z^2p_3(z)^2+2z^2p_3'(z)-4z^2p_2(z)-8}{4z^2}\alpha_1 - \frac{zp_3(z)-4}{2z}\alpha_1' \qquad &&\text{for} \ n=3\nonumber \, .
\end{align}


\section{Differential equations for deformed CY operators}
\label{sec:general_eps_fac}

In section~\ref{sec:roadmap}, based on the discussion in section~\ref{main}, we have proposed a roadmap of how the method of ref.~\cite{Gorges:2023zgv} for transforming a system of differential equations into canonical form can be extended to non-trivial geometries beyond the elliptic case. We apply these ideas to one of the simplest classes of higher-dimensional geometries, namely the one-parameter families of CY varieties reviewed in the previous section. We start by showing that we can apply the method of ref.~\cite{Gorges:2023zgv} successfully to those cases, and we show that the resulting canonical form agrees with the one presented in the literature for banana integrals using an alternative technique~\cite{Pogel:2022yat, Pogel:2022vat, Pogel:2022ken}. We then use our explicit results to understand the systematic properties of the resulting canonical differential equations. In particular, this allows us to present further evidence for the conjectured properties from section~\ref{sec:roadmap_summary} that we expect canonical differential equations to satisfy.

\subsection{A class of deformed CY operators}
\label{sec:eps_deformation}

A cornerstone of our method to obtain an $\eps$-factorised system of differential equations is to choose an initial basis that is compatible with the MHS attached to the geometry connected to the maximal cuts. 
In this section, we analyse in detail the case of a sector whose MHS is associated to a one-parameter family of CY $n$-folds with a MUM-point at $z=0$ and whose Hodge filtration satisfies eq.~\eqref{eq:Hodge_numbers_CY_ops}.
We pick a master integral from the sector, say $I_1(z,\eps)$, such that 
its integrand on the maximal cut for $\eps = 0$ 
corresponds to the holomorphic $(n,0)$-differential form $\Omega$
on the corresponding Calabi-Yau geometry, see the discussion at the beginning of~\cref{eq:CY_ops}.\footnote{
The integral representation of the maximal cut will, 
in general, depend on more integration variables than expected for the associated CY variety. This means that, before getting to the holomorphic period, the extra integration variables have to be constrained by taking all available residues, i.e., by considering the leading singularities of the maximal cut. 
Following ref.~\cite{Gorges:2023zgv}, we always require that there are only simple poles in each extra integration variable.}
We recall that, for a particular choice of the contour, this integral can be associated with the period holomorphic at the MUM-point.

From Griffiths transversality, we know that we can choose $n$ additional master integrals
\beq
\label{eq:derbasisn}
I_{k+1}(z,\eps) = \partial_z^{k}I_1(z,\eps) \, , \qquad k=1,\dots,n \, ,
\eeq
compatible with the Hodge filtration coming from the CY $n$-fold, cf.~\cref{eq:CY_Hodge}. The integrals $I_{1},\dots, I_{n+1}$ then span the sector corresponding to the one-parameter family of CY $n$-folds under consideration. At the same time, as explained in the previous sections, this basis automatically respects the pure Hodge structure on the middle cohomology of the CY variety.

From here on, we focus on the homogeneous part of the differential equations, i.e.,~we work on the maximal cut neglecting all sub-topologies, and call $\cL_{\eps}^{(n+1)}$ the $(n+1)^{\textrm{th}}$-order Fuchsian differential operator that annihilates the $\epsilon$-dependent maximal cuts of $I_1(z,\eps)$~\cite{Primo:2016ebd, Primo:2017ipr}. Note that the higher-order differential equation given by this operator is equivalent to the homogeneous part of the first-order system of the considered sector. 
By construction, its coefficients are rational functions in both $z$ and $\eps$, and $\cL_{0}^{(n+1)} = \cL_{\eps}^{(n+1)} \big|_{\eps=0} $ is a CY operator as discussed in section~\ref{eq:CY_ops}, see also~\cref{eq:CYop} and~\cref{eq:pfcy}. This is also the operator that annihilates the leading singularities of the maximal cut of the integral, i.e.,~the periods of the one-parameter family of CY $n$-folds.

We have not yet addressed in more detail the $\eps$-dependence of the coefficients in $\cL_{\eps}^{(n+1)}$, which is of course fixed, in general, by the fact that the operator annihilates the $\epsilon$-dependent maximal cut of the integral $I_1(z,\eps)$. In the following, we will restrict our considerations to the class of differential operators of the form
\beq\label{eq:L_eps_def}
    \mathcal L^{(n+1)}_{\eps} = \mathcal L^{(n+1)}_0 + \mathcal L^{(n)}_1 \, , 
\eeq
where $\mathcal L^{(n)}_1$ is an $n^{\textrm{th}}$-order differential operator whose coefficient of the $i^{\textrm{th}}$-derivative $\partial_z^i$ is a \textit{polynomial} in $\eps$ of degree $n+1-i$,
\beq
    \mathcal L^{(n)}_1 = \epsilon \sum_{i=0}^{n}q^{(1)}_i(z)\, \partial_z^i+\epsilon^2 \sum_{i=0}^{n-1}q^{(2)}_i(z)\, \partial_z^i+\ldots+\epsilon^{n+1} q^{(n+1)}_{0}(z) \, .
\label{eq:Ltot}
\eeq
The dependence on the kinematic parameter $z$ remains rational, i.e.,~the functions $q^{(j)}_i(z)$ are rational functions. Note that $\cL_{1}^{(n)} \big|_{\eps=0} =0$ such that $\cL_{\eps}^{(n+1)} \big|_{\eps=0} = \cL_{0}^{(n+1)}$ reduces to the CY operator.

The motivation to consider this restricted class of differential operators comes from the observation that it captures all known examples of Feynman integrals in dimensional regularisation associated to CY operators. 
In particular, it captures the maximal cuts of equal-mass banana integrals to any loop order~\cite{Bonisch:2020qmm, Bonisch:2021yfw}, 
as well as the homogeneous differential equations associated to Feynman integrals appearing in calculations for gravitational wave production in black-hole scattering~\cite{Klemm:2024wtd}.\footnote{The question of which classes of problems fall in this category is an interesting one with many ramifications. We do not elaborate on this topic further, as it eludes the scope of the present paper.} 
Furthermore, operators in this category also annihilate interesting classes of hypergeometric
functions associated with CY operators deformed by some parameter $\eps$. 
In particular, consider a hypergeometric function of the form
\beq\bsp\label{eq:hyp_geo_def}
{}_{n+1}F_n&(a_1,\ldots,a_{n+1};b_1,\ldots,b_n;z) = \sum_{k=0}^{\infty}\frac{(a_1)_k\cdots (a_{n+1})_k}{(b_1)_k\cdots (b_n)_k}\frac{z^k}{k!}\\
\,&=\frac{\Gamma(b_n)}{\Gamma(a_{n+1})\Gamma(b_n-a_{n+1})}\int_0^1\rd x\,x^{a_{n+1}-1}(1-x)^{b_n-a_{n+1}-1}\\
&\,\phantom{\frac{\Gamma(b_n)}{\Gamma(a_{n+1})\Gamma(b_n-a_{n+1})}\int_0^1\rd x\,}\times{}_{n}F_{n-1}(a_1,\ldots,a_{n};b_1,\ldots,b_{n-1};xz)\,,
\esp\eeq
with $(a)_k = \frac{\Gamma(a+k)}{\Gamma(a)}$ the Pochhammer symbol. This function is annihilated by the differential operator
\begin{equation}\label{eq:HF_DEQ}
    z\prod_{k=1}^{n+1} (\theta_z+a_k)-\theta_z\prod_{k=1}^n(\theta_z+b_k-1) \, .
\end{equation}
It is easy to check that if the $a_i$ and $b_i$ are linear in $\eps$, i.e.,
\beq
\label{eq:indexepsdep}
a_i = r_i+s_i\eps \textrm{~~~and~~~} b_i = q_i+t_i\eps\,,
\eeq
with $r_i,s_i,t_i,q_i\in\mathbb{Q}$, then the differential operator in eq.~\eqref{eq:HF_DEQ} takes the form~\eqref{eq:L_eps_def} when normalised by its discriminant $(z-1)z^{n+1}$. Further, the operator $\mathcal{L}_0^{(n+1)}$ has a MUM-point at $z=0$ with all indicials equal to $0$, if it is of the form $-\theta_z^{n+1} +  \mathcal O(z)$. This implies that $q_i = 1$ if we require $z=0$ to be a MUM-point for $\eps=0$. 
In general, the differential operator in eq.~\eqref{eq:HF_DEQ} will not be a CY operator for $\eps=0$. There are, however, several infinite families\footnote{Infinite in the order of the CY operator.} of hypergeometric CY operators for specific choices of the $r_i$, cf.,~e.g.,~refs.~\cite{Batyrev:1993wa, Doran:2005gu}. The values of the $r_i$ are constrained, for example, by the requirement that $\mathcal{L}_0^{(n+1)}$ be essentially self-adjoint. Moreover, for the differential equation~\cref{eq:defnptcoupling} for the $n$-point coupling to admit a rational solution, we find that the following additional constraint on the sum of the $r_i$ must be fulfilled:
\beq
\sum_{i=1}^{n+1}r_i=\frac{n+1}{2}\,.
\eeq
 We have explicitly considered several examples, which we summarise in~\cref{tab:hypgeo}.\footnote{Note that in~\cref{tab:hypgeo} we only discuss examples with $n\ge2$. Examples with $n=1$ can be found, e.g., in ref.~\cite{Gorges:2023zgv}.} An explicit example from the context of Feynman integrals corresponding to this case of hypergeometric functions is the $\eps$-dependent maximal cut of the corner integral associated with the equal-mass sunrise graph, i.e.,~\cref{def:sunrise} with $m^2=M^2$, $\nu_1=\nu_2=\nu_3=1$ and $\nu_4=\nu_5=0$~\cite{Tarasov:2006nk}. 

We now show that up to $n=3$, we can always bring the first-order system obtained from a differential operator of this class into $\eps$-factorised form using the method of ref.~\cite{Gorges:2023zgv} by following the roadmap laid out in section~\ref{sec:roadmap}. We conjecture that this is possible for all $n$. The proofs are based on explicit computations, which we present in the following.

\begin{table}[!th]
\centering
\begin{tabular}{|c l  l|} 
\hline
CY $n$-fold & $(r_1,\hdots, r_{n+1})$ & Comments\\
\hline\hline
\multirow{2}{*}{$2$}   & \multirow{2}{*}{$\frac12, \frac12, \frac12$} & Symmetric square of the $\Gamma(2)$ (Legendre) family,     \\ 
                       &                                              & K3 surface appearing at 4 and 5PM 1SF~\cite{Klemm:2024wtd} \\

    $2$ & $\frac14, \frac24, \frac34$                                 &Symmetric square of the $\Gamma_0(2)$  family               \\[0.5ex]
\hline
    $3$ & $\frac12, \frac12, \frac12, \frac12$        & AESZ 3, CY three-fold appearing at 5PM 1SF~\cite{Klemm:2024wtd}            \\
    $3$ & $\frac12, \frac12, \frac13, \frac23$        & AESZ 5           \\
    $3$ & $\frac12, \frac12, \frac14, \frac34$        & AESZ 6           \\
    $3$ & $\frac12, \frac12, \frac16, \frac56$        & AESZ 14          \\
    $3$ & $\frac13, \frac13, \frac23, \frac23$        & AESZ 4           \\
    $3$ & $\frac13, \frac23, \frac14, \frac34$        & AESZ 11          \\
    $3$ & $\frac13, \frac23, \frac16, \frac56$        & AESZ 8           \\
    $3$ & $\frac14, \frac14, \frac34, \frac34$        & AESZ 10          \\
    $3$ & $\frac14, \frac34, \frac16, \frac56$        & AESZ 12          \\
    $3$ & $\frac15, \frac25, \frac35, \frac45$        & AESZ 1           \\
    $3$ & $\frac16, \frac16, \frac56, \frac56$        & AESZ 13          \\
    $3$ & $\frac18, \frac38, \frac58, \frac78$        & AESZ 7           \\
    $3$ & $\frac1{10}, \frac3{10}, \frac7{10}, \frac9{10}$    & AESZ 2   \\
    $3$ & $\frac1{12}, \frac5{12}, \frac7{12}, \frac{11}{12}$ & AESZ 9   \\ [0.5ex]
\hline
    $4$ & $\frac12, \frac12, \frac12, \frac12, \frac12$  & Four-loop traintrack integral in 2 dimensions~\cite{Duhr:2022pch,Duhr:2023eld}     \\
    $4$ & $\frac16, \frac26, \frac36, \frac46, \frac56$    &   \\
\hline
\end{tabular}
\caption{Hypergeometric CY examples for which we have derived an $\eps$-form. The AESZ classification follows ref.~\cite{Almkvist:2005qoo}.}
\label{tab:hypgeo}
\end{table}

\subsection{$\eps$-factorisation for $n=1$}
\label{epsfactn=1}

We start by discussing the case $n=1$, which includes, as a special case, the equal-mass case of the sunrise graph considered in~\cref{main}. 
Finding the kernel of a second-order differential operator $\cL_{\eps}^{(2)}$ satisfying our assumptions is equivalent to solving the homogeneous first-order linear system\footnote{We only focus on maximal cuts, so we can work modulo subtopologies.}
\beq
    \partial_z\!\begin{pmatrix} I_1 \\  I_2  \end{pmatrix}      = \text{GM}_1(z;\epsilon)\begin{pmatrix} I_1 \\  I_2  \end{pmatrix}\, ,
    \eeq
    with
\begin{equation}
    \text{GM}_1(z;\epsilon) = \begin{pmatrix}
                0 & 1 \\
                -p_0(z)-\epsilon \, q_0^{(1)}(z)-\epsilon^2 \, q_0^{(2)}(z) \ \ & -p_1(z)-\epsilon \, q_1^{(1)}(z)
            \end{pmatrix}\,.
\label{eq:gmell}
\end{equation}
We now show how to transform this system into canonical form. Examples related to elliptic curves have been discussed at length in ref.~\cite{Gorges:2023zgv}. We repeat the main steps without specifying the rational functions $p_i$ and $q_i^{(j)}$, proving that, indeed, our method allows one to transform any system of this type into canonical form. 

We define the new basis
\beq\label{eq:general_n=2}
\begin{pmatrix}
J_1\\J_2
\end{pmatrix} = T_1(z,\eps) \begin{pmatrix}
I_1\\I_2
\end{pmatrix} \,,
\eeq
with
\begin{equation}
    T_1(z;\epsilon) = \begin{pmatrix}
        1 & 0 \\ t_{21}(z) & 1
    \end{pmatrix}D_1(\eps) \, W_1^{\textrm{ss}}(z)^{-1}\,,
\label{eq:rotell}
\end{equation}
where $D_1(\eps)$ was defined in eq.~\eqref{eq:D_n_def} and $W_1^{\textrm{ss}}(z)$ is the semi-simple part of the period matrix of the family of elliptic curves attached to CY operator $\cL_{\eps}^{(2)}$, see also eq.~\eqref{eq:SeSiExplicit}. The function $t_{21}(z)$ is determined by requiring that the vector $(J_1,J_2)^T$ satisfies an $\eps$-factorized system of differential equations
\beq\label{eq:ell_general_eq}
\partial_z\!\begin{pmatrix}J_1\\J_2\end{pmatrix} = \left[ \mathcal{A}_1(z) +\eps\,\mathcal{B}_1(z)\, \right] \begin{pmatrix}J_1\\J_2\end{pmatrix} \, ,
\eeq
where
\begin{align}
    \mathcal{A}_1(z)+\epsilon \mathcal{B}_1(z) &= \left(T_1(z;\epsilon) \text{GM}_1(z;\epsilon) + \partial_z T_1(z;\epsilon) \right) T_1(z;\epsilon)^{-1} \, .
\end{align}
We see that~\cref{eq:ell_general_eq} is $\eps$-factorised provided that $\mathcal{A}_1(z)$ vanishes. Indeed, after an explicit computation, one finds that the rotation in eq.~\eqref{eq:rotell} gives
\beq\label{eq:A1_def}
    \mathcal{A}_1(z) = \begin{pmatrix}
        0 & 0 \\
        t_{21}'(z)-\frac{q_0^{(1)}(z) \varpi_0(z)^2+q_1^{(1)}(z) \varpi_0(z) \varpi_0'(z)}{C_1(z)} & 0
    \end{pmatrix} \,,
    \eeq
where $C_1(z)$ is the one-point coupling (see~\cref{eq:defnptcoupling}) associated to the CY operator $\cL_0^{(2)}$. From the condition that $\mathcal{A}_1(z)$ has to vanish, we get a defining differential equation for $t_{21}(z)$,
\begin{equation}\label{eq:t21_DEQ}
    t_{21}'(z)=\frac{q_0^{(1)}(z) \varpi_0(z)^2+q_1^{(1)}(z) \varpi_0(z) \varpi_0'(z)}{C_1(z)}.
\end{equation}
We can massage this further: the term containing $\varpi_0'(z)$ can be traded for a total derivative,
\begin{equation}
     t_{21}'(z)=\partial_z\left(\frac{q_1^{(1)} \varpi_0^2}{2 C_1(z)}\right)+\varpi_0^2\,\frac{ 2 q_0^{(1)}(z)-p_1(z) q_1^{(1)}(z)-\partial_zq_1^{(1)}(z)}{2 C_1(z)} \, .
\end{equation}
Hence, it is natural to redefine
\beq
    t_{21}(z) = G_1(z) + \frac{q_1^{(1)}(z)}{2C_1(z)}\varpi_0^2\,,
\eeq 
where $G_1(z)$ is given by the integral
\begin{equation}
    G_1(z) =  \int_0^z r_1(x)\varpi_0^2(x)\, \mathrm dx \,,
    \label{eq:newgell}
\end{equation}
where we defined the rational function
    \beq
    r_1(x) = \frac{2q_0^{(1)}(x)-p_1(x)q_1^{(1)}(x)-\partial_xq_1^{(1)}(x)}{2C_1(x)}\,.
\end{equation}
After the redefinition in~\cref{eq:newgell}, we end up with
\beq
\partial_z\!\begin{pmatrix}J_1\\J_2\end{pmatrix} = \eps\,\mathcal{B}_1(z)\,\begin{pmatrix}J_1\\J_2\end{pmatrix}\,,
\eeq
where
\beq
    \mathcal{B}_1(z) = \begin{pmatrix}
        -\frac12q_1^{(1)}(z)-C_1(z)\frac{G_1(z)}{\varpi_0^2} & \frac{C_1(z)}{\varpi_0^2} \\
        \frac{q_1^{(1)}(z)^2-4q_0^{(2)}(z)}{4C_1(z)}\varpi_0^2-C_1(z)\frac{G_1(z)^2}{\varpi_0^2} & -\frac12q_1^{(1)}(z)-C_1(z)\frac{G_1(z)}{\varpi_0^2}
    \end{pmatrix} \,.
\label{eq:epsell}
\eeq
We see that besides the coefficients $q_i^{(j)}$ of the differential equation~\eqref{eq:gmell}, also the period $\varpi_0$ in eq.~\eqref{eq:frob}, the one-point coupling $C_1$ in eq.~\eqref{eq:defnptcoupling}, and the function $G_1$ in eq.~\eqref{eq:newgell} appear. The previous formulas hold for any operator $\cL_\eps^{(2)}$, as long as it satisfies the assumptions introduced earlier in this section. This shows that the method of ref.~\cite{Gorges:2023zgv} can be used to obtain an $\eps$-form for elliptic blocks of large classes of integrals. Note that $G_1(z)$ is only defined up to a constant, which we can fix, for example, by requiring $G_1(0)=0$, tacitly assuming the absence of a singularity in the point $z=0$.\footnote{If $G_1(z)$ has a simple pole at $z=0$, we can also interpret the lower integration boundary as being regulated by a tangential base-point prescription, cf.~ref.~\cite{Brown:mmv}.}

One may wonder if it is possible to perform the integral in eq.~\eqref{eq:newgell} in closed form. When expressed in terms of the canonical variable $t$ as defined in eq.~\eqref{eq:canvar}, $\varpi_0(t)$ is an Eisenstein series of weight one for some congruence subgroup $\Gamma$ of $\mathrm{SL}_2(\mathbb{Z})$, and the mirror map $z(t)$ is a modular function for $\Gamma$. If $\partial_{t}z = \rho(t)\varpi_0(t)^2$ denotes the jacobian of the change of variables from $z$ to $t$, then it follows that the integrand $r_1(t)\rho(t)\varpi_0(t)^4$ is a meromorphic modular form of weight four for $\Gamma$. Here it suffices to recall that a meromorphic modular form of weight $k$ for a finite-index subgroup of $\mathrm{SL}_2(\mathbb{Z})$ is a meromorphic function $f:\mathfrak{H}\to \mathbb{C}$ (with $\mathfrak{H}$ denoting the complex upper half-plane) such that
\beq
f\left(\tfrac{at+b}{ct+d}\right) = (ct+d)^kf(t)\,,\textrm{~~~for all } \left(\begin{smallmatrix} a&b\\c&d\end{smallmatrix}\right)\in\Gamma\,.
\eeq
The question of whether it is possible to perform the integral in eq.~\eqref{eq:newgell} in closed form is then equivalent to asking if there is a meromorphic modular form $f(t)$ such that $r_1(t)\rho(t)\varpi_0(t)^4 = \partial_tf(t)$. The answer to this question depends on the subgroup $\Gamma$ and the location and the order of the poles of the integrand. For more details, we refer to refs.~\cite{matthes_AMS, Broedel:2021zij}. However, we note here that for all examples of Feynman integrals that we have studied, we observed that $r_1(z)=0$ such that it was possible to choose $G_1(z)=0$.

\subsection{$\eps$-factorisation for $n=2$}

Let us now discuss the case of the third-order differential operator $\cL_\eps^{(3)}$. Finding its kernel is equivalent to solving the homogeneous first-order linear system for $I(z,\eps) = \big(I_1(z,\eps), I_2(z,\eps), I_3(z,\eps)\big)^T$ with
\beq\label{eq:general_GM_K3}
\partial_z I(z,\eps) = \textrm{GM}_2(z,\eps) I(z,\eps)\,,
\eeq
where the explicit expression for $ \textrm{GM}_2(z,\eps)$ can immediately be read off from the coefficients of $\cL_{\eps}^{(3)}$, 
\begin{equation}
    \text{GM}_2(z;\epsilon) = \begin{pmatrix}
                0 & 1 & 0 \\
                0 & 0 & 1 \\
                -p_0(z)- \sum\limits_{j=1}^{3} \epsilon^j q_0^{(j)}(z) \ \ & -p_1(z)- \sum\limits_{j=1}^{2} \epsilon^j q_1^{(j)}(z) \ \ & -p_2(z)-\epsilon q_2^{(1)}(z)
            \end{pmatrix}\,.
\label{eq:gmK3}
\end{equation}
We define the new basis
\beq
J(z,\eps) = T_2(z,\eps)I(z,\eps)\,, \textrm{~~~with~~~} T_2(z,\eps) = \begin{pmatrix} 1&0&0 \\ t_{21}(z,\eps)&1&0\\ t_{31}(z,\eps)&t_{32}(z,\eps)&1\end{pmatrix}D_2(\eps)W_2^{\textrm{ss}}(z)^{-1}\,,
\label{eq:rotK3}
\eeq
where $D_2(\eps)$ was defined in eq.~\eqref{eq:D_n_def} and $W_2^{\textrm{ss}}(z)$ is given in~\cref{eq:SeSiExplicit}.
Requiring that $J(z,\eps)$ satisfies a homogeneous differential equation of the form
\beq
\partial_z J(z,\eps) = \eps \, \mathcal{B}_2(z)J(z,\eps)\,,
\eeq
with
\begin{align}\label{eq:A2_T3_gauge}
    \eps \, \mathcal{B}_2(z) &= \left(T_2(z;\epsilon) \text{GM}_2(z;\epsilon) + \partial_z T_2(z;\epsilon) \right) T_2(z;\epsilon)^{-1} \, , 
\end{align}
we obtain differential equations that allow us to determine the three functions $t_{ij}(z,\eps)$. Explicitly, we find
\begin{equation}
\begin{split}
t_{21}(z,\eps) \,=& \phantom{-}G_2(z)+G_3(z) + \frac{q_2^{(1)}(z)}{3\sqrt{C_2(z)}}\varpi_0(z)\,\,,\\
t_{32}(z,\eps) \,=& -G_2(z)+G_3(z) + \frac{2 q_2^{(1)}(z)}{3\sqrt{C_2(z)}}\varpi_0(z)\,,\\
t_{31}(z,\eps) \,=& \frac{1}{\eps} \left( G_1(z) + \frac{9q_1^{(1)}(z)-4p_2(z)q_2^{(1)}(z)-3\partial_z q_2^{(1)}(z)}{18 C_2(z)}\varpi_0(z)^2 + \frac{q_2^{(1)}(z)}{3 C_2(z)}\varpi_0(z) \varpi_0 '(z)\right) \\
&- G_4-\frac12G_2^2+G_2G_3+\frac12G_3^2-\frac{q_2^{(1)}(z)}{3\sqrt{C_2(z)}} (G_2(z)-G_3(z)) \varpi_0(z) \\
&+ \frac{9q_1^{(2)}(z)-q_2^{(1)}(z)}{18 C_2(z)}\varpi_0(z)^2 \,,
\end{split}
\end{equation}
where the functions $G_1, \hdots, G_4$ can be written as iterated integrals containing besides themselves the coefficients of the differential equation $q_i^{(j)}$~\eqref{eq:gmK3}, the period $\varpi_0$~\eqref{eq:frob}, and the two-point coupling $C_2$~\eqref{eq:defnptcoupling},
\begin{align}
    G_1(z) &= \int_0^z \frac{r_1(x)}{18C_2(x)}\varpi_0(x)^2\,  \mathrm dx \, , \nonumber \\
    G_2(z) &= \int_0^z \frac{\sqrt{C_2(x)}}{\varpi_0(x)}G_1(x)\, \mathrm dx \,,  
    \label{eq:gk3}\\
    G_3(z) &= \int_0^z \frac{3q_1^{(1)}(x)-2p_2(x)q_2^{(1)}(x)-3\partial_xq_2^{(1)}(x)}{6\sqrt{C_2(x)}}\varpi_0(x)\, \mathrm dx \, ,\nonumber \\
    G_4(z) &= \int_0^z \left[\frac{r_2(x)}{18C_2(x)}\varpi_0(x)^2 + \frac{3q_1^{(1)}(x)-2p_2(x)q_2^{(1)}(x)-3\partial_xq_2^{(1)}(x)}{3\sqrt{C_2(x)}}G_2(x)\varpi_0(x)\right]  \mathrm dx \, . \nonumber 
\end{align}
We have again chosen an initial condition such that $G_i(0)=0$ and we defined the rational functions
\begin{align}
    r_1(x) &= 18 q_0^{(1)}-6 p_2 q_1^{(1)}-6 p_1 q_2^{(1)}+4 p_2^2 q_2^{(1)}+6 q_2^{(1)} \partial_xp_2'-9 \partial_xq_1^{(1)}+6 p_2
   \partial_xq_2^{(1)}+3 \partial_x^2q_2^{(1)} \, ,  \nonumber \\
   r_2(x) &= 6 q_1^{(1)} q_2^{(1)}-4 p_2 q_2^{(1)2}-18 q_0^{(2)}+6 p_2 q_1^{(2)}-6 q_2^{(1)} \partial_xq_2^{(1)}+9 \partial_xq_1^{(2)} \, . 
\end{align}
This shows that the method of ref.~\cite{Gorges:2023zgv} can be used to obtain an $\eps$-form for blocks in the differential equation associated to one-parameter families of CY two-folds, or equivalently one-parameter families of K3 surfaces. Let us make some comments about these results. First, note that while $G_1(z)$ and $G_3(z)$ are one-fold integrals, $G_2(z)$ and $G_4(z)$ are iterated integrals of length two. Second, analysing the $\eps$-dependence of eqs.~\eqref{eq:gmK3} and~\eqref{eq:rotK3}, one observes that only the functions $G_2(z)$, $G_3(z)$ and $G_4(z)$ enter the final expression for $\mathcal{B}_2(z)$, while $G_1(z)$ drops out. Later in this section, we will demonstrate that this is a special case of a more general pattern. Second, one may again ask if it is possible to perform some of the integrations in eq.~\eqref{eq:gk3} in closed form. This question can be answered in a very similar manner as in the elliptic case $n=1$. Indeed, every one-parameter CY operator of order three is the symmetric square of a CY operator of order two~\cite{Doran:1998hm, BognerThesis, BognerCY}. Since the solutions of the latter can be expressed in terms of modular forms in the canonical variable $t$, the same is true for the solutions of $\cL_0^{(3)}$. The question of whether some of the integrals in eq.~\eqref{eq:gk3} can be performed in closed form again boils down to the question of whether we can write a modular form as a total derivative, which can be answered using the results of refs.~\cite{matthes_AMS, Broedel:2021zij}. 
Finally, while $\cL_0^{(3)}$ is always the symmetric square of an operator of degree 2, $\cL_{\eps}^{(3)}$ is, in general, not. 
In~\cref{app:Clausen}, we discuss for the hypergeometric cases the precise conditions when the full operator $\cL_{\eps}^{(3)}$ is a symmetric square.

\subsubsection{Analytic continuation for $n=2$}
\label{sec:anal_cont_n2}

Let us give a concrete example of the transformation behaviour of the semi-simple part of the Wronskian in the case of $n=2$. We will focus on the hypergeometric example $(r_1,r_2,r_3)=(1/2,1/2,1/2)$ listed in~\cref{tab:hypgeo}. At $z=0$, the corresponding differential equation has a MUM-point, and the corresponding Frobenis basis is given by
\begin{align}
    \varpi_0 &= 1+8 z+216 z^2+8000 z^3+343000 z^4 + \mathcal O(z^5) \, , \\
    \varpi_1 &= \varpi_0 \log(z) + S_1 = \varpi_0 \log(z) + 24 z+756 z^2+29600 z^3+1305850 z^4 + \mathcal O(z^5) \, , \nonumber \\
    \varpi_2 &= \frac12\varpi_0 \log^2(z) + S_1\log(z) + 288 z^2+15840 z^3+807240 z^4 + \mathcal O(z^5) \, . \nonumber
\end{align}
In this basis, the quadratic relations are given by
\begin{align}
    W \Sigma W^T = \begin{pmatrix}
         0 & 0 & \frac{1}{(1-64 z) z^2} \\
 0 & -\frac{1}{(1-64 z) z^2} & \frac{1-96 z}{(1-64 z)^2 z^3} \\
 \frac{1}{(1-64 z) z^2} & \frac{1-96 z}{(1-64 z)^2 z^3} & -\frac{(1-128 z) (1-80 z)}{(1-64 z)^3 z^4}
    \end{pmatrix} \, .
\label{eq:Griffithsn2}
\end{align}
The hypergeometric differential operator in~\cref{eq:HF_DEQ} has another singular point at $z=1/4^3$. We can take the following basis of solutions close to that singular point,
\begin{align}
\label{eq:newlocalbasis}
    \varpi_0^{[u]} &=  1-16 u+\frac{1792 u^2}{3}-\frac{81920 u^3}{3}+\frac{86179840 u^4}{63}+ \mathcal O(u^5) \, , \\
    \varpi_1^{[u]} &= 16 i \sqrt{u} \left(1-32 u+\frac{20992 u^2}{15}-\frac{344064 u^3}{5}+\frac{125960192 u^4}{35}+ \mathcal O(u^5)\right)    \, , \nonumber \\
    \varpi_2^{[u]} &= -128 u+6144 u^2-\frac{1572864 u^3}{5}+\frac{84410368 u^4}{5}   + \mathcal O(u^5) \, , \nonumber
\end{align}
expressed through the local variable $u=z-1/4^3$.
Notice that this basis was chosen such that the quadratic relations in~\cref{eq:Griffithsn2} are also satisfied for our new basis after we replace the periods $(\varpi_0,\varpi_1,\varpi_2)\rightarrow(\varpi_0^{[u]},\varpi_1^{[u]},\varpi_2^{[u]})$ and the local variables $z \rightarrow u=z-1/4^3$. We notice that the choice of basis in~\cref{eq:newlocalbasis} is not the only one that leaves the quadratic relations invariant modulo the aforementioned replacements. As argued in~\cref{sec:commentsana}, this reflects the fact that a canonical basis is not unique at a non-MUM-point.

With these two bases, it is not hard to work out the proper analytic continuation. We find
\begin{align}
    (\varpi_0,\varpi_1,\varpi_2)^T = 
    \begin{pmatrix}
         \frac{A}{\pi} & \frac{1}{\pi } & \frac{B}{ \pi ^3} \\
 -8 A & 0 & \frac{B}{\pi ^2} \\
 \frac{ \pi A }{2} & -\frac{\pi }{2} & \frac{B}{2 \pi }
    \end{pmatrix}(\varpi_0^{[u]},\varpi_1^{[u]},\varpi_2^{[u]})^T \, .
\end{align}
with
\beq
A=\frac{8 \Gamma \left(\frac{5}{4}\right)^2}{\Gamma \left(\frac{3}{4}\right)^2} \textrm{~~~and~~~} B=\frac{1}{24} \Gamma \left(-\tfrac{1}{4}\right)^2 \Gamma \left(\tfrac{3}{4}\right) \Gamma \left(\tfrac{7}{4}\right)\,.
\eeq
From this, we can compute the matrix $R^{[u]}(u)$ that determines the transformation behaviour under analytic continuation of the semi-simple and unipotent parts of the Wronskian, see~\cref{eq:Wu_cont}. We find
\begin{align}
    R^{[u]}(u) = \begin{pmatrix}
         \frac{A}{\pi }+\frac{t^{[u]} }{\pi }+\frac{B t^{[u]2}}{2 \pi ^3} & 0 & 0 \\
 \frac{1}{\pi }+\frac{B t^{[u]} }{\pi ^3} & 1 & 0 \\
 \frac{\left(\pi ^2+B t^{[u]} \right)^2}{2 A \pi ^5+\pi ^3 t^{[u]}  \left(2 \pi ^2+B t^{[u]} \right)} & \frac{2 \left(\pi ^2+B t^{[u]} \right)}{2 A
   \pi ^2+t^{[u]}  \left(2 \pi ^2+B t^{[u]} \right)} & \frac{1}{\frac{A}{\pi }+\frac{t^{[u]} }{\pi }+\frac{B t^{[u]2}}{2 \pi ^3}}
    \end{pmatrix} \,,
\end{align}
with 
\beq
A=\frac{8 \Gamma \left(\frac{5}{4}\right)^2}{\Gamma \left(\frac{3}{4}\right)^2} \textrm{~~~and~~~} B=\frac{1}{24} \Gamma \left(-\tfrac{1}{4}\right)^2 \Gamma \left(\tfrac{3}{4}\right) \Gamma \left(\tfrac{7}{4}\right)\,.
\eeq
Notice that $R^{[u]}(u)$ only contains the unipotent quantity $t^{[u]} = \varpi_1^{[u]}/\varpi_0^{[u]}$. It is easy to check, that $R^{[u]}(u)$ indeed satisfies~\cref{eq:randsigma}.

\subsection{$\eps$-factorisation for $n\ge 3$}
From the previous discussions for $n=1$ and $n=2$, we can start to see a pattern emerge. Finding the kernel of $\cL_\eps^{(n+1)}$ is equivalent to solving the homogeneous first-order linear system 
\beq\label{eq:general_GM}
\partial_z I(z,\eps) = \textrm{GM}_{n}(z,\eps)I(z,\eps)\,,\qquad I(z,\eps) = \big(I_1(z,\eps),\ldots,I_{n+1}(z,\eps)\big)^T\,.
\eeq
We define the new basis
\beq\label{eq:Jc_def}
J(z,\eps) = T_{n}(z,\eps)I(z,\eps)\,,
\eeq
with 
\beq\label{eq:T_def}
T_{n}(z,\eps) = \mathbb{T}_{n}(z,\eps)D_{n}(\eps)W_n^{\textrm{ss}}(z)^{-1}\,,
\eeq
where $D_{n}(\eps)$ was defined in~\cref{eq:D_n_def},
and $\mathbb{T}_{n}(z,\eps) = \big(t_{ij}(z,\eps)\big)_{1\le i,j\le n+1}$ is a lower-triangular matrix whose entries are polynomials in $\tfrac{1}{\eps}$,
\beq
\label{eq:tijexp}
t_{ij}(z,\eps) = \left\{\begin{array}{ll}
\sum_{k=0}^{i-j-1}\tfrac{1}{\eps^k}\,t_{ij}^{(k)}(z)\,, &  i>j\,,\\
1\,,&i=j\,,\\
0\,& i<j\,.
\end{array}\right.
\eeq
We can then fix the functions $t_{ij}^{(k)}(z)$ in such a way that $J(z,\eps)$ satisfies a homogeneous differential equation of the form
\beq\label{eq:general_eps_fac}
\partial_zJ(z,\eps) = \eps \,\mathcal{B}_{n}(z)J(z,\eps)\,,
\eeq
with
\begin{align}\label{eq:An_Tn_gauge}
    \epsilon\, \mathcal{B}_{n}(z) &= \left( T_{n}(z,\epsilon) \text{GM}_{n}(z,\epsilon)+ \partial_z T_{n}(z,\epsilon) \right)T_{n}(z,\epsilon)^{-1}  \, .
\end{align}
In practice, this can be done as follows:
Let us define an intermediate basis
\beq
\widetilde{J}(z,\eps) = D_n(\eps)W_n^{\textrm{ss}}(z)^{-1}I(z,\eps)\,,
\label{eq:basisafterrescaling}
\eeq
which satisfies a differential equation whose homogenous part is not in $\eps$-form,
\beq
\label{eq:remainingterms}
\partial_z\widetilde{J}(z,\eps) = \left[\frac{1}{\eps^{n-1}}\widetilde{C}_{1-n}(z)+\ldots+\frac{1}{\eps}\widetilde{C}_{-1}(z)+\widetilde{A}(z)+\eps \,\widetilde{B}(z)\right]\widetilde{J}(z,\eps)\,.
\eeq
Comparing~\cref{eq:basisafterrescaling} to~\cref{eq:T_def}, we clearly have ${J}(z,\eps)=\mathbb{T}_{n}(z,\eps)\widetilde{J}(z,\eps)$. Given the structure in~\cref{eq:tijexp}, we can write
\beq \label{eq:epsexpT}
\mathbb{T}_{n}(z,\epsilon) = \frac{1}{\eps^{n-1}}\mathbb{T}_{n}^{(1-n)}(z)+\ldots+\frac{1}{\eps}\mathbb{T}_{n}^{(-1)}(z)+\mathbb{T}_{n}^{(0)}(z)\,,
\eeq
with 
\begin{equation}
\label{eq:Tcomp}
    \left(\mathbb{T}_{n}^{(-k)}(z)\right)_{i,j}=t_{ij}^{(k)}(z)\, .
\end{equation}
Upon performing the remaining rotation from the basis $\widetilde{J}(z,\eps)$ to $J(z,\eps)$ via the transformation $\mathbb{T}_{n}(z,\epsilon)$, one observes that the terms in $\widetilde{C}_{k}(z)$ in~\cref{eq:remainingterms} can only be cancelled by the functions contained in the $\mathbb{T}_{n}^{(l)}(z)$ with $l\leq k$. This allows us to solve for the $t_{ij}^{(k)}(z)$ iteratively: first, we fix the $t_{ij}^{(n-1)}(z)$ such that they cancel $\widetilde{C}_{1-n}(z)$. Then, we fix the $t_{ij}^{(n-2)}(z)$ such that they cancel $\widetilde{C}_{2-n}(z)$ and so on, until in the last step, we fix the $t_{ij}^{(0)}(z)$ such that they cancel $\widetilde{A}(z)$. Concretely, at step $k$, we obtain unipotent first-order linear differential equations for $t_{ij}^{(n-k)}(z)$, whose inhomogeneous terms involve only $t_{ij}^{(l)}(z)$ with $l>k$ and functions from the field $\cA^{\textrm{ss}}$.

We have explicitly checked this procedure for $n=3$, which establishes that the method of ref.~\cite{Gorges:2023zgv} can also be applied to one-parameter families of CY three-folds, though the final expressions for the functions $t_{ij}^{(k)}(z)$ for $n=3$ are too lengthy to be shown here explicitly. Based on our general 
findings for $n\le 3$, supplemented by the explicit $n=4$ cases that have been worked out 
and some hypergeometric functions presented in table \ref{tab:hypgeo}, we conjecture that it works for all $n$. 

Let us make some additional comments about the functions $t_{ij}^{(k)}(z)$. As they are obtained by iteratively solving unipotent first-order linear differential equations, they can always be expressed as iterated integrals over functions from the field of functions $\cA^{\textrm{ss}}$ attached to the CY operator $\cL_{0}^{(n+1)}$. Unlike for $n\le 2$, however, to the best of our knowledge, it is not understood how one can identify if a function from $\cA^{\textrm{ss}}$ is a total derivative. As a consequence, from $n\ge 3$, it is not obvious how one can decide if the integrals defining the functions $t_{ij}^{(k)}(z)$ can be evaluated in closed form in terms of functions from $\cA^{\textrm{ss}}$. Finally, we mention that only the functions $t_{ij}^{(0)}(z)$ enter the differential equation matrix $\mathcal{B}_n(z)$, which can be seen from the discussion above by noticing that $\mathcal{B}_{n}(z)$ is given by
\beq
\mathcal{B}_{n}(z) = \mathbb{T}_{n}^{(0)}(z)\widetilde{B}(z)\mathbb{T}_{n}^{(0)}(z)^{-1}\,.
\label{eq:finalDEmatrixn}
\eeq

\subsubsection{Analytic continuation for $n=3$}
\label{sec:anal_cont_n3}

Also for $n=3$, let us give an explicit example where we compute the matrix $R^{[u]}(u)$ from~\cref{eq:Wu_cont}. We consider the hypergeometric CY operator defined by $(r_1,r_2,r_3,r_4)=(1/2,1/2,1/2,1/2)$ listed in~\cref{tab:hypgeo}. We begin by stating the Frobanius basis at the MUM-point $z=0$,
\begin{align}
    \varpi_0 &= 1+16 z+1296 z^2+160000 z^3+24010000 z^4 + \mathcal O(z^5) \, , \\
    \varpi_1 &= \varpi_0 \log(z) + S_1 = \varpi_0 \log(z) + 64 z+6048 z^2+\frac{2368000 z^3}{3}+\frac{365638000 z^4}{3} + \mathcal O(z^5) \, , \nonumber \\
    \varpi_2 &= \frac12\varpi_0 \log^2(z) + S_1\log(z) + S_2 \nonumber\\
    &= \frac12\varpi_0 \log^2(z) + S_1\log(z)
    + 32 z+5832 z^2+\frac{8182400 z^3}{9}+\frac{1374099650 z^4}{9} + \mathcal O(z^5) \, \nonumber \\
    \varpi_3 &= \frac16\varpi_0 \log^3(z) + \frac12S_1\log^2(z) + S_2\log(z) \nonumber \\
    &\quad -64 z-4296 z^2-\frac{10334080 z^3}{27}-\frac{1110845155 z^4}{27} + \mathcal O(z^5) \, . \nonumber
\end{align}
From this, we find the quadratic relations
\begin{align}
     W\Sigma W^T = \begin{pmatrix}
    0 & 0 & 0 & \frac{1}{(1-256 z) z^3} \\
 0 & 0 & -\frac{1}{(1-256 z) z^3} & \frac{3-1024 z}{(1-256 z)^2 z^4} \\
 0 & \frac{1}{(1-256 z) z^3} & 0 & -\frac{1-640 z}{(1-256 z)^2 z^5} \\
 -\frac{1}{(1-256 z) z^3} & -\frac{3-1024 z}{(1-256 z)^2 z^4} & \frac{1-640 z}{(1-256 z)^2 z^5} & 0
    \end{pmatrix} \, .
\label{eq:Griffihtsn3}
\end{align}
At $z=1/4^4$ the hypergeometric differential operator in~\cref{eq:HF_DEQ} has another singular point. A local Frobenius basis in $u=z-1/4^4$ that is compatible with the quadratic relations in~\cref{eq:Griffihtsn3} is given by
\begin{align}
\label{eq:newlocalbasisn3}
    \varpi_0^{[u]} &=  1-640 u+\frac{837632 u^2}{3}-\frac{226361344 u^3}{3}+\frac{511063359488 u^4}{27}+ \mathcal O(u^5) \, , \\
    \varpi_1^{[u]} &= u-192 u^2+\frac{118784 u^3}{3}-\frac{25690112 u^4}{3} + \mathcal O(u^5)    \, , \nonumber \\
    \varpi_2^{[u]} &= 65536\varpi_1^{[u]}\log(u) -\frac{39936}{229}+\frac{430973124608 u^3}{2061}-\frac{188136747433984 u^4}{2061} + \mathcal O(u^5) \, , \nonumber \\
    \varpi_3^{[u]} &= \frac{312}{1145}+\frac{49152 u^2}{5}-\frac{3629121536 u^3}{1145}+\frac{2930878971904 u^4}{3435}   + \mathcal O(u^5) \, . \nonumber
\end{align}

We can compute again the analytic continuation between the two different bases. Compared to the lower-dimensional CY cases, it is in general very hard to give analytic expressions (see ref.~\cite{Bonisch:2022mgw}) for the entries of the analytic continuation matrix $T$ defined through
\begin{equation}
    (\varpi_0,\varpi_1,\varpi_2,\varpi_3)^T = T \, (\varpi_0^{[u]},\varpi_1^{[u]},\varpi_2^{[u]},\varpi_3^{[u]})^T \, .
\end{equation}
Therefore, we just give the (approximate) numerical values of $T$
\begin{align}
    T = \left(
    \begin{array}{rrlrr}
        -0.35336    & -361.69984    &+81.48733 i     & 0.00040       & 5.65534    \\
        1.81436     & 1396.19731    &                & 0             & -27.47401  \\
        -4.78627    & -3584.51882   &-536.16515 i   & -0.00260      & 69.92130   \\
        8.53992     & 8234.95848    &+ 783.61927 i  & 0.00381       & -126.38742
    \end{array}\right) \, .
\end{align}
With this input, we can compute $R^{[u]}(u)$ also for a Calabi-Yau three-fold example. The explicit expression for $R^{[u]}(u)$ is too lengthy to be expressed here explicitly. We include 
$R^{[u]}(u)$ as an ancillary file with the arXiv subrmission. We note, however, that as expected, $R^{[u]}(u)$ only depends on the unipotent quantity $t^{[u]} = \varpi_1^{[u]}/\varpi_0^{[u]}$. From this form, one can then also easily verify~\cref{eq:randsigma}.

\subsection{Equivalence to other methods}
\label{sec:equivalence}

In the previous subsection, we have argued that we can apply the method of ref.~\cite{Gorges:2023zgv} to obtain a system of differential equations in $\eps$-form for certain classes of deformed CY operators. In ref.~\cite{Pogel:2022yat, Pogel:2022vat, Pogel:2022ken} an alternative method was introduced and applied to equal-mass banana integrals, and later also extended to Feynman integrals that arise from precision computations for gravitational waves~\cite{Frellesvig:2024rea}. It is then natural to ask what the relation between the canonical bases produced by the two methods is. In the following, we argue that the methods of ref.~\cite{Gorges:2023zgv} and refs.~\cite{Pogel:2022yat, Pogel:2022vat, Pogel:2022ken} are fully equivalent, at least for the class of deformed CY operator introduced in the previous section (which is the class of examples that have been considered in refs.~\cite{Pogel:2022yat, Pogel:2022vat, Pogel:2022ken}).

We start by giving a very brief review of the method of refs.~\cite{Pogel:2022yat, Pogel:2022vat, Pogel:2022ken}. We use the notations from the previous sections, and we start from a deformed CY operator $\cL_{\eps}^{(n+1)}$. 
Also, in this case, we start from a distinguished integral $I_1(z,\eps)$, such that its maximal cut (or its leading singularities) computes the period $\varpi_0(z)$ that is holomorphic at the MUM-point $z=0$. 
One then inductively defines a new set of master integrals $\big(M_1(z,\eps),\ldots,M_{n+1}(z,\eps)\big)^T$ by 
\beq\bsp\label{eq:Weinzierl_review}
M_1(z,\eps) &\,= \epsilon^{n}\frac{I_1(z,\eps)}{\varpi_{0}(z)}\,,\\
M_{i+1}(z,\eps)&\, = \,\frac{1}{Y_{i-2}(q)}\left(\frac{1}{\eps}\theta_qM_i(z,\eps)+\sum_{j=1}^{i}F_{ij}(z)M_j(z,\eps)\right)\,,\qquad 1\leq i\leq n\,.
\esp\eeq
At this point, the functions $F_{ij}(z)$ are undetermined and are specified by the requirement that $\theta_qM_{n+1}$ is in $\epsilon$-form.
Moreover, it is easy to see by using the ansatz recursively, that $\theta_qM_{n+1}\propto \theta_q^{n+1}M_1$. This means, since we are disregarding possible subtopologies, that the integral $I_1(z,\eps)$ satisfies a homogeneous differential equation of order $(n+1)$.
We now show that this method produces the same basis of master integrals 
as the one introduced in the previous section (possibly up to some constant rotation).

To get started, we focus on a particular homogeneous solution of eq.~\eqref{eq:general_GM}, namely the one where $I_1(z,\eps) = \varpi_n(z) + \ord(\eps)$, i.e., where $I_1(z,\eps)\sim \log^nz+\ord(z,\eps)$ diverges like the $n^{\textrm{th}}$ power of a logarithm close to the MUM-point. The resulting derivative basis from~\cref{eq:derbasisn} then corresponds to the $n^{\textrm{th}}$ column of the Wronskian in eq.~\eqref{eq:CY_Wronskian}, up to terms of order $\eps$. From section~\ref{sec:SS_U} we know that we can write
\beq
I(z,\eps) = W^{\textrm{ss}}_n(z)u(q) + \ord(\eps)\,,
\eeq
 where $u(q) = $ denotes the vector
\beq
u(q) = \begin{pmatrix} I(1,Y_1,\ldots, Y_1,1;q) \\
I(Y_1,\ldots, Y_1,1;q) \\
\vdots\\
I(1;q) \\1\end{pmatrix} = 
\cD_{q}I(1,Y_1,\ldots, Y_1,1;q)\,,
\eeq
and $\cD_q$ denotes the vector-valued differential operator
\beq
\cD_q = \begin{pmatrix} 1 \\
\theta_q \\
\vdots\\
\frac{1}{Y_1}\theta_q\cdots \theta_q\frac{1}{Y_1}\theta_q^2 \\\theta_q\frac{1}{Y_1}\theta_q\cdots \theta_q\frac{1}{Y_1}\theta_q^2\end{pmatrix}\,.
\eeq
We then define a new basis of master integrals
\beq
N(z,\eps) = \cD_q\frac{I_1(z,\eps)}{\varpi_0(z)}\,.
\eeq
Using the identity $\theta_q= \alpha_1(z)\theta_z$, we can easily see that there is a matrix $R(z)$ such that
\beq
I(z,\eps) = R(z)N(z,\eps) = R(z)u(q) + \ord(\eps)\,,
\eeq
which in turn implies
\beq
R(z)u(q) = W^{\textrm{ss}}_n(z)u(q)\,.
\eeq
We can repeat exactly the same argument for all other columns of the Wronskian $W_n(q)$, and since the latter are linearly independent, we find that $R(z)=W^{\textrm{ss}}_n(z)$, and so 
\begin{equation}
\label{eq:basisaftersesi}
    N(z,\eps) = W^{\textrm{ss}}_n(z)^{-1}I(z,\eps) \, .
\end{equation}
In other words, we see that inductively defining a new basis of master integrals by differentiating with $\theta_q$ and rescaling by the Yukawa couplings, which is part of the method of ref.~\cite{Pogel:2022yat, Pogel:2022vat, Pogel:2022ken}, cf.~eq.~\eqref{eq:Weinzierl_review}, is equivalent to dividing by the semi-simple part of the period matrix.

Next, applying the $\eps$-rescalings to~\cref{eq:basisaftersesi}, it is easy to see that the components of the basis $\widetilde{J}(z,\eps)$ in \cref{eq:basisafterrescaling} satisfy the recursion
\beq
\widetilde{J}_{i+1}(z,\eps) = \frac{1}{\eps}\frac{1}{Y_{i-2}(q)}\theta_q\widetilde{J}_i(z,\eps)\,.
\eeq
Finally, from eqs.~\eqref{eq:Jc_def} and~\eqref{eq:T_def}, we see that the canonical basis obtained by the method of ref.~\cite{Gorges:2023zgv} is given by
\beq\label{eq:Jc_recursion}
J(z,\eps) = \mathbb{T}_{n}(z,\eps)\widetilde{J}(z,\eps)\,,
\eeq
and since $ \mathbb{T}_{n}(z,\eps)$ is lower triangular, we have the recursion
\beq
J_{i+1}(z,\eps) = \frac{1}{\eps}\frac{1}{Y_{i-2}(q)}\theta_qJ_i(z,\eps) + \sum_{j=1}^it_{ij}(z,\eps)J_j(z,\eps)\,.
\eeq
Comparing eqs.~\eqref{eq:Weinzierl_review} and~\eqref{eq:Jc_recursion}, we see that the bases obtained from the two different methods satisfy a recursion of the same type, with the same initial condition $M_1(z,\eps) = J_1(z,\eps)$. The unknown functions $F_{ij}(z)$ and $t_{ij}(z,\eps)$ are fixed such that the vectors $M(z,\eps)$ and $J(z,\eps)$ satisfy a system of differential equations in $\eps$-form. The solutions for these functions are unique, up to fixing initial conditions for the differential equations they satisfy. We therefore conclude that the bases obtained from the two methods are equivalent, up to possibly choosing different initial conditions. 

Let us finish by making a comment about the functions $t_{ij}(z,\eps)$. From the previous discussion, it immediately follows that we can express the $t_{ij}(z,\eps)$ in terms of the $F_{ij}(z),\,\varpi_0$, the Yukawa couplings and algebraic functions of $z$. Since only the functions $t_{ij}^{(0)}(z)$ enter the differential equation matrix $\mathcal{B}_n(z)$, we can express the $t_{ij}^{(0)}(z)$ in terms of polynomials in $F_{ij}(z)$. The functions $t_{ij}^{(k)}(z)$ with $k>0$ must cancel out, and by relating them to $F_{ij}(z)$, we see that the $t_{ij}^{(k)}(z)$ with $k>0$ can be expressed in terms of the functions $t_{ij}^{(0)}(z)$ and their derivatives.

\subsection{Some properties of deformed CY operators}
\label{sec:properties}

We have derived the $\eps$-form of the differential equations for equal-mass banana integrals with up to five loops and for all the hypergeometric functions listed in~\cref{tab:hypgeo}.
For all of these examples, we obtain a deformed CY operator of the form discussed earlier. The considered examples range from CY two-folds up to CY four-folds. Based on these examples, we can observe some general recurrent features discussed in the subsequent paragraphs, which are shared by all examples. Some of these properties had already been observed for the equal-mass banana integrals in refs.~\cite{Pogel:2022yat, Pogel:2022vat, Pogel:2022ken}. We expect these features to apply to all examples from our class of deformed CY operators.

\subsubsection{Integrality of the $q$-expansion}
\label{sec:integer}

It follows from the mirror symmetry conjecture that the expansion close to the MUM-point of the holomorphic period~\cref{eq:frob}, the Yukawa or $n$-point coupling~\cref{eq:defnptcoupling}, and the mirror map~\cref{eq:mirrormap} in the canonical coordinate $q$ in~\cref{eq:canvar} have integer coefficients (possibly up to overall normalisation). This property is preserved by differentiation, and it follows that all elements from the algebra $\cA^{\textrm{ss}}$ have integer $q$-expansions.

The entries of the differential equation matrix $\mathcal{B}_{n}$ (see~\cref{eq:finalDEmatrixn}), however, are not all from $\cA^{\textrm{ss}}$, but they may contain iterated integrals over functions from $\cA^{\textrm{ss}}$. These iterated integrals do not necessarily admit integer coefficients for their $q$-expansion close to the MUM-point. In refs.~\cite{Pogel:2022yat, Pogel:2022vat, Pogel:2022ken} it was observed that in the case of equal-mass banana integrals, the iterated integrals that enter the differential equation matrix admit integral $q$-expansions. We observe that the same property extends to all examples that we have considered, and we conjecture that this holds for all our deformed CY operators. 

We stress that the fact that the iterated integrals that enter $\mathcal{B}_{n}$ have an integral $q$-expansion is highly non-trivial! To corroborate this point, let us focus on the case $n=2$. We have already mentioned that, in this case, the CY operator $\cL_0^{(3)}$ is the symmetric square of a Picard-Fuchs operator describing a family of elliptic curves, and so the periods and the mirror map admit a modular parametrisation and the functions from~\cref{eq:gk3} reduce to iterated integrals of (meromorphic) modular forms~\cite{ManinModular, Brown:mmv, matthes_AMS}. The integrality of the $q$-expansion of these iterated integrals of modular forms then translates into certain congruence properties for the coefficients of the $q$-expansion of those modular forms. Specifically, we discover instances of \emph{magnetic modular forms}~\cite{magnetic1, magnetic2, magnetic3, magnetic4, Bonisch:2024nru}. Loosely speaking, a modular form $f(t) = \sum_{n\ge 1}a_nq^n$ is called \emph{magnetic} of depth $d\ge 0$ if $a_n$ is divisible by $n^d$. This property ensures that the first $d$ primitives of $f(t)$ have an integral $q$-expansion. There are not many known examples of magnetic modular forms, especially not for depth $d\ge 2$. The canonical forms of the differential equations obtained from our deformed CY operators appear to provide an interesting source of magnetic modular forms of weights 4 and 6 and depth $1$ and $2$, respectively. Some of the authors have used this property to obtain novel results for magnetic modular forms for the congruence subgroups $\Gamma(2)$, $\Gamma_0(2)$ and $\Gamma_1(6)$~\cite{Bonisch:2024nru} (see also ref.~\cite{Pogel:2022yat}). In the future, it would be interesting to see if one can use our deformed CY operators to discover more instances of magnetic modular forms, including examples attached to the higher-dimensional cases, i.e., $n>2$.

\subsubsection{Independence of the integration kernels}

We expand the differential equation matrix $\mathcal{B}_{n}(q)$ into a basis of functions (see~\cref{eq:mat_basis_dec})\footnote{The matrix $\mathcal{B}$ in~\cref{eq:mat_basis_dec} is a matrix of one-forms. For one-variable cases like those considered in this section, it is more convenient to consider the matrix $\mathcal{B}_{n}(q)$ a matrix of functions. The relationship is $\mathcal{B}_{n} = \tfrac{\rd q}{q}\,\mathcal{B}_{n}(q)$.}
\beq
\mathcal{B}_{n}(q) = \sum_{i=1}^{d}B_i\,f_i(q)\,,
\eeq
where $d$ is a finite integer and the $B_i$ are constant matrices. Notice that the number of independent $f_i$ is bounded by the number of entries in the matrix,  $d\le (n+1)^2$.

The solution to the differential equation will then involve iterated integrals over the letters $f_i(q)$. 
In section~\ref{sec:roadmap_summary} we have conjectured that the functions $f_i(q)$ are independent up to total derivatives. In the present case,~\cref{eq:independence_general} reduces to the statement that
\beq\label{eq:independence_eq}
\sum_{i=1}^dc_i\,f_i(q) = \theta_qg(q)\,,\qquad c_i\in\mathbb{C}, \qquad g\in\cA^{\textrm{ss}},
\eeq
admits only the trivial solution $c_i=0$, for all $1\le i\le d$ (in which case $g(q)$ must be a constant). 
Answering the independence question, however, can be a complicated task, because one needs to exclude the existence of a function $g$ such that eq.~\eqref{eq:independence_eq} admits a non-trivial solution. 

In the previous section we have argued that for our class of deformed CY operators, the $f_i(q)$ admit an integral $q$-expansion. In that case, one can exclude the existence of a function $g(q)$ with high confidence, provided that enough coefficients in the $q$-expansion of the $f_i(q)$ are known. To see how this works, assume  the $q$-expansions of the $f_i(q)$ and $g(q)$ are given by,
\beq
f_i(q) = \sum_{k=0}^{\infty}a_{ik}q^k\,,\qquad g(q) = \sum_{k=0}^\infty b_{k}q^k\,.
\eeq
Note that $b_k\in\mathbb{Z}$, because $g\in\cA^{\textrm{ss}}$.
Then eq.~\eqref{eq:independence_eq} is equivalent to the infinite linear system with integer coefficients for the unknowns $c_{i}$ and $b_k$,
\beq
\sum_{i=1}^dc_ia_{ik} = k\,b_k\,,\qquad k\in\mathbb{Z}_{\ge 0}\,.
\eeq
Since the system is defined over the integers, the coefficients $c_i$ can be chosen to be integers, and reducing the $k^{\textrm{th}}$ equation modulo $k$, we see that the unknowns $b_k$ drop out:
\beq
\sum_{i=1}^dc_ia_{ik} = 0\!\!\mod k\,,\qquad k\in\mathbb{Z}_{\ge 0}\,.
\eeq
We see that we can reduce eq.~\eqref{eq:independence_eq} to an infinite system of congruence conditions for the coefficients $c_i$. In particular, the dependence on the coefficients $b_k$, which parametrise the unknown function $g(q)$, has completely dropped out.
In applications, we will typically have access to the first $p>d$ terms in the $q$-expansion of the $f_i(q)$. The resulting system of $p$ congruence conditions can be solved using modern computer algebra systems, and we obtain a solution for the coefficients $c_i$ modulo $K=\textrm{lcm}(1,\ldots,p)$, where lcm$(a_1,\ldots,a_p)$ denotes the least common multiple of $a_1,\ldots,a_p$. If $p$ is large enough, we can exclude the existence of a non-trivial solution with $|c_i|<K$, and so if enough terms in the $q$-expansion are known, then we can exclude a non-trivial solution with a high-level of confidence. 

We have applied this method to our examples of deformed CY operators, and in all cases, we find that we can exclude the existence of a non-trivial solution with a high level of confidence, or equivalently, we believe with a high level of confidence that the letters $f_i(q)$ are independent. We conjecture that this holds in all cases. This would be in line with the situation of $\eps$-factorised systems in dlog-form, where the resulting letters are always independent.

\subsubsection{Logarithmic singularities}

Pure functions, both in the polylogarithmic case~\cite{Arkani-Hamed:2010pyv} and beyond~\cite{Broedel:2018qkq}, are expected to be closely connected to functions with at most logarithmic singularities. It is thus natural to ask if the class of iterated integrals that solve the $\eps$-factorised differential equation has this property. Since the notion of transcendental weight is defined locally close to the MUM-point at $z=0$, we will analyse the behaviour of the resulting functions, or equivalently of the matrix of differential forms, close to this MUM-point. We will find that, as conjectured in section~\ref{sec:roadmap_summary}, the differential equation only has simple poles at the MUM-point.

We start by analyzing hypergeometric CY operators (see~\cref{eq:hyp_geo_def}). In other words, we focus on Fuchsian differential operators of the form
\beq\label{eq:again}
\cL_{\eps}^{(n+1)} =  z\prod_{k=1}^{n+1} (\theta_z+r_k+s_k\eps)-\theta_z\prod_{k=1}^n(\theta_z+t_k\eps) \,,
\eeq
where $s_k$ and $t_k$ are constants, and the $r_k$ are rational numbers such that $\cL_{0}^{(n+1)}$ is a CY operator. We can see $\cL_{\eps}^{(n+1)}$ as a polynomial in two variables $z$ and $\theta_z$. The constant term in $\theta_z$ is
\begin{equation}
    \left.\cL_{\eps}^{(n+1)}\right|_{\theta_z=0} = z\,\prod_{i=1}^{n+1} (r_k+s_k\eps)\,.
\end{equation}
It is easy to see that the discriminant of the differential operator (i.e., the coefficient of $\partial_z^{n+1}$) is $(z-1)z^{n+1}$. 
It follows that, if we normalise the operator in~\cref{eq:again} by the discriminant, the term in $\cL_{\eps}^{(n+1)}\big|_{\theta_z=0}$ without any derivatives has a pole of order (at most) $n$ at $z=0$. We stress that for a Fuchsian operator of order $(n+1)$, the constant term in $\theta_z$ could, in principle, have a pole of order $(n+1)$. 

We know that finding the kernel of $\cL_{\eps}^{(n+1)}$ is equivalent to solving the linear system in~\cref{eq:general_GM}, and we can transform this system into the $\eps$-factorised form in~\cref{eq:general_eps_fac}. We now argue that the matrix $\mathcal{B}_n(z)$ in~\cref{eq:general_eps_fac} has at most simple poles at the MUM-point $z=0$. We discuss the case $n=1$ in detail, but we have checked that the same kind of argument can be applied for higher values of $n$, and we expect that the conclusions hold for all values of $n$.

In the case $n=1$, finding the kernel of the operator is equivalent to solving the first-order linear system defined by the matrix $\text{GM}_1(z;\epsilon)$ in~\cref{eq:gmell}. From the previous discussion, it follows that the functions $p_0(z)$, $q_0^{(1)}(z)$ and $q_0^{(2)}(z)$ that enter $\text{GM}_1(z;\epsilon)$ have at most a simple pole at $z=0$. Our goal is now to trace how the order of the pole evolves when performing the sequence of gauge transformations in~\cref{eq:rotell} required to bring the system into $\eps$-factorised form. 

Recall that after the sequence of gauge transformations in~\cref{eq:rotell}, the matrix of the system of differential equations is $\mathcal{A}_1(z)+\eps\mathcal{B}_1(z)$, where $\mathcal{A}_1(z)$ and $\mathcal{B}_1(z)$ are defined in eqs.~\eqref{eq:A1_def} and~\eqref{eq:epsell}, and $\mathcal{A}_1(z)$ vanishes after we impose~\cref{eq:t21_DEQ}. From~\cref{eq:epsell} we see that the matrix $\mathcal{B}_1(z)$ involves the one-point coupling $C_1(z)$ and the new function $t_{21}(z)$ defined by~\cref{eq:t21_DEQ}. Hence, in order to understand the behaviour of $\mathcal{B}_1(z)$ close to $z=0$, we need to understand the behaviour of $C_1(z)$ and $t_{21}(z)$. 

Let us start by discussing the behaviour of $C_1(z)$ close to $z=0$. The one-point coupling is a rational solution to the differential equation in~\cref{eq:defnptcoupling}. Since $\mathcal{L}_0^{(n+1)}$ is Fuchsian, the rational function $p_n(z)$ on the right-hand side of~\cref{eq:defnptcoupling} can at most have a simple pole at $z=0$. In all cases considered, $C_1(z)$ has at most a simple pole at $z=0$. From~\cref{eq:t21_DEQ} we then see that $t_{21}(z)$ is regular at $z=0$. Finally, using the fact that $\mathcal{L}_{\eps}^{(n+1)}$ is Fuchsian and the functions $q_m^{(k)}$ have at most simple poles, we arrive at the conclusion that $\mathcal{B}_1(z)$ has at most a simple pole at $z=0$. 

We have performed the same analysis also for the cases $n=2$ and $n=3$. In both cases, we checked that the functions introduced in~\cref{eq:gk3} and the $t^{(0)}_{ij}(z)$ from~\cref{eq:Tcomp} for $n=3$ are holomorphic around $z=0$ and that the final differential equation matrix $\mathcal{B}_{n}$ has only simple poles at $z=0$. We conjecture that this pattern continues for $n>3$. Note that in this analysis, it was important to understand the behaviour of the $n$-point coupling $C_n(z)$, which can always be done by studying its defining differential equation~\cref{eq:defnptcoupling}. 

We can extend the previous analysis to the banana integrals, whose Picard-Fuchs operators are CY operators that are not of hypergeometric type. We do not have an all-loop form for these operators, but we can analyse them order by order. Methods for constructing the Picard-Fuchs operators of banana graphs in dimensional regularization can, for example, be found in refs.~\cite{Bonisch:2021yfw, Mishnyakov:2023wpd, delaCruz:2024xit}. We have checked the following arguments up to four loops, corresponding to $n \leq 3$. By explicit inspection with a choice of kinematic variable $z$ such that the MUM-point lies at $z=0$, we notice that the terms in the Picard-Fuchs operator containing no derivative take the form
\begin{equation}
    z\,\left(\sum_{i=0}^{n}\epsilon^i\,f_i(z)\right)+\epsilon^{n+1}\,\mathcal{O}(z^0) \, ,
\end{equation}
where the functions $f_i(z)$ are polynomials in $z$. As before, this implies that the operator normalised by its discriminant contains poles at $z=0$ of order at most $n$, with the exception of the highest appearing power of $\epsilon$, $\epsilon^{n+1}$,  which has the highest possible pole allowed for a Fuchsian differential operator of order $(n+1)$, i.e., a pole of order $(n+1)$. However, the term in the Picard-Fuchs operator proportional to $\epsilon^{n+1}$ only enters the term $\widetilde{B}$ in~\cref{eq:remainingterms}. Consequently, we observe similarly to the hypergeometric examples that the functions $t_{ij}^{(k)}$ are all holomorphic around $z=0$. Further, we checked that this term indeed only produces a simple pole in $\mathcal{B}_n$.

To conclude, we observe that for all the examples that we have considered, the canonical differential equations obtained for deformed CY operators using the method of ref.~\cite{Gorges:2023zgv} have at most simple poles at the MUM-point $z=0$, and, following the discussion in section~\ref{sec:roadmap_summary}, we expect that this property applies to all canonical differential equations.

\subsubsection{A comment on self-duality}

In refs.~\cite{Pogel:2022yat, Pogel:2022vat, Pogel:2022ken} it was observed that for equal-mass banana integrals, it is possible to find a basis of master integrals such that the differential equation matrix $\mathcal{B}_n(z)$ in eq.~\eqref{eq:An_Tn_gauge} is persymmetric, i.e., it satisfies the condition
\beq\label{eq:Weinzierl_SD}
\mathcal{B}_n(z)_{n+2-j,n+2-i} = \mathcal{B}_n(z)_{ij}\,.
\eeq
This property was attributed to `self-duality', which in ref.~\cite{Pogel:2024sdi} was defined as the property that the differential equation matrix is persymmetric. We observe that for some of the hypergeometric functions in~\cref{tab:hypgeo}, we are not able to find a basis such that the matrix $\mathcal{B}_n(z)$ is persymmetric for generic values of the $s_k$ and $t_k$, but persymmetry can in general only be achieved if the $s_k$ and $t_k$ satisfy certain additional linear relations. In the following, we explain why it is generically not possible to obtain a persymmetric matrix $\mathcal{B}_n(z)$ for a generic deformed CY operator, and we also comment on the situation of Feynman integrals. 

Let us start by discussing the situation for $\eps=0$.
In section~\ref{eq:CY_ops}, we have seen that a CY operator $\cL_0^{(n+1)}$ is by definition essentially self-adjoint. Essential self-adjointness is equivalent to the existence of a monodromy-invariant pairing compatible with Griffith transversality~\cite{BognerThesis, BognerCY}. Together with other defining properties of CY operators, like the existence of a MUM-point, this implies the symmetry property for the $Y$-invariant in~\cref{eq:Y_self-duality}~\cite{BognerThesis, BognerCY}, which in turn implies the persymmetry of the differential equation matrix $N_n(q)$ for the pure part of the Wronskian $W^{\textrm{u}}(q)$, see~\cref{eq:N_n(q)}. 

For $\eps\neq0$, however, the deformed operator $\cL_{\eps}^{(n+1)}$ does not need to be (essentially) self-adjoint, and even if it is, there is no apparent reason why its self-adjointness implies the persymmetry of the differential equations matrix in~\cref{eq:Weinzierl_SD} (because, e.g., for $\eps\neq0$ there is no MUM-point). Nevertheless, persymmetry is observed to hold for the maximal cuts of the equal-mass banana integrals.\footnote{On a case-by-case basis, assuming that the differential forms in the canonical differential equation matrix are independent, it is possible to prove it as a consequence of the constancy of the intersection matrix~\cite{Duhr:2024xsy}.}

In order to understand the connection between essential self-adjointness (for $\eps=0$), self-duality, and persymmetry of the differential equation matrix, we use the theory of twisted cohomology~\cite{aomoto_theory_2011, Mastrolia:2018uzb}, which comes with a natural notion of self-duality (which is distinct from the one introduced in refs.~\cite{Pogel:2022yat, Pogel:2022vat, Pogel:2022ken}).
That notion of self-duality is, in essence, the same as the notion of essential self-adjointness in~\cref{eq:essselfadj}, which means that $\cL_0^{(n+1)}$ and its dual share the same solution space, up to multiplication by an algebraic function $A(z)$. Equation~\eqref{eq:essselfadj}, however, is valid for the differential operator at $\eps=0$, and in order to understand self-duality or essential self-adjointness for the full integrals in dimensional regularization, we need to understand how eq.~\eqref{eq:essselfadj} extends to $\eps\neq0$. 

Loosely speaking, twisted cohomology provides a framework to study integrals of the form
\beq\label{eq:dual_ints}
\int_{\Gamma}\Phi \, \omega\,,
\eeq
where $\Gamma$ is a $n$-cycle and $\omega$ is a rational $n$-form. $\Phi$ is a multi-valued function called the \emph{twist}, which can be written as
\beq\label{eq:twist}
\Phi = \prod_{i}P_i(\underline{x},\underline{z})^{a_i}\,,
\eeq
where $P_i(\underline{x},\underline{z})$ are polynomials that depend both on the integration variables $\underline{x}$ and the external parameters $\underline{z}$, and $a_i$ are non-integer complex numbers. The dual integral to eq.~\eqref{eq:dual_ints} is then defined as the integral obtained by inverting the twist,
\beq\label{eq:dual_dual_ints}
\int_{\Gamma}\Phi^{-1} \, \omega\,.
\eeq
In this context, \emph{self-duality} means that eqs.~\eqref{eq:dual_ints} and~\eqref{eq:dual_dual_ints} define the \emph{same} class of integrals. In particular, all the hypergeometric functions that are solutions of CY operators have this property (see table~\ref{tab:hypgeo}). Self-duality implies that (in the case where the integrals depend on a single external parameter $z$) the differential operators that annihilate the integrals are essentially self-dual according to the definition in eq.~\eqref{eq:essselfadj}.

So far, the notion of self-duality applies to the situation where the exponents $a_i$ in eq.~\eqref{eq:twist} are rational numbers. We now consider the case where $a_i=r_i+s_i\eps$, which is the typical case encountered for our deformed CY operators. The integrals in eqs.~\eqref{eq:dual_ints} and~\eqref{eq:dual_dual_ints} are functions of $\eps$, and inverting the twist is clearly related to replacing $\eps$ by $-\eps$. An appropriate generalization of the notion of essential self-adjointness to our case is, therefore, that there is an algebraic function $A(z)$ such that
\begin{equation}
\label{eq:essentialSD_eps}
    \mathcal L_{-\eps}^{(n+1)} A(z) = A(z) \mathcal L_{\eps}^{*(n+1)} \, .
\end{equation}
Note that the previous equation reduces to eq.~\eqref{eq:essselfadj} for $\eps=0$, and if $ \mathcal L_{\eps}^{(n+1)}$ belongs to our class of deformed CY operators, then $A(z)$ is again given by the $n$-point coupling of the CY operator $\cL_{0}^{(n+1)}$. 
We stress that the essential self-adjointness of the CY operator $\cL_{0}^{(n+1)}$ does not imply eq.~\eqref{eq:essentialSD_eps}. 

Let us illustrate this in an example. Consider the hypergeometric function (which corresponds to the second entry in table~\ref{tab:hypgeo})
\beq
F(z) =   {_{3}F_{2}}\left(\tfrac{1}{4}+s_1\epsilon,\tfrac{2}{4}+s_2\epsilon,\tfrac{3}{4}+s_3\epsilon;1+t_1\epsilon,1+t_2\epsilon;z\right)\,.
  \eeq
It is easy to write down the third-order differential operator that annihilates $F(z)$:
\begin{align}
\cL_{\eps}^{(3)} =& (1- z) z^3 \partial_z^3-\frac{3}{2}(3z-2) z^2 \partial_z^2+(1-\frac{51}{16} z) z \partial_z-\frac{3}{32} z \nonumber\\
&+\epsilon\biggl[ (t_1+t_2- z\,(s_1+s_2+s_3) )z^2 \partial_z^2\nonumber\\
&+ (t_1+t_2-\frac{1}{4}\,z \,(9 s_1+8 s_2+7 s_3) )z \partial_z-\frac{1}{16} \,z\,(6 s_1+3 s_2+2 s_3)  \biggr]\nonumber\\
&+\epsilon^2\biggl[ (t_1 t_2- \,z(s_2 s_3+s_1 s_2+s_1s_3))z \partial_z-\frac{1}{4} \,z(s_2 s_3+3 s_2s_1 +2 s_3s_1)  \biggr]\nonumber\\
&-\epsilon^3 s_1 s_2 s_3 \,z \,.
\end{align}
It is known that $\cL_0^{(3)}$ is a CY operator, and so $\cL_0^{(3)}$ is essentially self-adjoint. $\cL_{\eps}^{(3)}$ belongs to our class of deformed CY operators. It is easy to check, however, that we cannot find a function $A(z)$ such that $\cL_{\eps}^{(3)}$ satisfies eq.~\eqref{eq:essentialSD_eps}. In other words, this is an example where $\cL_{0}^{(3)}$ is essentially self-adjoint, but $\cL_{\eps}^{(3)}$ is not. Moreover, 
we cannot find a canonical basis such that differential equation matrix $\mathcal{B}_2(z)$ satisfies eq.~\eqref{eq:Weinzierl_SD}. This is consistent with the results of ref.~\cite{Duhr:2024xsy}, where self-duality was an essential feature to arrive at a persymmetric differential equation matrix.
We may ask if we can find values for $(s_1,s_2,s_3,t_1,t_2)$ such that $\cL_{\eps}^{(3)}$ satisfies eq.~\eqref{eq:essentialSD_eps}. The answer is positive, and we see that for the choice $s_1=s_3$, $\mathcal{B}_2(z)$ is persymmetric. 

The previous example shows that essential self-adjointness for $\eps=0$ does not imply essential self-adjointness for $\eps\neq0$. One may wonder if we can find a criterion when the deformed operator is essentially self-adjoint for all values of $\eps$. In the following, we present a sufficient (but by no means necessary) condition~\cite{Duhr:2024rxe}. Indeed, it is easy to see that eqs.~\eqref{eq:dual_ints} and~\eqref{eq:dual_dual_ints} define the same class of integrals up to changing the sign of $\eps$ if there are integers $m_i$ such that the twist at $\eps=0$ satisfies
\beq
\Phi_{|\eps=0}^{-1} = \Phi_{|\eps=0} \prod_{i}P_i(\underline{x},\underline{z})^{m_i}\,.
\eeq
If this equation is satisfied, then for $\eps=0$, the twist and its inverse are identical, up to multiplication by a rational function.
This rational function in the right-hand side may then be absorbed into the rational differential form $\omega$, and so eqs.~\eqref{eq:dual_ints} and~\eqref{eq:dual_dual_ints} define the same class of integrals. Note that this sufficient condition is always satisfied if the exponents $a_i$ that define the twist are $a_i=\tfrac{m_i}{2}+\alpha_i\eps$, $m_i\in\mathbb{Z}$, $\alpha_i\in\mathbb{Q}$. From the Baikov representation for Feynman integrals, it is easy to see that this condition can always be satisfied for maximal cuts of Feynman integrals, and so maximal cuts of Feynman integrals are always self-dual~\cite{Duhr:2024rxe, Duhr:2024xsy}. 
As an example, we consider the hypergeometric function (which corresponds to the first entry in table~\ref{tab:hypgeo})
\begin{equation}\label{eq:3F2_Legendre}
    {_{3}F_{2}}\left(\tfrac{1}{2}+s_1\epsilon,\tfrac{1}{2}+s_2\epsilon,\tfrac{1}{2}+s_{3}\epsilon;1+t_1\epsilon,1+t_2\epsilon;z\right)\,.
\end{equation}
This function is annihilated by the differential operator
\begin{align}
\cL_{\eps}^{(3)} = & (1- z) z^3 \partial_z^3-\frac{3}{2}(3z-2) z^2 \partial_z^2+(1-\frac{13}{4} z) z \partial_z-\frac{1}{8} z \nonumber\\
&+\epsilon\biggl[ (t_1+t_2-\,z (s_1+s_2+s_3) )z^2 \partial_z^2\nonumber\\
&+ (t_1+t_2-2\,z (s_1+s_2+s_3) )z \partial_z-\frac{1}{4}\,z (s_1+s_2+s_3)  \biggr]\nonumber\\
&+\epsilon^2\biggl[ (t_1 t_2- \,z(s_2 s_3+s_1 s_2+s_1 s_3) )z \partial_z-\frac{1}{2}\,z (s_2 s_3+s_1 s_2+s_1s_3) \biggr]\nonumber\\
&-\epsilon^3s_1 s_2 s_3 \,z \biggr].
\end{align}
$\cL_0^{(3)}$ is a CY operator.
We can easily see from the integral representation in eq.~\eqref{eq:hyp_geo_def} that for $\eps=0$ all exponents that appear in the twist are either integer or half integers, and we have explicitly verified that $\cL_{\eps}^{(3)}$ is essentially self-adjoint for all values of $(s_1,s_2,s_3,t_1,t_2)$. We can also find a basis such that $\mathcal{B}_2(z)$ is persymmetric for all values of $(s_1,s_2,s_3,t_1,t_2)$.
\usetikzlibrary{decorations.pathmorphing}
\usetikzlibrary{decorations.markings}
\usetikzlibrary{positioning, shapes, snakes, arrows}

\tikzset{ 
	graviton/.style={line width=.8pt, -latex,decorate, decoration={snake, segment length=4pt,amplitude=1.8pt, pre length=.1cm, post length=.25cm}},
	worldline/.style={gray, line width=1pt},
	worldlineBold/.style={black, line width=.6pt},
        background/.style={black,dotted,line width=1pt},
	zUndirected/.style={line width=1pt},
	zParticle/.style={line width=1pt,postaction={decorate},decoration={markings,mark=at position .6 with {\arrow[#1]{latex}}}},
	zParticleF/.style={line width=1pt,postaction={decorate}},
	cscalar/.style={line width=1pt,postaction={decorate},decoration={markings,mark=at position .6 with {\arrow[#1]{latex}}}},
	cscalar2/.style={line width=1pt,postaction={decorate},decoration={markings,mark=at position .8 with {\arrow[#1]{latex}}}},
	photon/.style={line width =.8pt, decorate, decoration={snake, segment length=3pt, amplitude=1.8pt,  pre length=.1cm, post length=.1cm}},
	 mid arrow/.style={postaction={decorate,decoration={
        markings,
        mark=at position .5 with {\arrow[#1]{latex}}}}} ,
        worlddot/.style={dotted, line width=.8pt},
	worlddot2/.style={dotted, line width=1pt}   }

\section{Additional examples beyond pure CY operators}
\label{sec:additional_examples}

In the previous section, we have demonstrated that our method works for deformed CY operators. A key step was that we could easily choose the basis of integrals such that it aligns with the MHS, in line with the general roadmap from section~\ref{sec:roadmap}, because the CY operators describe pure Hodge structures whose Hodge numbers satisfy~\cref{eq:Hodge_numbers_CY_ops} (plus additional properties). 

For more general examples, it becomes increasingly more challenging to prove that our method works in general. In particular, identifying a starting basis compatible with the underlying MHS for $\eps=0$ may be a very challenging task. However, explicit calculations have already demonstrated that its applicability is not limited to the case of deformed CY operators.
In this section, we discuss further state-of-the-art examples, which, we believe, showcase important features of our approach. In addition, these examples allow us to illustrate that, even in cases where we cannot easily identify the complete MHS in question, we can nevertheless combine the general ideas presented in section~\ref{sec:roadmap} with intuition from physics to arrive at an $\eps$-factorised differential equation.
Central to our construction are the following steps:

\begin{enumerate}
\item First of all, once all geometries on the maximal cut have been identified (by which we mean, loosely speaking, the different graded pieces in the MHS), as the most important step, we start by selecting the Feynman integrals whose integrands correspond to the respective holomorphic $(n,0)$-differential forms. 
The remaining integrals, which often map to the independent forms with higher poles, can then be chosen as their derivatives.

\item Second, if present, one should always single out all Feynman integrals that are not independent for $\eps=0$ (and can, therefore, be decoupled from the homogeneous system of differential equations trivially for $\eps=0$) or whose maximal cuts in $\eps=0$ 
can be further localised by taking residues. 
We stress here that, depending on the problem at hand, it might be prohibitive to perform a complete integrand analysis to identify all forms with logarithmic singularities. Nevertheless, as we will see in two concrete examples in~\cref{sec:GWexamples}, this issue does not constitute a substantial bottleneck and can be circumvented by a case-by-case analysis.

\item Third, for each graded piece in the MHS, we decompose the corresponding Wronskian matrices on each maximal cut into a semi-simple and a unipotent part, we rotate away the semi-simple part, and finally rescale the integrals by appropriate powers of $\eps$. This step involves, in particular, the normalisation of each integrand such that all residues are constant (if there are any).
\end{enumerate}

In the following, we illustrate this procedure on three non-trivial state-of-the-art examples that go beyond the case of (deformed) CY operators.

\subsection{Ice cone graphs}
\label{reficecone}

In ref.~\cite{Duhr:2022dxb}, a subset of the present authors explored the class of ice cone Feynman integrals for arbitrary loop orders in exactly two dimensions. 
An integrand analysis in Baikov representation revealed that the maximal cut of the ice cone integral contains two distinct CY varieties. This observation was the starting point to demonstrate that in two dimensions, the ice cone integral can be expressed in terms of iterated integrals over CY periods associated with these two different geometries~\cite{Duhr:2022dxb}.

\begin{figure}[!h]
\centering
\begin{tikzpicture}
\coordinate (one) at (-2.5,0);
\coordinate (two) at (2.5,0);
\coordinate (One) at (-1.5,0);
\coordinate (Two) at (1.5,0);
\coordinate (Below) at (0,-2.5);
\coordinate (below) at (0,-3.5);
\begin{scope}[very thick,decoration={
    markings,
    mark=at position 0.6 with {\arrow{>}}}
    ] 
\draw [-, thick,postaction={decorate}] (one) to [bend right=0]  (One);
\draw [-, thick,postaction={decorate}] (two) to [bend right=0]  (Two);
\draw [-, thick,postaction={decorate}] (One) to [bend right=85]  (Two);
\draw [-, thick,postaction={decorate}] (One) to [bend right=25]  (Two);
\draw [-, thick,postaction={decorate}] (One) to [bend left=25]  (Two);
\draw [-, thick,postaction={decorate}] (One) to  [bend left=85] (Two);
\draw [-, thick,postaction={decorate}] (Below) to  [bend left=23] (One);
\draw [-, thick,postaction={decorate}] (Two) to  [bend right=-23] (Below);
\draw [-, thick,postaction={decorate}] (below) to  [bend left=0] (Below);
\end{scope}
\node (d1) at (0,1.1) [font=\scriptsize, text width=.2 cm]{$k_1$};
\node (d2) at (0,.6) [font=\scriptsize, text width=.2 cm]{$k_2$};
\node (p1) at (-2.2,.2) [font=\scriptsize, text width=1 cm]{$p_1^2=0$};
\node (p2) at (2.4,.2) [font=\scriptsize, text width=1 cm]{$p_2^2=0$};
\node (p3) at (1.3,-3) [font=\scriptsize, text width=2.4 cm]{$s=(p_1+p_2)^2 $};
\node (dlm1) at (0,-0.1) [font=\scriptsize, text width=.2 cm]{$k_3$};
\node (dl) at (0,-0.65) [font=\scriptsize, text width=.2 cm]{$k_4$};
\node (dlp1) at (-1.6,-1.5) [font=\scriptsize, text width=1.8 cm]{$\sum_i k_i-p_1$};
\node (dlp2) at (2.1,-1.4) [font=\scriptsize, text width=1.8 cm]{$\sum_i k_i +p_2$};
\end{tikzpicture}
\caption{The four-loop ice cone graph}
\label{fig:icecone}
\end{figure}
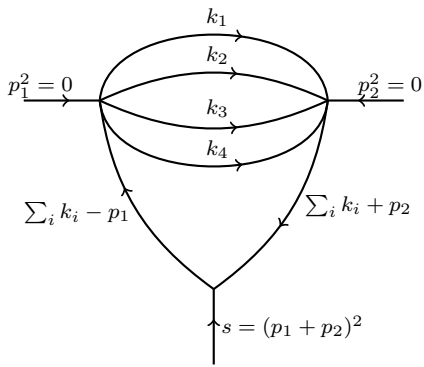

In this section, we consider the specific case of the four-loop ice cone integral. The full $\epsilon$-factorisation for the simpler, three-loop case had been worked out in ref.~\cite{Gorges:2023zgv}. 
We focus on the homogeneous part, and we extend the analysis of ref.~\cite{Duhr:2022dxb} to include its full $\epsilon$ dependence. 
We demonstrate how to transform the homogeneous system of differential equations into an $\epsilon$-factorised form following the approach explained in this paper. Despite the involved geometry at play, we can follow our prescription and select a starting basis that respects the MHS underlying the ice cone graph in $d=2$ dimensions. 
From the integrand analysis in $d=2$ dimensions from ref.~\cite{Duhr:2022dxb}, it follows that on the maximal cut, there is one master integral that vanishes for $\eps=0$. This integral will be discarded in the following, and it only enters as a trivial $1\times1$ block in the normalisation by the semi-simple parts. In addition, there are two distinct families of K3 surfaces that contribute at four loops, and each family is isomorphic to a family of K3 surfaces attached to an equal-mass three-loop banana integral. The second cohomology group of a family of K3 surfaces carries a pure Hodge structure of weight two. We thus conclude that the $W_2$ piece of the MHS contains the direct sum of these two pure Hodge structures of weight two, and we thus need to choose a basis of master integrals that is aligned with the two Hodge structures of the families of K3 surfaces. 
This is achieved, in practice, by selecting the same integrals as in ref.~\cite{Duhr:2022dxb}. In fact, this choice disentangles the two K3 geometries at $\epsilon=0$, which is key for our procedure to work. While we only discuss the four-loop case, based on the analysis in $d=2$ dimensions at higher loops, we anticipate that a similar approach can be applied to solve the ice cone integrals for any loop order.

To be self-contained, let us repeat the essential parts from ref.~\cite{Duhr:2022dxb}. In \cref{fig:icecone}, we have drawn the four-loop ice cone graph. The integral family associated with this graph is defined as in \cref{def:integralfamily} with the following propagators
\begin{equation}
\begin{aligned}
    D_i &= k_i^2 - m^2 \quad \text{for } i=1,\hdots,4 \, , \\
    D_5 &= \big( \sum_{i=1}^4 k_i - p_1 \big)^2 -m^2 \, , \quad D_6 = \big( \sum_{i=1}^4 k_i + p_2 \big)^2 -m^2 \, ,
\end{aligned}
\end{equation}
and the following irreducible scalar products
\begin{equation}
\begin{alignedat}{10}
    &N_1    &&= \big( \sum_{i=1}^4 k_i \big)^2 \, , \quad   &&N_2       &&= k_1\cdot k_3 \, , \quad &&N_3 &&= k_1\cdot k_4 \, , \quad &&N_4 &&= k_2\cdot k_3 \, , \quad &&N_5       &&= k_2 \cdot k_4 \, , \\
    &N_6    &&= k_3 \cdot k_4 \, , \quad                    &&N_7       &&= k_2\cdot p_1 \, , \quad &&N_8 &&= k_2\cdot p_2 \, , \quad &&N_9 &&= k_3\cdot p_1 \, , \quad &&N_{10}    &&= k_3\cdot p_2 \, , \\
    &N_{11} &&= k_4\cdot p_1 \, , \quad                     &&N_{12}    &&= k_4\cdot p_2 \, .       &&    &&                          &&    &&                          &&
    &&
\end{alignedat}
\end{equation}
The ice cone family defined in this way depends only on two parameters $s$ and $m^2$, with $s= (p_1+p_2)^2$ and $p_1^2 = p_2^2 = 0$. It is convenient to use the Landau variable $x$ defined through
\begin{equation}
    s = -m^2 \frac{(1-x)^2}{x} \, .
\end{equation}
We set $m^2=1$ in all subsequent formulas because the mass dependence can be restored by dimensional analysis. Our starting basis is\footnote{As in ref.~\cite{Duhr:2022dxb} we only write down the indices for the first two irreducible scalar products since the others are not necessary to define our basis.} 
\begin{equation}
\begin{alignedat}{3}
    I_1 &= -x\, I_{2,1,1,1,1,1;0,0} + I_{2,1,1,1,1,1;-1,0} \, , \quad &&I_2 = \partial_x I_1 \, , \quad &&I_3 = \partial_x^2 I_1 \, , \\
    I_4 &= -\frac1x\, I_{2,1,1,1,1,1;0,0} + I_{2,1,1,1,1,1;-1,0} \, , \quad &&I_5 = \partial_x I_4 \, , \quad &&I_6 = \partial_x^2 I_4 \, , \\
    I_7 &= \frac18 I_{1,1,1,1,1,1;0,0} -\frac{1-x+x^2}{8x}I_{1,1,1,1,1,1;-1,0} &&+ I_{1,1,1,1,1,1;-1,-1} \,,
\end{alignedat}
\label{eq:basisic}
\end{equation}
and, from here on, we discard all subtopologies and focus on the homogeneous part of the differential equations.
The only difference between~\cref{eq:basisic} and the choice of basis made in ref.~\cite{Duhr:2022dxb} is that here we are using the full $d$-dimesional integrals and not just their $\epsilon=0$ limit. 
More in detail, one can easily prove that $I_1$ and $I_4$ correspond to the holomorphic differential forms for the two families of K3 surfaces, while $I_2$, $I_3$, $I_5$ and $I_6$ are chosen as their derivatives.
Finally,  $I_7$ is obtained by multiplying the corner integral by the Gram determinant of the loop momentum flowing in the triangle and the two external momenta. This integral can be shown to be identically zero in $d=2$ dimensions~\cite{Duhr:2022dxb}.

In the basis in~\cref{eq:basisic}, the homogeneous differential equation at $\epsilon=0$ reads 
\begin{equation}
    \begin{pmatrix}
         0 & 1 & 0 & 0 & 0 & 0 & 0 \\
         0 & 0 & 1 & 0 & 0 & 0 & 0 \\
        \frac{1}{x^3 (1-16 x)} & -\frac{1-8 x+64 x^2}{x^2 (1-4 x) (1-16 x)} & \frac{6 (5-32 x)}{(1-4 x) (1-16 x)} & 0 & 0 & 0 & 0 \\
         0 & 0 & 0 & 0 & 1 & 0 & 0 \\
         0 & 0 & 0 & 0 & 0 & 1 & 0 \\
         0 & 0 & 0 & \frac{1}{(16-x) x^2} & -\frac{64-68 x+7 x^2}{(16-x) (4-x) x^2} & -\frac{6 \left(32-15 x+x^2\right)}{(16-x) (4-x) x} & 0 \\
         0 & 0 & 0 & 0 & 0 & 0 & 0 \\
    \end{pmatrix} \, ,  
\end{equation}
which makes the two CY blocks manifest. We notice that the two blocks are related by the simple $x \to 1/x$ transformation. Moreover, at $\epsilon=0$, these two blocks do not couple to each other, and the last integral decouples entirely. This basis choice reflects the MHS of the ice cone integrals: The MHS that we had identified consists of a direct sum of two K3 Hodge structures. Since the sum is direct, the two blocks corresponding to the K3 geometries are decoupled. In addition, there is a $1\times1$ block which completely decouples.
For convenience, let us explicitly write down the highest appearing $\epsilon$ orders of the entries of the differential equations matrix for this basis,
\begin{equation}
    \begin{pmatrix}
         - & 0 & - & - & - & - & - \\
         - & - & 0 & - & - & - & - \\
         3 & 2 & 1 & 3 & 2 & 1 & 1 \\
         - & - & - & - & 0 & - & - \\
         - & - & - & - & - & 0 & - \\
         3 & 2 & 1 & 3 & 2 & 1 & 1 \\
         1 & - & - & 1 & - & - & 1
    \end{pmatrix} \, ,
\end{equation}
where we use integers for the maximal power of $\epsilon$ in each entry and the symbol `$-$'
for the ones that are zero.

Before discussing the $\epsilon$-factorisation procedure, we focus on the two K3 geometries in the four-loop ice cone graph. These geometries were found to be the ones also appearing in the three-loop banana graph, which was intensively studied in ref.~\cite{Bloch:2014qca, Bloch:2016izu, Klemm:2019dbm, Bonisch:2020qmm, Bonisch:2021yfw}, evaluated at $x$ and $1/x$ respectively. The corresponding CY operator in the variable $x$ is given by
\begin{equation}
    \mathcal L_\text{3$l$ ban} = 1-4x - (3-18)\theta_x +(3-30x)\theta_x^2 -(1-4x)(1-16x)\theta_x^3 \,.
\end{equation}
The Frobenius basis around $x=0$ and $1/x=0$ both expressed in the variable $x$ can be chosen as
\begin{align}
    \varpi_0^+ &= x+4 x^2+28 x^3+256 x^4+2716 x^5 + \mathcal O(x^6) \, , \\
    \varpi_1^+ &= \varpi_0^+ \log(x)+S_1^+=\varpi_0^+ \log(x) +6 x^2+57 x^3+584 x^4+\frac{13081 x^5}{2} + \mathcal O(x^6) \, , \nonumber\\
    \varpi_2^+ &= \frac12\varpi_0^+ \log^2(x) + S_1^+ \log(x) + 18 x^3+270 x^4+\frac{7089 x^5}{2}\mathcal O(x^6) \, , \nonumber\\
    \varpi_0^- &= 1+\frac{x}{16}+\frac{7 x^2}{1024}+\frac{x^3}{1024}+\frac{679 x^4}{4194304}+\frac{1969 x^5}{67108864} + \mathcal O(x^6) \, , \\
    \varpi_1^- &= \varpi_0^- \log(x)+S_1^- \nonumber \\
                &= \varpi_0^- \log(x) +\frac{3 x}{32}+\frac{57 x^2}{4096}+\frac{73 x^3}{32768}+\frac{13081 x^4}{33554432}+\frac{195503 x^5}{2684354560} + \mathcal O(x^6) \, , \nonumber\\
    \varpi_2^- &= \frac12\varpi_0^- \log^2(x) + S_1^- \log(x) +\frac{9 x^2}{2048}+\frac{135 x^3}{131072}+\frac{7089 x^4}{33554432}+\frac{46185 x^5}{1073741824} + \mathcal O(x^6) \, , \nonumber
+\end{align}
where we used the symbol $\varpi_j^+$ for the elements of the basis at $x=0$ and $\varpi_j^-$ for the ones at ${x} = \infty$. 
Following the discussion in~\cref{sec:CY}, we find the two individual semi-simple parts of the Wronskians
\begin{align}
    W^{\text{ss}}_{\star} = \begin{pmatrix}
                        {\varpi_0^\star} & 0 & 0 \\
                        {\varpi_0^\star}' & {\scriptstyle \sqrt{C_2^\star}} & 0 \\
                        {\varpi_0^\star}'' & {\scriptstyle \sqrt{C_2^\star}} \frac{{\varpi_0^\star}'}{\varpi_0^\star} + \partial_x {\scriptstyle \sqrt{C_2^\star}} & \frac{C_2^\star}{\varpi_0^\star}
                    \end{pmatrix} \, ,
\end{align}
for $\star = +,-$ and with 
\begin{equation}
    C_2^+ =\frac{1}{(1-4x)(1-16x)}\,, \qquad C_2^- = \frac{64}{x^2(4-x)(16-x)}\,.    
\end{equation}
Moreover, we can express ${\varpi_0^{\star}}''$ through $\varpi_0^\star$ and ${\varpi_0^\star}'$ as in~\cref{eq:quadexplicitly} due to the quadratic relations given in~\cref{eq:qual_rel_ss}. The whole Wronskian is then the direct sum
\begin{equation}
    W^\text{ss}_\text{4$l$ ice} = \begin{pmatrix}
                                    W^\text{ss}_+ & 0 & 0 \\
                                    0 & W^\text{ss}_- & 0 \\
                                    0 & 0 & 1
                                \end{pmatrix} \, .
\end{equation}
The fact that the Wronskian is block-diagonal is a direct reflection that the Hodge structure is a direct sum.
We can now apply our $\epsilon$-factorisation method. After we have gauge-transformed our basis with the inverse of the semi-simple Wronskian $W^\text{ss}_\text{4$l$ ice}$, we perform the following $\epsilon$-rescaling 
\begin{equation}
   D(\epsilon) = \text{diag}(\epsilon^2,\epsilon,1,\epsilon^2,\epsilon,1,\epsilon^2) =  \begin{pmatrix}
                                    D_2(\eps) & 0 & 0 \\
                                    0 & D_2(\eps)& 0 \\
                                    0 & 0 & \eps^2
                                \end{pmatrix}\, ,
\end{equation}
which realigns the transcendental weight of the master integrals at the MUM-point.
Next, as the last step, we $\epsilon$-factorise the resulting equations, starting from the highest negative power of $\epsilon$ to the constant order. At every power, we divide this task into two steps. First, we aim to eliminate ${\varpi_0^+}'$ and ${\varpi_0^-}'$ from the differential equations by suitable total derivatives of functions constructed from the two periods $\varpi_0^+$ and $\varpi_0^-$ and algebraic functions. This is possible for many entries of the differential equation. For the remaining ones, one has to decide whether ${\varpi_0^+}'$ or ${\varpi_0^-}'$ should be removed. As a choice, we prioritize the removal of ${\varpi_0^-}'$. Second, to achieve the full $\epsilon$-factorisation of the homogeneous system, we introduce eight new functions to remove the remaining terms. For compactness, we do not provide the full basis here. Instead, we only present the new functions $G_i$ as entries of a transformation matrix that one would have to apply on the basis resulting after having rotated away the semi-simple part, performed the $\eps$-rescalings and removed the terms corresponding to total derivatives of the periods and algebraic functions as described above,
\begin{equation}
\begin{small}
   \begin{pmatrix}
        1 & 0 & 0 & 0 & 0 & 0 & 0 \\
 G_2^+ & 1 & 0 & -64 G_5 & 0 & 0 & 0 \\
 -\frac{1}{2} \left((G_2^-)^2+64 G_4^2\right) & -G_2 & 1 & -64 (G_2^- G_4-G_2 G_5+G_6) & -64 G_4 & 0 & 0 \\
 0 & 0 & 0 & 1 & 0 & 0 & 0 \\
 G_4 & 0 & 0 & G_2^- & 1 & 0 & 0 \\
 G_6 & G_5 & 0 & -\frac{1}{2} \left((G_2^-)^2+64 G_5^2\right) & -G_2^- & 1 & 0 \\
 0 & 0 & 0 & 0 & 0 & 0 & 1
   \end{pmatrix}  \begin{pmatrix}
        1 & 0 & 0 & 0 & 0 & 0 & 0 \\
 0 & 1 & 0 & 0 & 0 & 0 & 0 \\
 \frac{G_1^+}{\epsilon } & 0 & 1 & \frac{64 G_3}{\epsilon } & 0 & 0 & 0 \\
 0 & 0 & 0 & 1 & 0 & 0 & 0 \\
 0 & 0 & 0 & 0 & 1 & 0 & 0 \\
 \frac{G_3}{\epsilon } & 0 & 0 & \frac{G_1^-}{\epsilon } & 0 & 1 & 0 \\
 0 & 0 & 0 & 0 & 0 & 0 & 1
   \end{pmatrix}  \, .
\end{small}
\end{equation}
The full transformation is provided in the ancillary files. Here, we only aim at highlighting the new functions, and we stress that they are defined by the first-order differential equations provided in~\cref{app:icecone}.

Before we close this section, let us comment on the expansion of the new functions we had to introduce, particularly the integrality of their expansion coefficients. In the previous section, we have seen that in the case where the maximal cut of a given sector is related to a single CY geometry, we expect the functions $G_i$ to have expansions with integral coefficients. In our case, the new functions $G_1^\star(x),G_2^\star(x)$ for $\star=+,-$, have already appeared in the $\epsilon$-factorisation of the three-loop banana family~\cite{Pogel:2022yat, Pogel:2022vat} which has only a single CY geometry. Therefore, their integrality properties are not new. For the other four functions $G_3(x),\hdots, G_6(x)$, the situation is different. These functions have been introduced to $\epsilon$-factorise the mixings between the two different Calabi-Yau geometries. Thus, it is not immediately evident that these functions also have integral expansions. We have checked up to $\mathcal O(x^{1000})$ modulo the rescaling $x \rightarrow 2^6 x$ that these functions also have an integral expansion.

\subsection{Gravitational scattering at 5PM-2SF}
\label{sec:GWexamples}

As a second example, we consider two graphs that are relevant to the perturbative modelling of the scattering of two black holes in general relativity~\cite{Kovacs:1978eu, Westpfahl:1979gu, Bel:1981be, Damour:2017zjx, Hopper:2022rwo}.
These calculations play a crucial role in the theoretical prediction of gravitational waveforms coming from black hole binary systems. When looking at these scattering processes, one usually performs a weak field expansion in Newtons constant $G$ (Post-Minkowskian (PM) expansion) in an effective field theory framework~\cite{Goldberger:2004jt, Porto:2016pyg, Mogull:2020sak} in which the black holes are modelled as point particles. Additionally, one can also
perform a so-called self-force (SF) expansion~\cite{Mino:1996nk, Poisson:2011nh, Barack:2018yvs, Gralla:2021qaf} in the mass ratio of both black holes $\nu=(m_1 m_2)/(m_1+m_2)^2$. 
As it turns out, already at 4PM-1SF ($\mathcal{O}(G^4 \nu)$) a period over a K3 surface contributes to the scattering angle~\cite{Bern:2021dqo, Bern:2021yeh, Dlapa:2021npj, Dlapa:2022lmu, Jakobsen:2023ndj, Damgaard:2023ttc}, while one order higher, at 5PM-1SF ($\mathcal{O}(G^5 \nu)$), a period over a CY three-fold and related functions contribute to the radiated energy \cite{Driesse:2024feo, Driesse:2024xad, Klemm:2024wtd}. 
Finally, at 5PM-2SF, two new geometries appear~\cite{Klemm:2024wtd, Frellesvig:2024ymq, Frellesvig:2023bbf}, which we will discuss here. 

The problems presented here are state-of-the-art and quite involved. 
In particular, in both cases one finds many master integrals on the maximal cut. 
Moreover, the high number of loops and the complexity of the geometry render a complete integrand analysis less straightforward. 
This is not only a practical problem due to the proliferation of terms but also a conceptual one, as it is related to the appropriate mathematical definition of differential forms with non-vanishing residues on arbitrary geometries. 
While this is well understood for elliptic and hyperelliptic cases, it becomes more delicate for higher-dimensional varieties.
Nevertheless, we have chosen them since they prove an important point: as long as we are able to identify the integrands corresponding to the holomorphic forms, our procedure can succeed even if not all integrals corresponding to the forms with additional poles with non-vanishing residues are properly selected.
In particular, as we will see explicitly below, it is typically enough to select either integrals with contaminations from double poles or supplement the basis with a few integrals found by a trial-and-error search. 
As long as the forms of the first and second kind have been correctly chosen, this often only generates some minor pollution in the $\mathcal{O}(\epsilon)$ terms of the differential equations, which can be removed with a case-by-case analysis.

Specifically, we consider the following family of planar integrals\footnote{With respect to the conventions established in~\cref{def:integralfamily} we refrain here from defining ISPs $N_i$ as we consider two different sectors within this family.}
\begin{align}
  I_{n_1,\hdots,n_{22}}
  =
  \int
 \left(\prod_{j=1}^4\frac{\mathrm d^dk_j}{i\pi^{d/2}}\right) \frac{
    \delta^{(n_1\!-\!1)}(k_1\cdot v_1)\delta^{(n_2\!-\!1)}(k_2\cdot v_1)\delta^{(n_3\!-\!1)}(k_3\cdot v_2)\delta^{(n_4\!-\!1)}(k_4\cdot v_2)
    }{
    \prod_{i=5}^{22}D_{i}^{n_{i}}
    }\,,
\end{align}
where $\delta^{(n)}(x)$ with $n>0$, denotes the $n$-th derivative of the delta function and the propagators are defined as
\begin{equation}
\begin{alignedat}{8}
    &D_5    \,&&= k_1 \cdot v_2      \, ,\quad &&D_6      &&= k_2 \cdot v_2    \, ,\quad &&D_7      \,&&= k_3 \cdot v_1    \, ,\quad    &&D_8    \,&&= k_4 \cdot v_1    \, ,\quad \\
    &D_9    \,&&= (k_{13})^2         \, ,\quad &&D_{10}   &&= (k_{23})^2       \, ,\quad &&D_{11}   \,&&= (k_{14})^2       \, ,\quad    &&D_{12} \,&&= (k_{24})^2       \, ,\quad \\
    &D_{13} \,&&= (k_{12})^2         \, ,\quad &&D_{14}   &&= (k_{34})^2       \, ,\quad &&D_{15}   \,&&= (k_1+q)^2        \, ,\quad    &&D_{16} \,&&= (k_2+q)^2        \, ,\quad \\
    &D_{17} \,&&= (k_3+q)^2          \, ,\quad &&D_{18}   &&= (k_4+q)^2        \, ,\quad &&D_{19}   \,&&= k_1^2            \, ,\quad    &&D_{20} \,&&= k_2^2            \, ,\quad \\
    &D_{21} \,&&= k_3^2              \, ,\quad &&D_{22}   &&= k_4^2            \, ,\quad &&         \,&&                                &&       \,&&    
\end{alignedat}
\end{equation}
with $k_{ij}=k_i-k_j,\, v_i^2=1,\, v_i \cdot q=0$ and $d=4-2\epsilon$. The only dimensionless scale is given by the scalar product of the initial velocities of both black holes $v_1 \cdot v_2=(x+1/x)/2$. We use mostly minus metric and set $q^2=-1$ for the space like vector $q$. The dependence on this variable can be reconstructed by dimensional analysis when needed. We will only look at the maximal cut of a given sector and neglect all contributions from subsectors.

Two comments are in order. First, as defined, the integral family splits into two families depending on whether the sum of the first eight propagator powers is even or odd. The even family contributes to the scattering angle, and the odd family contributes to the radiated energy and momentum. Both sectors we will look at are even sectors.

The second comment is about the choice of $i0$ prescription for the propagators. It turns out that one can use Feynman propagators when neglecting radiative effects, while retarded/advanced ones have to be used to properly include them. This difference is important when building waveform models out of these results. In particular, when solving IBPs using  {\tt KIRA}~\cite{Maierhofer:2017gsa, Klappert:2020nbg} and constructing the system of differential equations, the choice of retarded/advanced propagators does not allow for the use of symmetry relations. The sign of the $i0$ prescription of retarded/advanced propagators is sensitive to the sign of the momentum
\begin{align}
    \dfrac{1}{(k^0+i0)^2-\vec{k}^2} \xrightarrow{k \rightarrow -k} \dfrac{1}{(k^0-i0)^2-\vec{k}^2} \, .
\end{align}
This breaks some symmetries of the integrals that are present when using a Feynman prescription. These missing relations in the IBP reduction result in an increased number of master integrals. 

Though using the relation
\begin{align}
        \dfrac{1}{(k^0\pm i0)^2-\vec{k}^2}=\dfrac{1}{k^2+i0} + 2 \pi i \delta(k^2) \theta( \mp k^0) \, ,
\end{align}
we can split our retarded propagators into Feynman propagators and a cut to connect both $i0$ choices \cite{Jakobsen:2023oow, Damgaard:2023ttc}.
In this paper, we will compute all integrals only with Feynman propagators, making use of all symmetry relations. The results we present here will contribute to both conservative and radiative calculations. The latter will need additional cut contributions, which we will neglect here.

\subsubsection{A sector related to a K3 surface}

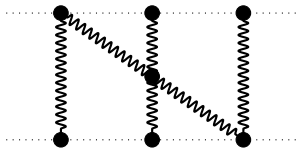
\begin{figure}[b]
    \centering
\begin{tikzpicture}[baseline={([yshift=-1ex]current bounding box.south)},scale=1.2]
  \coordinate (inA) at (0.4,.7);
  \coordinate (outA) at (3.6,.7);
  \coordinate (inB) at (0.4,-.7);
  \coordinate (outB) at (3.6,-.7);
  \coordinate (xA) at (1,.7);
  \coordinate (xxA) at (1.825,.7) ;
  \coordinate (xxB) at (1.825,-.7) ;
  \coordinate (yA) at (1.60+0.4,.7);
  \coordinate (yyA) at (2.375,0) ;
  \coordinate (zA) at (3,.7);
  \coordinate (zzA) at (3.75,.7) ;
  \coordinate (xB) at (1,-.7);
  \coordinate (yB) at (1.6+0.4,-.7);
  \coordinate (zB) at (3,-.7);
  \coordinate (zzB) at (3.75,-.7);
  \coordinate (xM) at (1,0);
  \coordinate (yM) at (1.6+0.4,0);
  \coordinate (zM) at (3,-0);
  \draw [dotted] (inA) -- (outA);
  \draw [dotted] (inB) -- (outB) node [midway, below]{} ;
  \draw [draw=none] (xA) to[out=40,in=140] (zA);
  \draw [photon] (xA) -- (xB);
  \draw [photon] (yA) -- (yM) -- (yB);
  \draw [photon] (zA) -- (zB);
  \draw [photon] (xA) -- (yM) -- (zB);
  \draw [fill] (yM) circle (.08);

  \draw [fill] (xA) circle (.08);
  \draw [fill] (yA) circle (.08);
  \draw [fill] (zA) circle (.08);
  \draw [fill] (xB) circle (.08);
  \draw [fill] (yB) circle (.08);
  \draw [fill] (zB) circle (.08);
\end{tikzpicture}
    \caption{A K3 sector at 5PM-2SF. The dotted lines represent the worldlines and translate on the upper line to $\delta(k_i \cdot v_1)$ and on the bottom line to $\delta(k_i \cdot v_2)$. The wiggly lines represent gravitons and translate to the propagators $D_9-D_{22}$. }
    \label{fig:K3GW}
\end{figure}

The first interesting sector is pictured in~\cref{fig:K3GW} and has nine master integrals on the maximal cut. At this point, we have to make an important remark. When analysing families of Feynman integrals appearing in black hole scattering processes at second self-force order, it has been observed that in some special cases, if one chooses the right basis of master integrals, certain sectors on their maximal cut naturally split into two separate blocks even for arbitrary values of $\epsilon$. In other words, one of the blocks only appears as an inhomogeneous term of the other, as if the latter were a top sector of the former.
To find a basis where such a splitting is manifest is typically non-trivial, and there is currently no known method for finding it. Often, it can only be uncovered by trial and error. While this is an interesting problem in general, its solution goes beyond the scope of our paper. For the present analysis, it suffices to say that a possible hint to test whether a given sector splits into smaller subsectors, according to the definition above, is to check whether the corresponding corner integral and its derivatives generate the whole sector for generic $\epsilon$. For our problem, it turns out that this is not the case, and in fact, the sector splits into two sub-sectors of size $7 \times 7$ and $2 \times 2$, respectively, where the latter contains the former in its inhomogeneous terms.
If we now consider the smaller $2\times2$ sector as a top sector of the $7 \times 7$ one, it is easy to see that its homogeneous second-order system can be put in dlog-form. This indicates that the two corresponding integrals can be easily obtained by iterated integrations over the solution of the $7 \times 7$ sector.
For this reason, and to simplify our treatment, we ignore it for the subsequent discussion, but we stress that it constitutes no problem to find an $\epsilon$-factorised basis for the full $9 \times 9$ system, and in fact, we provide the full $\epsilon$-factorised basis for the $9 \times 9$ problem in the ancillary files.

Let us now focus on the $7\times7$ sector. We follow our general strategy and try first to find a good choice of starting basis. For this, we start by analysing the corner integral $I_{1,1,1,1,0,0,0,0,1,0,0,1,1,1,1,0,0,0,0,0,0,1}$ using a loop-by-loop Baikov representation~\cite{Frellesvig:2017aai, Frellesvig:2024ymq}. Its maximal cut is given as an integral in four unconstrained variables. By direct computation, one can prove that two of these integrations give rise to dlogs. After this step, one is left with one single leading singularity that still involves the remaining two integrations
\begin{align}
    \text{LS}(I_{1,1,1,1,0,0,0,0,1,0,0,1,1,1,1,0,0,0,0,0,0,1})=\frac{  x^2}{x^2-1} \int \dfrac{\mathrm dt_1 \mathrm dt_2}{\sqrt{P_{4,4}(t_1,t_2)}} \,,
\label{eq:K3LS}
\end{align}
with
\begin{equation}\bsp
    P_{4,4}(t_1,t_2) &= (1-t_1)(1+t_1)(1-t_2)(1+t_2) \\
                    &\quad\times (1-2x(1-2t_1t_2)+x^2)(1+2x(1+2t_1t_2)+x^2) \, . 
\esp\end{equation}
Notice that the polynomial $P_{4,4}$ has a total degree eight in the variables $t_1, t_2$, whereas the individual variables appear at most with degree four.
In this form, the integrand seems to correspond to the holomorphic $(2,0)$-form of an elliptically fibred K3 geometry. 
However, this geometry is singular\footnote{The variety defined by the constraint $W=y^2 - P_{4,4}(t_1,t_2)=0$ is singular, meaning that $W$ and its Jacobian have common zeros. Due to the factorised form of the polynomial $P_{4,4}$, it is not hard to find such common zeros.} and one has to be careful when identifying it with a smooth K3 surface. One way of showing that, indeed, the integral in~\cref{eq:K3LS} gives rise to a smooth K3 surface is to resolve all its singularities. 
This was done in refs.~\cite{Klemm:2024wtd, MR0749676}. 
Another approach, which suffices for our purposes, is to examine the corresponding Picard-Fuchs operator and check that it satisfies all properties of a CY operator of order three. We can directly derive the Picard-Fuchs operator from the representation in~\cref{eq:K3LS}. 
A standard way to do this is to first compute explicitly the holomorphic period integral and then use its explicit expression to infer the irreducible differential operator that annihilates it. 

It is easy to obtain a Laurent series for the holomorphic solution at $x=0$ as follows.
We know that each homogeneous solution can be obtained by evaluating the integral in~\cref{eq:K3LS} along one of the independent integration cycles~\cite{Primo:2016ebd, Primo:2017ipr}. 
To select the one that corresponds to the holomorphic solution at $x=0$, we start by expanding the integrand around $x=0$.\footnote{It can be argued that the integration cycle we are choosing
is given by a two-dimensional torus $T^2$. Therefore, the integral we are computing is also known as 
\emph{torus period}, and its computation is a special type of residue computation, see refs.~\cite{Klemm:2024wtd, Duhr:2023eld, Duhr:2024hjf} and references therein for further examples of this kind of calculation.} 
In the language of expansion by regions, this corresponds to focussing on the hard branch only.
In this limit, the integral becomes
\begin{align}
    \oint \dfrac{\mathrm dt_1 \mathrm dt_2}{\sqrt{P_{4,4}(t_1,t_2)}}  &= 
    \sum_{n=0,2,4,...}^\infty x^n \, \oint \dfrac{\mathrm dt_1 \mathrm dt_2}{\sqrt{(1-t_1^2)(1-t_2^2)}}\left[ \sum_{j=0,2,4...}^n c_{j}^{(n)} (t_1 t_2)^{j}  \right]\,, \nonumber \\
    &=     \sum_{n=0,2,4,...}^\infty x^n \,\sum_{j=0,2,4...}^n c_{j}^{(n)} \left[I(j) \right]^2\,,
\end{align}
where the $c_j^{(n)}$ are numerical coefficients and we have defined
\begin{equation}
I(j) = \oint \dfrac{\mathrm dt\, t^{j}}{\sqrt{(1-t^2)}}\,, \qquad j\geq 0\,.
\end{equation}
Notice that the sums run only over even numbers. This is because, by symmetry, $I(j)$ is non-zero only for even powers of $t$, which in turn are only produced for $x^n$ with $n$ even. 

For each $j\geq 0$, $I(j)$ always has a branch cut, which can be chosen to be for $-1<t<1$. 
Moreover, $I(j)$ has a pole at infinity with non-vanishing residue, and the order of the pole increases with $j$.
The integral around the branch cut must be proportional to the residue at infinity if it exists. 
To study the behaviour of the integral at $t=\infty$, we send $t \to 1/u$ and expand the integrand for $u\to0$ to find
\begin{equation}
I(j) \sim \oint \mathrm du    \left[ \sum_{i=-1}^\infty a_i^{(j)} u^{i-j} \right]\,,\qquad j\geq 0\,,
\end{equation}
where $a_i^{(j)}$ are rational numbers. This shows, again, that only integrals with $j$ even can produce a single pole at infinity, which is consistent with the fact that all integrals with $j$ odd are zero by symmetry, as anticipated above. 
In this limit, there is, therefore, only one independent cycle that can be chosen as
\begin{equation}
    \oint \dfrac{\mathrm dt\, t^{j}}{\sqrt{(1-t^2)}} \propto \int_{-1}^{1} \dfrac{\mathrm dt\, t^{j}}{\sqrt{(1-t^2)}} \propto \oint_{\mathcal{C}_\infty}\dfrac{\mathrm dt\, t^{j}}{\sqrt{(1-t^2)}} = \left\{ \begin{array}{ll} 
    2 \pi i\, a^{(j)}_{j-1} & \quad \mbox{for } j \mbox{ even,}  \\ 
    0 & \quad \mbox{for } j \mbox{ odd,} \end{array} \right.  
\end{equation}
where $\mathcal{C}_\infty$ is the contour that picks up the pole at infinity.
Using this, modulo an overall prefactor, the integral becomes
\begin{align}
    \oint \dfrac{\mathrm dt_1 \mathrm dt_2}{\sqrt{P_{4,4}(t_1,t_2)}}  &\propto 
1 + 5 x^2 + 73 x^4 + 1445 x^6 + 33001 x^8 + \mathcal O(x^{10}) \, .
\end{align}

Having obtained sufficient orders in $x$, one can construct the corresponding Picard-Fuchs operator, which in our case is given by
\begin{equation}
\begin{aligned}
    \mathcal{L}^{(3)}_0 &= \left(1-34 x^2+x^4\right)\theta ^3-6 x^2 \left(17-x^2\right)\theta ^2 -12 (3-x) (3+x) x^2 \theta -8 \left(5-x^2\right) x^2 \, .
\end{aligned}
\label{eq:aperyop}
\end{equation}
This operator is well-known in the mathematical literature~\cite{MR3890449, MR0749676}, because of its appearance in Ap\'ery's proof of the irrationality of $\zeta(3)$~\cite{MR3363457}, and it is, in fact, a CY operator of order three. The full Frobenius basis to the Ap\'ery operator in \cref{eq:aperyop} is
\begin{align}
    \varpi_0(x)&=1+5 x^2+73 x^4 +1445 x^6+\mathcal{O}(x^8) \, ,\\
    \varpi_1(x)&=\varpi_0(x) \log(x)+S_1 = \varpi_0(x) \log(x) + 6 x^2+105 x^4 +2219 x^6+\mathcal{O}(x^8) \, , \nonumber\\
    \varpi_2(x)&=\dfrac{1}{2} \varpi_0(x) \log^2(x) +S_1 \log(x)+ 18 x^4 +540 x^6+\mathcal{O}(x^8) \, . \nonumber
\label{eq:frobK3BK}
\end{align}
We conclude that the corner integral, modulo the rational function $x^2/(x^2-1)$, contains in its integrand the holomorphic differential form of a K3 geometry, and the corresponding torus period relates to the Ap\'ery operator. Based on this, we select this integral as the first element of our starting basis. As previously explained, we can also incorporate the first and second derivatives of this integral into our starting basis, thereby completing the integrals for the pure K3 block.

In addition to these first three integrals, we need to select four more. As anticipated above, the integrand analysis for these integrals turns out to be non-trivial. 
One reason for this complexity is that it is unclear how the differential forms with additional non-vanishing residues look on a general K3 surface. These differential forms would generalise the differentials of the third kind on an elliptic curve. 
Without delving too deeply into this topic, we will employ a more heuristic approach here to choose a suitable basis.
To this end, let us list the criteria we impose on the differential equations of a good starting basis. First, we do not want to have any dependence on $\eps$ in the denominators of the differential equations. Secondly, if we set $\epsilon=0$, we want the additional four integrals to decouple from the block describing the periods of the K3 surface, such that we recover the Picard-Fuchs operator from~\cref{eq:K3LS}. Furthermore, the $4\times 4 $ block formed by the integrals that do not belong to the block describing the K3 periods should already be in $\eps$-form. The latter requirement also fixes the normalisation of these master integrals. 
In addition, we managed for one out of the four remaining integrals to perform an integrand analysis in a loop-by-loop Baikov representation. This integral is $I_{1,1,1,1,0,0,0,0,1,0,0,2,1,1,1,0,0,0,0,0,0,1}$, and is thus suited as a starting integral.
For the additional three missing basis integrals we made a scan through a list of potential candidate integrals. We checked whether the completion of our basis by three candidate integrals out of this list fulfils all of our requirements listed above. Here, it is essential to check that the three new integrals $\epsilon$-factorise immediately the $4\times4$ block of residue integrals. In this way, we found our initial basis
\begin{align}
    I_1 =& \frac{x^2-1  }{x^2}I_{1,1,1,1,0,0,0,0,1,0,0,1,1,1,1,0,0,0,0,0,0,1} \, , \quad I_2 =   \partial_x I_1 \, , \quad I_3 = \partial_x^2 I_1 \, , \nonumber\\
    I_4 =& \frac1\epsilon  \frac{\left(x^2-1\right)^2  }{x^2}I_{1,1,1,1,0,0,0,0,1,0,0,2,1,1,1,0,0,0,0,0,0,1}\, , \nonumber\\
    I_5 =& \dfrac{1+2 \epsilon}{\epsilon^2} I_{1,1,1,1,0,0,0,0,1,0,0,1,1,1,1,0,0,0,0,0,0,2}\, , \\
    I_6 =& \frac1{\epsilon^2}\frac{ x^2 }{\left(x^2-1\right)^2}I_{1,1,1,1,-1,0,0,-1,1,0,0,1,1,1,2,0,0,0,0,0,0,2}\, , \nonumber\\
    I_7 =& \frac1{\epsilon^2}\frac{x^2-1}{x} \big[ I_{1,1,1,2,0,0,0,-1,1,0,0,2,1,1,1,0,0,0,0,0,0,1} +I_{1,1,1,2,0,0,0,-1,2,0,0,1,1,1,1,0,0,0,0,0,0,1} \big]\nonumber\\
         & \quad + \frac4\epsilon  \frac{x^2-1  }{x^2}I_{1,1,1,1,0,0,0,0,1,0,0,2,1,1,1,0,0,0,0,0,0,1}\, . \nonumber
\label{eq:basisK3}
\end{align}
With this choice of integrals $I=(I_1,\hdots, I_7)^T$, the differential equation is given as
\begin{align}
    \partial_x I=M(x,\epsilon) I \,.
\end{align}
The full matrix $M(x,\epsilon)$ for this choice of basis is given in the ancillary file. 
Of particular interest for us, is the part constant in $\epsilon$, which is given by
\begin{align}
 M(x;,0)=  \left(
\begin{array}{ccccccc}
 0 & 1 & 0 & 0 & 0 & 0 & 0 \\ 
 0 & 0 & 1 & 0 & 0 & 0 & 0 \\ 
 \frac{8 \left(5-x^2\right)}{x (1-34 x^2+x^4)} & -\frac{1-244x^2+19 x^4}{x^2 (1-34 x^2+x^4)} & -\frac{3 \left(1-68x^2+3 x^4\right)}{x (1-34 x^2+x^4)} & 0 & 0 & 0 & 0 \\ 
 0 & 0 & 0 & 0 & 0 & 0 & 0 \\ 
 \frac{10 x \left(17-x^2\right)}{3 (1+x) (1-x)} & -\frac{5 \left(1-34 x^2+x^4\right)}{3 (1+x) (1-x)} & 0 & 0 & 0 & 0 & 0 \\ 
 \frac{5 x^2 \left(5+x^2\right)}{3 (1-x)^2 (1+x)^2} & -\frac{5 x \left(1+x^2\right)}{6 (1-x) (1+x)} & 0 & 0 & 0 & 0 & 0 \\ 
 0 & 0 & 0 & 0 & 0 & 0 & 0 \\ 
\end{array}
\right) \, . \label{eq:K32sfeps0}
\end{align}
Additionally, it is useful to highlight the highest power of $\epsilon$ in the individual matrix entries
\begin{align}
    \left(
\begin{array}{ccccccc}
 -  & 0  & -  & -  & -  & -  & -  \\ 
 -  & -  & 0  & -  & -  & -  & -  \\ 
 3  & 2  & 1  & 3  & 3  & 3  & 3  \\ 
 -  & -  & -  & 1  & -  & -  & 1  \\ 
 1  & 0  & -  & 1  & 1  & 1 & 1   \\ 
 1  & 0  & -  & 1  & 1  & 1 & -   \\ 
 1  & -  & -  & 1  & 1  & -  & 1  \\ 
\end{array}
\right) \, ,
\label{eq:K3eps}
\end{align}
where each integer indicates the highest power of $\eps$ in that entry of the matrix and, for ease of reading, 
we marked by `$-$' an entry that is zero.
In the following, we will now demonstrate that even with this heuristic choice of initial basis, our method works well and produces an $\epsilon$-factorised basis.

Before rotating our basis by the inverse of the semi-simple part, it is useful to try to improve the integrals $I_{4-7}$ by a simple rotation. By this, we mean a basis where these four integrals couple at most to the first integral $I_1$. This additional rotation simplifies the later steps, although our method also works without it. Concretely, the rotation is given by
\begin{align}
    T=\left(
\begin{array}{ccccccc}
 1 & 0 & 0 & 0 & 0 & 0 & 0 \\ 
 0 & 1 & 0 & 0 & 0 & 0 & 0 \\ 
 0 & 0 & 1 & 0 & 0 & 0 & 0 \\ 
 0 & 0 & 0 & 1 & 0 & 0 & 0 \\ 
 \frac{5 \left(1-34x^2+x^4\right)}{3 \left(1-x^2\right)} & 0 & 0 & 0 & 1 & 0 & 0 \\ 
 \frac{5 \left(1+x^2\right)x}{6 \left(1-x^2\right)} & 0 & 0 & 0 & 0 & 1 & 0 \\ 
 0 & 0 & 0 & 0 & 0 & 0 & 1 \\ 
\end{array}
\right) \, .
\end{align}
After this, we proceed to rotate the $3 \times 3$ K3 block by the inverse of the semi-simple part of the Wronskian (see~\cref{eq:SeSiExplicit,eq:quadexplicitly})
\begin{align}
    {W}^\text{ss}=\left(
\begin{array}{ccc}
 \varpi_0 & 0 & 0 \\
 \varpi_0' & \sqrt{C_2} & 0 \\
 \varpi_0'' & \sqrt{C_2}\frac{\varpi_0'}{\varpi_0} + \partial_x\sqrt{C_2} & \frac{C_2}{\varpi_0} \\
\end{array}
\right)  \; \text{with} \quad   C_2 = \frac{1}{x^2 \left(1-34 x^2+x^4\right)} \, .
\label{eq:Grav_K3_W}
\end{align}
We then rescale the basis with:
\begin{align}
    D(\epsilon) = \text{diag}(\epsilon^2,\epsilon,1,\epsilon^2,\epsilon^2,\epsilon^2,\epsilon^2,\epsilon^2) \, .
    \label{eq:Grav_K3_D}
\end{align}

After that, we can systematically remove terms that are not in $\epsilon$-form, starting with those proportional to $1/\epsilon$ and proceeding to the constant terms in $\epsilon$. For this, we identify, as before, total derivatives in the differential equations that can be easily removed through an ansatz of the form
\begin{align}
    \hat T_{i,j}=\sum_{k=-1}^0  \big(f_{i,j,k}(x)\,  \varpi^2_0(x)+g_{i,j,k}(x)\,  \varpi_0(x)\varpi'_0(x)+h_{i,j,k}(x)\,  \varpi_0(x)\big)\epsilon^k \, ,
\label{eq:rottotderiv}
\end{align}
where $f_{i,j,k}(x), g_{i,j,k}(x), h_{i,j,k}(x)$ are algebraic functions in $x$.
This nearly eliminates all non-$\epsilon$-factorised terms in the differential equations. To obtain a fully $\epsilon$-factorised system of differential equations, we are forced to introduce new transcendental functions. 
In our specific case, one has to introduce four of them, which can be done through the following rotation
\begin{align}
    \tilde{T}=
\begin{pmatrix}
 1 & 0 & 0 & 0 & 0 & 0 & 0 \\ 
 G_4 & 1 & 0 & 0 & 0 & 0 & 0 \\ 
 \tilde{T}_{3,1} & -G_4 & 1 & -64 G_2 & -\frac{896}{3} G_2 & 256 G_3 & 0 \\ 
 0 & 0 & 0 & 1 & 0 & 0 & 0 \\ 
 \frac{160 }{3}G_2 & 0 & 0 & 0 & 1 & 0 & 0 \\ 
 -\frac{5}{6}  G_3 & 0 & 0 & 0 & 0 & 1 & 0 \\ 
 0 & 0 & 0 & 0 & 0 & 0 & 1 \\ 
\end{pmatrix}\,,
\label{eq:K3G}
\end{align}
with
\begin{align}
    \tilde{T}_{3,1} = \frac1\epsilon G_1 -\frac{71680}{9}G_2^2-\frac{320}{3}G_3^2-\frac12 G_4^2 \, .
\end{align}
This last rotation $\tilde{T}$ is applied on the basis after the rotation in~\cref{eq:rottotderiv}.
The differential equations defining the $G_i(x)$ functions are as follows
\begin{equation}
\begin{aligned}
   G_1'(x) &= -\frac{16 x \left(1+x^2\right)p_1(x)}{9 (1-x)^3 (1+x)^3 \left(1-34 x^2+x^4\right)^2}\varpi_0(x)^2 \, ,\\
   p_1(x)  &= 115+4062 x^2-80259 x^4+1495652 x^6-80259 x^8+4062 x^{10}+115 x^{12} \, ,\\
   G_2'(x) &= \frac{x \left(1+x^2\right) }{(1-x)^2 (1+x)^2}\varpi_0(x) \, ,\\
   G_3'(x) &= \frac{\left(1+14x^2+x^4\right) }{(1-x)^2 (1+x)^2}\varpi_0(x) \, ,\\
   G_4'(x) &= \frac{1}{x \sqrt{(1-34 x^2+x^4)}} \frac{G_1(x)}{\varpi_0(x)} \, ,
\end{aligned}
\end{equation}
and the full $\epsilon$-factorised differential equations together with all rotations starting from our initial basis in~\cref{eq:basisK3} are given in the ancillary file.

Even though the size of the sector is quite large, we stress that the transformations needed are still quite simple, as long as a good initial basis is identified. In total, we had to correct our initial basis in the following way: After the gauge transformation with the inverse semi-simple part of the Wronskian in~\cref{eq:Grav_K3_W} and the rescaling in $\epsilon$ in~\cref{eq:Grav_K3_D}, only $I_2$ and $I_3$ got corrections of the form \eqref{eq:rottotderiv}. Moreover, we have introduced four new transcendental functions $G_i$. The integrals of $I_4-I_7$ only needed corrections from $I_1$ through some of the new functions $G_i$. In particular, our choice of $I_4$ and $ I_7$ does not need corrections at all, and this proves that their choice was appropriate.
This shows that even though a full integrand analysis is non-trivial for this example, the heuristic approach we used here led to a good choice of starting basis.

Let us close this section with a final remark. We have again checked whether the new functions $G_i$ have an expansion in $x$ with integer coefficients. For $G_1$ and $G_4$, this is expected from the discussion in section~\ref{sec:integer}. 
But also $G_2$ and $G_3$ have an integral expansion, which we have checked up to $\mathcal O(x^{1000})$.

\subsubsection{A sector related to a Calabi-Yau three-variety}

\begin{figure}[h]
    \centering
\begin{tikzpicture}[baseline={([yshift=-1ex]current bounding box.south)},scale=1.2]
  \coordinate (inA) at (0.4,.7);
  \coordinate (outA) at (3.6,.7);
  \coordinate (inB) at (0.4,-.7);
  \coordinate (outB) at (3.6,-.7);
  \coordinate (xA) at (1,.7);
  \coordinate (xxA) at (1.825,.7) ;
  \coordinate (xxB) at (1.825,-.7) ;
  \coordinate (yA) at (1.60+0.4,.7);
  \coordinate (yyA) at (2.375,0) ;
  \coordinate (zA) at (3,.7);
  \coordinate (zzA) at (3.75,.7) ;
  \coordinate (xB) at (1,-.7);
  \coordinate (yB) at (1.6+0.4,-.7);
  \coordinate (zB) at (3,-.7);
  \coordinate (zzB) at (3.75,-.7);
  \coordinate (xM) at (1,0);
  \coordinate (yM) at (1.6+0.4,0);
  \coordinate (zM) at (3,-0);
  \draw [dotted] (inA) -- (outA);
  \draw [dotted] (inB) -- (outB) node [midway, below]{};
  \draw [draw=none] (xA) to[out=40,in=140] (zA);
  \draw [photon] (xA) -- (xB);
  \draw [photon] (yA) -- (xM);
  \draw [photon] (zA) -- (zB);
  \draw [photon] (xM) -- (yM) -- (zM);
  \draw [photon] (yB) -- (zM);
  \draw [fill] (xM) circle (.08);
   \draw [fill] (zM) circle (.08);

  \draw [fill] (xA) circle (.08);
  \draw [fill] (yA) circle (.08);
  \draw [fill] (zA) circle (.08);
  \draw [fill] (xB) circle (.08);
  \draw [fill] (yB) circle (.08);
  \draw [fill] (zB) circle (.08);
\end{tikzpicture}
    \caption{The graphs related to a family of CY threefolds.}
    \label{fig:CY3GW}
\end{figure}
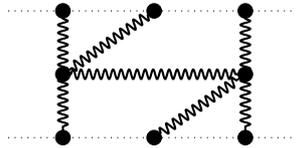

As a last example, we  consider a sector that contains a non-trivial 
 CY three-fold. The corresponding graph is shown in~\cref{fig:CY3GW}.
The Picard-Fuchs equation related to this geometry was first discussed in ref.~\cite{Frellesvig:2023bbf}, and its geometry was fully classified in ref.~\cite{Klemm:2024wtd}.
Recently this sector was also brought into $\epsilon$-form using a different method, which introduces apparent singularities \cite{Frellesvig:2024rea}. In contrast to this, we will show that an $\epsilon$-factorisation can be achieved without any such apparent singularities, which simplifies the discussion substantially. As in the previous example, choosing a suitable starting basis is crucial. We use the following integrals
\begin{equation}
\begin{aligned}
I_1&=\dfrac{1}{x} I_{1,1,1,1,0,0,0,0,0,0,1,0,1,1,1,0,1,0,0,1,0,1} \, ,\quad 
I_2=\partial_x I_1      \, ,\quad
I_3=\partial_x^2 I_1    \, ,\quad
I_4=\partial_x^3 I_1    \, ,\\
I_5&=(1-6 \epsilon ) \frac{x^2-1 }{4x} I_{1,1,1,1,0,0,0,0,0,0,1,0,1,1,1,0,1,0,-1,1,0,1} \, .
\label{eq:startCY3}
\end{aligned}
\end{equation}
Let us argue now why this choice of basis is well-suited. It was already observed using a Baikov analysis in refs.~\cite{Frellesvig:2023bbf, Klemm:2024wtd} that the leading singularity of the first integral $I_1$ is given by
\begin{align}
\text{LS}(I_1) = &\oint \dfrac{\mathrm dt_1 \mathrm dt_2 \mathrm dt_3 }{\sqrt{P}} 
\end{align}
with
\begin{align}
P =&64\, t_1^2 t_2^2 t_3^2 x^2 (t_1+t_2+1)+(t_1+1)^2 (t_2+1)^2 \left(x^2-1\right)^2 \left(t_1 t_2+t_3^2\right)^2 \, .
\end{align}
Moreover, it was shown, that from this representation one can define a singular CY three-variety $X$ as a double cover of $\mathbb P^3$, i.e.,
\begin{equation}
   X = \left\{t_4^2 = P(t_1,t_2,t_3)\right\}  \, .
\end{equation}
This construction is a generalisation of describing an elliptic curve through a double cover of $\mathbb P^1$, i.e., $\mathcal E=\{y^2 = P(x)\}$ with $P(x)$ a polynomial of degree three or four in $x$, cf. ref.~\cite{Duhr:2024hjf} and references therein. The Picard-Fuchs operator of this family of CY three-varieties is a CY operator of order four and reads
\begin{align}
    \mathcal{L}^{(4)}_0=& (1-x)^3 (1+x)^3 \left(1+x^2\right)^2 \theta_x ^4-8 x^2 (1-x)^2 (1+x)^2 \left(1+x^2\right) \left(2+x^2\right)\theta_x ^3\nonumber \\
    &-4 x^2 (1-x) (1+x) \left(7-12 x^2-17 x^4-6 x^6\right)\theta_x ^2\\
\nonumber    &-8 x^2 \left(3-15 x^2+3 x^4+9 x^6+4 x^8\right) \theta_x -x^2 \left(9-71 x^2+39 x^4+39 x^6+16 x^8\right) \, .
\end{align}
Its Frobenius basis can be chosen as
\begin{align}
    \varpi_0(x)=&1+\frac{9 x^2}{16}+\frac{1681 x^4}{4096}+\frac{21609 x^6}{65536}+\mathcal{O}(x^8) \, ,\\
    \varpi_1(x)=&\varpi_0(x) \log(x)+S_1(x) = \varpi_0(x) \log(x) +\frac{3 x^2}{8}+\frac{1517 x^4}{4096} +\frac{22295 x^6}{65536}+\mathcal{O}(x^8) \, , \nonumber \\
    \varpi_2(x)=&\dfrac{1}{2} \varpi_0(x) \log^2(x) +S_1(x) \log(x)+S_2(x) \nonumber \\
                &= \dfrac{1}{2} \varpi_0(x) \log^2(x) +S_1(x) \log(x)+ \frac{5 x^2}{32}+\frac{7289 x^4}{32768}+\frac{1123325 x^6}{4718592}+\mathcal{O}(x^8) \, , \nonumber \\
    \varpi_3(x)=&\dfrac{1}{6} \varpi_0(x) \log^3(x) +\dfrac{1}{2}S_1(x) \log^2(x)+S_2(x) \log(x) \nonumber \\
    &-\frac{5}{32}  x^2 -\frac{7865 x^4}{65536}-\frac{2448943 x^6}{28311552}+\mathcal{O}(x^8) \, . \nonumber
\end{align}
The details of the corresponding CY three-variety, including the computation of its torus period, were discussed in ref.~\cite{Klemm:2024wtd}. 

Once we have identified the integral of the first kind $I_1$, we know from the dimensionality of the CY variety that we have to choose three extra integrals $I_2-I_4$ as derivatives of $I_1$. This fixes four out of five master integrals.
The fifth master, $I_5$,
is chosen such that its leading singularity has an additional pole at infinity in $t_1$ and reads
\begin{equation}
\begin{aligned}
    \text{LS}(I_5)\propto  &(x^2-1) \int \dfrac{\mathrm dt_1 \mathrm dt_2 \mathrm dt_3  \, t_1}{\sqrt{P}}\\
    \propto                & (x^2-1) \int \dfrac{\mathrm d\widetilde{t}_1 \mathrm dt_2 \mathrm dt_3}{\widetilde{t}_1 t_2 (1+t_2) (x^2-1)}+\mathcal{O}\left(\widetilde{t}_1^0\right)\, ,
\end{aligned}
\end{equation}
where $\widetilde{t}_1=1/t_1$ and we have ignored numerical prefactors.
By taking the two additional residues in $t_2$ and $\widetilde{t}_1$ and we are then left with
\begin{align}
    \text{LS}(I_5)\propto\int \mathrm{d}t_3=\int \dfrac{ \mathrm{d} \widetilde{t}_3}{\widetilde{t}_3^2} \, ,
\end{align}
which shows that this integral has a single double pole. This indicates a weight drop of integral $I_5$ compared to $I_1$ which can be compensated by a reduced $\epsilon$-rescaling.

In this basis, the homogeneous part of the differential equations reads
\begin{align}
    \partial_x I =M(x,\epsilon) I
\end{align}
with its $\epsilon=0$ part given by
\begin{align}
 \resizebox{\textwidth}{!}{$\displaystyle
 M(x,0)=  \left(
\begin{array}{ccccc}
 0 & 1 & 0 & 0 & 0 \\
 0 & 0 & 1 & 0 & 0 \\
 0 & 0 & 0 & 1 & 0 \\
 \frac{9-71 x^2+39 x^4+39 x^6+16 x^8}{x^2(1-x)^3  (1+x)^3 \left(1+x^2\right)^2} & -\frac{1-69 x^2+202 x^4+22 x^6-123 x^8-65 x^{10}}{x^3(1-x)^3  (1+x)^3 \left(1+x^2\right)^2} & -\frac{7-76 x^2+10 x^4+116 x^6+55 x^8}{x^2(1-x)^2 
   (1+x)^2 \left(1+x^2\right)^2} & -\frac{2 \left(3-8 x^2-7 x^4\right)}{x(1-x)  (1+x) \left(1+x^2\right)} & 0 \\
 -\frac{2 x \left(1-2 x^2\right)}{1+x^2} & \frac{(1-x) (1+x) \left(1-5 x^2\right)}{1+x^2} & \frac{x(1-x)^2  (1+x)^2}{1+x^2} & 0 & 0 \\
\end{array}
\right) \, .
   $ }
\end{align}

We again see that the choice of the fifth integral leads to its decoupling from the first four integrals for $\epsilon=0$. Furthermore, the rescaling with $(1-6 \epsilon)$ removes all $\epsilon$ dependent denominators.\footnote{The need to introduce an $\eps$-dependent prefactor might be interpreted as a sign of the fact that the initial basis has not been optimally chosen. It would be interesting to see if a better choice exists where no $\eps$-dependent prefactor is required and the integrand analysis shows no double pole. We do not elaborate on this further here.}
Note that the $x$-dependent prefactor is needed to properly normalise the leading singularity and, in this way, bring $M_{5,5}$ into $\epsilon$-form. 
Also, in this case, it is useful to showcase the dependence on $\epsilon$ as follows
\begin{align}
    \left(
\begin{array}{ccccc}
 -  & 0 & -  & -  & -  \\
 -  & -  & 0 & -  & -  \\
 -  & -  & -  & 0 & -  \\
 4 & 3 & 2 & 1 & 3 \\
 2 & 1 & 0 & -  & 1 \\
\end{array}
\right) \, ,
\end{align}
where again we used an integer to indicate the highest power of $\epsilon$ in each entry
and `$-$' if the entry is zero.
We see that the fifth integral couples now to three integrals instead of two, as in the K3 case.

As for the K3 case analysed before, it is convenient to improve our starting basis before rotating with the inverse of the semi-simple part of the Wronskian. In particular, we would like the coupling of $I_5$ to $I_2$ and $I_3$ to vanish for $\epsilon=0$, which can be achieved with the following rotation
\begin{align}
    T=\left(
\begin{array}{ccccc}
 1 & 0 & 0 & 0 & 0 \\
 0 & 1 & 0 & 0 & 0 \\
 0 & 0 & 1 & 0 & 0 \\
 0 & 0 & 0 & 1 & 0 \\
 \frac{(1-x) (1+x) \left(1+6 x^2-3 x^4\right)}{4 \left(1+x^2\right)^2} & \frac{(1-x)^2 (1+x)^2 x}{2 \left(1+x^2\right)} & 0 & 0 & 1 \\
\end{array}
\right) \, .
\end{align}
Now we can further rotate this basis with the inverse of the semi-simple part of the Wronskian (see also the general form in~\cref{eq:SeSiExplicit})
\begin{align}
    {W}^\text{ss}=\left(
\begin{array}{cccc}
 \varpi_0 & 0 & 0 & 0 \\
 \varpi_0' & \frac{\varpi_0}{x \alpha_1} & 0 & 0 \\
 \varpi_0'' & \frac{2\varpi_0'}{x\alpha_1}-\left( \frac{1}{x^2\alpha_1} + \frac{\alpha_1'}{x\alpha_1^2} \right)\varpi_0 & \frac{xC_3\alpha_1}{\varpi_0} & 0 \\
 \varpi_0''' &{W}^\text{ss}_{4,2}& {W}^\text{ss}_{4,3} & \frac{C_3}{\varpi_0} \\
\end{array}
\right) \, ,
\end{align}
where
\begin{align}
    {W}^\text{ss}_{4,2} =& \frac{2{\varpi_0'}^2}{x\alpha_1\varpi_0} - \frac{\alpha_1'\varpi_0'}{x\alpha_1^2} - \frac{\varpi_0''}{x\alpha_1} +  \frac{2 \left(1+5 x^2-7 x^4-3 x^6\right)}{(1-x)^2 (1+x)^2 \left(1+x^2\right) x^3}\frac{\varpi_0}{\alpha_1} \nonumber \\
            & \quad - \frac{\left(1-3 x^2\right) \left(7+5 x^2\right)}{(1-x) (1+x) \left(1+x^2\right) x^2}\frac{\varpi_0'}{\alpha_1} + \frac{3-8 x^2-7 x^4}{(1-x) (1+x) \left(1+x^2\right) x^2}\frac{\alpha_1'\varpi_0}{\alpha_1^2} \, ,\\
    {W}^\text{ss}_{4,3} =& \left(\frac{x\alpha_1\varpi_0'}{\varpi_0^2} -\frac{3-8 x^2-7 x^4}{(1-x) (1+x) \left(1+x^2\right)}\frac{\alpha_1}{\varpi_0}\right)C_3 \, . \nonumber
\end{align}
The structure series $\alpha_1(x)$ in~\cref{eq:defu} and the three-point coupling $C_3(x)$ in~\cref{eq:defnptcoupling} are
\begin{equation}
    \alpha_1 = \frac{\varpi_0^2}{x(\varpi_0\varpi_1'-\varpi_0'\varpi_1)} \quad\text{and}\quad C_3 = \frac{1+x^2}{(1-x)^3 (1+x)^3 x^3} \, .
\end{equation}
We rescale with
\begin{align}
    D(\epsilon) = \text{diag}(\epsilon^3,\epsilon^2,\epsilon,1,\epsilon^2) \, ,
\end{align}
where the reduced rescaling by $\epsilon^2$ of $I_5$ is due to the weight drop discussed above. After this step, we can now successively remove all terms in the differential equations that are not in $\epsilon$-form, starting with the ones proportional to $1/\epsilon^2$ and continuing up to $\epsilon^0$. Large portions of these terms are again identified as total derivatives that can be removed through an ansatz containing products of $\alpha_1(x), \varpi_0(x)$, their derivatives, and algebraic functions in $x$ (compare also with~\cref{eq:rottotderiv} in the K3 case).
Notice that in this step, we have decided to remove all terms containing $\alpha_1'(x)$, but we could have equally well chosen to remove terms containing $\varpi_0' (x)$. It is a priori unclear if some choice is advantageous over the other. To derive the fully $\epsilon$-factorised system, we have to introduce seven new transcendental functions $G_i(x)$ that eliminate the still non-$\epsilon$-factorised terms. Having performed the previous rotations, these new functions can be introduced in the following way
\begin{align}
    \tilde{T}=\left(
\begin{array}{ccccc}
 1 & 0 & 0 & 0 & 0 \\
 G_6 & 1 & 0 & 0 & 0 \\
\tilde{T}_{3,1} & 20 G_4^2 & 1 & 0 & 40 G_4 \\
 \tilde{T}_{4,1} & \frac{G_2}{\epsilon }+G_7 & -G_6 & 1 & \frac{40G_1}{\epsilon}-40G_4G_6+40G_5 \\
 -\frac{G_1}{\epsilon }+G_5 & G_4 & 0 & 0 & 1 \\
\end{array}
\right),
\label{eq:finalrotCY3}
\end{align}
where
\begin{equation}
\begin{aligned}
    \tilde{T}_{3,1}=&-\frac{40 G_1 G_4}{\epsilon }+\frac{G_2}{\epsilon }+40 G_4 G_5-20 G_4^2 G_6-G_7 \, , \\
    \tilde{T}_{4,1}=&-\frac{20G_1^2}{\epsilon^2} + \frac{40G_1G_4G_6}{\epsilon} - \frac{40G_1G_5}{\epsilon} - \frac{G_2G_6}{\epsilon} + \frac{G_3}{\epsilon} \\
    &\quad + 20 G_4^2G_6^2 - 40 G_4G_5G_6 +20 G_5^2 + G_6G_7 \, .
\end{aligned}
\end{equation}
The full rotation into $\epsilon$-form beginning from our starting basis in~\cref{eq:startCY3} is given in an ancillary file, whereas the defining first-order differential equations for the new functions $G_i(x)$ for $i=1,\hdots,7$ are given in~\cref{app:bh}. 

We see that, compared to the K3 case in \cref{eq:K3G}, the rotations are more involved. 
For instance, the fifth integral gets corrections proportional to the first two integrals instead of just one, and $I_3$ and $I_4$ get corrections proportional to $I_5$.  
In comparison, in the K3 case, only one integral of the K3 block got corrections from the additional residue integrals. We note, though, that in both cases, the filtration of the MHS is preserved.

Let us again conclude with some comments about the expansions of the new functions $G_i(x)$ for $i=1,\hdots,7$. Since the new functions appear at multiple places in the rotation in~\cref{eq:finalrotCY3}, it is not that easy to relate them just to the pure CY block or to the additional residue integral alone. Therefore, it is not clear anymore whether we expect them to have an integral expansion. 
We have tested up to $\mathcal O(x^{1000})$ that all eight new functions have indeed an integral expansion after one takes 
into account the rescaling $x\rightarrow 2^4x$.


\section{Conclusions}
\label{conclusions}

In this paper, we have elaborated on the method to
find $\epsilon$-factorised differential equations proposed in ref.~\cite{Gorges:2023zgv}. In particular, we generalised its scope and
discussed various aspects of canonical differential equations for multi-loop Feynman integrals. 
Indeed, the method of ref.~\cite{Gorges:2023zgv} had originally been introduced for Feynman integrals whose associated geometry is a family of elliptic curves, with the only
exception of the case of the three-loop equal-mass banana integral, which is associated to a one-parameter family of K3 surfaces, i.e., a symmetric square of a family of elliptic curves. After its original formulation, its applicability to
more complicated geometries was demonstrated in a case-by-case approach, both for families of Calabi-Yau geometries emerging from the calculation of gravitational waves emission in black hole scattering and, more recently, for a genus-two problem. Motivated by these findings, in this paper, we have proposed a roadmap for extending the procedure of ref.~\cite{Gorges:2023zgv} from elliptic curves to larger classes of geometries that may arise from the leading singularities associated with the maximal cuts of the corresponding Feynman integrals. 
Central to our proposal is the idea that a good starting basis for transforming the system of differential equations into canonical form is aligned with the mixed Hodge structure (MHS) underlying the geometry. This idea is compatible with the known polylogarithmic cases, as well as with the cases where the associated geometry is a family of elliptic curves. Indeed, in the former case, the MHS captures the fact that one can associate the canonical master integrals with independent leading singularities. In contrast, in the latter case, the MHS allows one to choose the starting basis of master integrals in such a way that on the maximal cut in integer dimensions, the integrands reduce to differentials of the first, second, and third kinds on the elliptic curve. 

Once a suitable starting basis has been identified, the method of ref.~\cite{Gorges:2023zgv} proceeds via algorithmic steps. First, for each block (or more precisely, each weight-graded piece from the MHS), we normalise the integrals by the inverse of the semi-simple part of the corresponding period matrix, as prescribed by ref.~\cite{Broedel:2018qkq}. This step is followed by a rescaling by powers of $\eps$, such as to realign the transcendental weights and the $\eps$-expansion. The $\eps$-factorisation is then achieved by performing a rotation of the basis integrals. The entries in the rotation matrix are fixed by solving first-order linear differential equations. In~\cref{tab:applications}, we summarise the cases where the method of ref.~\cite{Gorges:2023zgv} has already been successfully applied, together with the geometries underlying these integrals. This list illustrates the flexibility and the power of the method and shows that it can be used as a practical tool for state-of-the-art computations in quantum field theory and gravitational wave physics.

\begin{table}[h]
\begin{center}
\resizebox{\textwidth}{!}{
\begin{tabular}{|l|l|l|}
\hline
Description & References & Geometry \\
\hline\hline
Equal-mass banana graphs            & \cite{Gorges:2023zgv}, this paper   & CY 2-, 3- and 4-folds                   \\
\hline
Single scale triangle graphs        & \cite{Gorges:2023zgv}               & Elliptic curve                          \\
\hline
3-loop corrections to the electron  & \cite{Duhr:2024bzt,Forner:2024ojj}  & Sunrise elliptic curve,                 \\
and photon self-energies in QED     &                                     & banana K3 surface                       \\
\hline
3- and 4-loop ice cone integrals    & \cite{Gorges:2023zgv}, this paper   & Two copies of sunrise elliptic          \\
                                    &                                     & curve and banana K3 surface             \\
\hline
Deformed CY operators               & this paper                          & CY 2-, 3- and 4-folds                   \\
\hline
Equal-mass banana graphs with       & unpublished                         & CY 1-, 2-, 3-folds                      \\
one massless propagator             &                                     &                                         \\
\hline
Gravitational scattering            & \cite{Klemm:2024wtd, Driesse:2024feo, Driesse:2024xad}  & Sym. square of Legendre curve,  \\
at 5PM-1SF                          &                                     & CY 3-fold AESZ 3                        \\
\hline
Gravitational scattering            & this paper                          & Ap\'ery family of K3 surfaces,          \\
at 5PM-2SF                          &                                     & CY 3-fold                               \\
\hline
Generic three-mass sunset           & \cite{Gorges:2023zgv}               & Elliptic curve                          \\
\hline
2-loop 3-point integrals for $gg\rightarrow H$ & \cite{Marzucca:2025eak}  & Two-mass sunrise elliptic curve         \\
\hline
2-parameter triangle graph          & \cite{Gorges:2023zgv}               & Elliptic curve                          \\
\hline  
2-loop 4-point integrals for Bhabha  & unpublished                     & Elliptic curve                          \\
and M\o ller scattering & & \\
\hline
2-loop 4-point integrals for diphoton & \cite{Becchetti:2025rrz}          & Elliptic curve                          \\
\hline
2-loop 4-point acnode integral      & unpublished                         & Elliptic curve                          \\
(diagonal box) & & \\
\hline
3-parameter double box              & \cite{Gorges:2023zgv}               & Elliptic curve                          \\
\hline
2-loop 5-point integrals for $t\bar{t}+$jet & \cite{Becchetti:2025oyb}    & Elliptic curve                          \\
\hline
3-loop two-mass banana graph        & unpublished                         & K3 surface                              \\
\hline
4-loop two-mass banana graph        & unpublished                         & CY 3-fold                               \\
\hline
Maximal cut of a non-planar         & \cite{Duhr:2024uid}                 & Hyperelliptic curve of genus 2          \\
double box                          &                                     &                                         \\
\hline
\end{tabular}
}
\caption{\label{tab:applications}Summary of the different applications of the method of ref.~\cite{Gorges:2023zgv}. We intend to release the unpublished results listed above as part of future works in several different collaborations.
}
\end{center}
\end{table}

In the future, besides applying our method to obtain new results for multi-loop integrals, it would be interesting to study more deeply the properties of the canonical differential equations and their solutions. For example, we have observed that, at least for the examples of deformed CY operators considered in section~\ref{sec:general_eps_fac}, the differential forms entering the final canonical differential equations are linearly independent and only have at most simple poles at the MUM-point.
We expect these properties to be generic and to extend beyond the examples that we have considered in this paper. It would be interesting to see if these properties can be rigorously proven. Another important question is to more systematically understand the new functions one has to introduce to bring the system into $\eps$-factorised form, in particular, if and when they can be explicitly evaluated in terms of other functions. One approach could be to exploit the fact that if the differential forms are independent, then the intersection matrix in the canonical basis must be constant~\cite{Duhr:2024xsy}. In ref.~\cite{Duhr:2024uid}, it was shown that this puts strong constraints on the new functions and sometimes even fixes them completely. Finally, in a series of explicit calculations of scattering amplitudes and correlators in realistic gauge theories~\cite{Duhr:2024bzt, Forner:2024ojj, Becchetti:2025rrz}, it has been observed that large sets of differential forms
appearing in the canonical differential equations, consistently drop in the finite remainder of the relevant physical quantities.
 We leave these questions for future work.

\acknowledgments

We thank Felix Forner, Albrecht Klemm, Cesare Mella, Nikolaos Syrrakos and Johann Usovitsch for collaboration on related projects
and useful comments on the manuscript.
This work was supported in part by the Deutsche Forschungsgemeinschaft (DFG, German Research Foundation) through the Excellence Cluster ORIGINS under Germany’s Excellence Strategy – EXC-2094-390783311 (CN, LT, FW) and through Projektnummer 417533893/GRK2575 ``Rethinking Quantum Field Theory'' (BS), and in part by the European Research Council (ERC) under the European Union’s research and innovation program grant agreements 949279 (ERC Starting Grant HighPHun (CN, LT)) and 101043686 (ERC Consolidator Grant LoCoMotive (CD, SM)).
Views and opinions expressed are, however, those of the author(s) only and do not necessarily reflect those of the European Union or the European Research Council. Neither the European Union nor the granting authority can be held responsible for them.

\appendix


\section{Short review of mixed Hodge structures}
\label{app:mhs}

This section provides a brief overview of pure and mixed Hodge structures. For an introduction to the subject, we refer to refs.~\cite{Durfee1981ANG, Filippini2015}.

\subsection{Pure Hodge structures}
A pure Hodge structure of weight $n$ is a finite-dimensional $\mathbb{Z}$-module $V_{\mathbb{Z}}$ together with a decomposition of its complexification $V_{\mathbb{C}} = V_{\mathbb{Z}}\otimes \mathbb{C}$ such that
\beq
V_{\mathbb{C}} = \bigoplus_{p+q=n}V^{p,q}\,,\qquad V^{q,p} = \overline{V^{p,q}}\,.
\eeq
Note that we could replace $V_{\mathbb{Z}}$ by a rational vector space $V_{\mathbb{Q}}$ (in which case we would talk about a \emph{rational} Hodge structure).
The \emph{Hodge numbers} are defined as $h^{p,q} = \dim V^{p,q}$. It is easy to see that if $n$ is odd, then the dimension of $V_{\mathbb{Z}}$ (respectively $V_{\mathbb{Q}}$) must necessarily be even.
The existence of a Hodge decomposition is equivalent to the existence of a Hodge filtration on $V_{\mathbb{C}}$, defined by
\beq
F^p = \bigoplus_{k=p}^nV^{k,n-k}\,.
\eeq
The Hodge filtration satisfies $F^p\oplus \overline{F^{n-p+1}} = V_{\mathbb{C}}$. 

In the case where $V_{\mathbb{Z}}$ is one-dimensional, one can easily classify all pure Hodge structures. Indeed, it is easy to see that we need to have $n=-2m$ even, and $V_{\mathbb{C}} = V^{-m,-m}$. The unique one-dimensional pure Hodge structure of weight $-2m$ is denoted by $\mathbb{Z}(m)$ (or $\mathbb{Q}(m)$ in the case of a rational Hodge structure). 
Note that all constructions from linear algebra can be applied to Hodge structures. In particular, we may define tensor products and direct sums of Hodge structures. 

So far, all considerations have been general. We now focus on the case where $V_{\mathbb{Q}} = H^n(X,\mathbb{Q})$ is the $n^{\textrm{th}}$ cohomology group of some algebraic variety $X$. It is known that, if $X$ is smooth and projective, then $H^n(X,\mathbb{Q})$ carries a pure Hodge structure of weight $n$ (cf.,~e.g.,~ref.~\cite{GriffithsHarris}). The Hodge decomposition corresponds, in this case, simply to the decomposition of differential forms into holomorphic and antiholomorphic forms:
\beq
H^n(X,\mathbb{C}) = \bigoplus_{p+q=n} H^{p,q}(X)\,,
\eeq
where $H^{p,q}(X)$ is the complex vector space generated by all cohomology classes of $(p,q)$-forms on $X$.

If $X$ is not smooth and/or not projective, pure Hodge structures are insufficient to capture the structure of the cohomology groups. One easy way to see this is that all pure Hodge structures of odd weight are even-dimensional, but there are varieties that have odd-dimensional cohomology groups $H^n(X,\mathbb{Q})$ for $n$ odd. As an example, take $X = \mathbb{C}^{\times} = \mathbb{C}\setminus\{0\}$. $X$ is not projective, and $H^1(\mathbb{C}^{\times},\mathbb{Q})$ is one-dimensional.

\subsection{Mixed Hodge structures}

A \emph{mixed Hodge structure} is a rational vector space $V_{\mathbb{Q}}$ together with an increasing filtration, called the \emph{weight filtration},
\beq
0\subseteq\ldots\subseteq W_r \subseteq W_{r+1}\subseteq \ldots \subseteq W_{N} = V_{\mathbb{Q}}\,,
\eeq
and a decreasing filtration called the \emph{Hodge filtration},
\beq
0\subseteq \ldots\subseteq F^p \subseteq F^{p-1}\subseteq \ldots \subseteq F^0 = V_{\mathbb{Q}}\,,
\eeq
such that the Hodge filtration induces on the graded quotient $\Gr_k^W = W_{k}/W_{k-1}$ a pure Hodge structure of weight $k$.
Note that every pure Hodge structure of weight $n$ is a MHS, with the weight filtration concentrated in weight $n$, $0=W_{n-1}\subseteq W_n = V_{\mathbb{Q}}$.

Deligne proved that the cohomology of every variety $X$ carries a MHS~\cite{PMIHES_1971__40__5_0, PMIHES_1974__44__5_0}. The weight filtration on $H^n(X,\mathbb{Q})$ has the form
\beq
0=W_{-1} \subseteq W_0\subseteq W_1\subseteq \ldots\subseteq W_{2n} = H^n(X,\mathbb{Q})\,.
\eeq
There are interesting special cases.
\begin{itemize}
\item If $X$ is smooth and projective, then we recover the previous result that $H^n(X,\mathbb{Q})$ carries a pure Hodge structure of weight $n$:
\beq
0=W_{n-1}\subseteq W_n = H^n(X,\mathbb{Q}) \,.
\eeq
\item If $X$ is smooth but not necessarily projective, then the weight filtration takes the form
\beq
0=W_{n-1}\subseteq W_n\subseteq \ldots\subseteq W_{2n} = H^n(X,\mathbb{Q}) \,,
\eeq
and if $\overline{X}$ is any smooth compactification of $X$, then $W_{n} = H^n(\overline{X},\mathbb{Q})$.
\item If $X$ is projective but not necessarily smooth, we have
\beq
0=W_{-1}\subseteq W_0\subseteq \ldots\subseteq W_{n} = H^n(X,\mathbb{Q}) \,.
\eeq
\end{itemize}
In the general case, for example, when $X$ is quasi-projective (meaning that $X=\overline{X}\setminus Y$, where $\overline{X}$ and $Y$ are projective), then all components of the weight filtration may be non-trivial.


\section{Symmetric squares from Clausen's formula}
\label{app:Clausen}

Let $\mathcal{L}^{(3)}$ be a linear differential operator of order three, and let us denote by $\Sol(\mathcal{L}^{(3)})$ its solution space, which is a complex three-dimensional vector space,
\beq
\Sol(\mathcal{L}^{(3)}) \coloneqq \big\{f:\mathbb{C}\to\mathbb{C}: \mathcal{L}^{(3)}f=0\big\}\,.
\eeq
$\mathcal{L}^{(3)}$ is the \textit{symmetric square} of a second-order operator $\mathcal{L}^{(2)}$ if we can find a basis of solutions $\{f_1^2,f_1f_2,f_2^2\}$ of $\Sol(\mathcal{L}^{(3)})$ with $\mathcal{L}^{(2)}f_{1,2}=0$. Said differently, we have
\beq
\Sol(\mathcal{L}^{(3)}) = \Sym^2\Sol(\mathcal{L}^{(2)}) \,.
\eeq
It is well known that every CY operator of order three is the symmetric square of a CY operator of order two~\cite{Doran:1998hm, BognerCY, BognerThesis}. This begs the natural question of whether this result can be extended to the deformed CY operators of order three introduced in section~\ref{sec:general_eps_fac}. In general, this will not be the case. The following discusses when a deformed CY operator of hypergeometric type of order three is a symmetric square.

The holomorphic solution of a (deformed) CY operator of hypergeometric type of order $n$ is a generalised hypergeometric function ${}_{n+1}F_n$ (cf.~eq.~\eqref{eq:hyp_geo_def}). Hence, if the deformed CY operator $\mathcal{L}_\eps^{(3)}$ is the symmetric square of  $\mathcal{L}_\eps^{(2)}$, then there must be a corresponding identity of hypergeometric functions expressing a ${}_{3}F_2$ function as the square of a ${}_2F_1$ function. Such an identity is given by \emph{Clausen's formula}~\cite{Clausen1828}, 
\begin{equation}
    \,_3F_2\left(2a,2b,a+b;a+b+\tfrac{1}{2},2a+2b;z\right)={}_2F_1\left(a,b;a+b+\tfrac{1}{2};z\right)^2\,.
\end{equation}

Let us start by discussing the undeformed case.
We have checked that the known examples of CY operators of order three of hypergeometric type arise through Clausen's formula. Combining with the constraints on the indices of the hypergeometric function discussed in section~\ref{sec:general_eps_fac}, we see that the most general hypergeometric solution of a CY operator or order three that arises from Clausen's formula takes the form
\beq\label{eq:Clausen_CY}
{}_3F_2\big(r,1-r,\tfrac{1}{2};1,1;z\big) = {}_2F_1\big(\tfrac{r}{2},\tfrac{1-r}{2};1;z\big)^2\,,
\eeq
where $0<r\le \tfrac{1}{2}$ is a rational number. It is easy to see that the examples from~\cref{tab:hypgeo} correspond to $r=\tfrac{1}{2}$ and $r=\tfrac{1}{4}$.

Let us now consider the $\eps$-deformed case. The most general hypergeometric solution to a deformed CY operator of order three that reduces to eq.~\eqref{eq:Clausen_CY} for $\eps=0$ then takes the form
\beq
{}_3F_2\big(r+s_1\eps,1-r+s_2\eps,\tfrac{1}{2}+s_3\eps;1+t_1\eps,1+t_2\eps;z\big)\,.
\eeq
We immediately see that such a function satisfies~\cref{eq:Clausen_CY} for $\eps=0$, but it does not for generic values of $s_i$, $t_i$, and $\eps$. Instead, Clausen's formula is satisfied, provided that
\beq
2s_3 = 2t_1=t_2= s_1+s_2 \eqqcolon s_{12}\,,
\eeq
in which case Clausen's formula reduces to
\beq\bsp
{}_3F_2\big(r+s_1\eps,&1-r+s_2\eps,\tfrac{1}{2}+\tfrac{1}{2}s_{12}\eps;1+\tfrac{1}{2}s_{12}\eps,1+s_{12}\eps;z\big) = \\
&\,={}_2F_1\big(\tfrac{r}{2}+\tfrac{1}{2}s_{1}\eps,\tfrac{1-r}{2}+\tfrac{1}{2}s_{2}\eps;1+\tfrac{1}{2}s_{12}\eps;z\big)^2\,.
\esp\eeq
For example, for the two cases discussed in~\cref{tab:hypgeo}, we have
\begin{equation}\begin{split}
    _3F_2\!\left(\tfrac{1}{2}+s_1\epsilon,\right.&\!\!\left.\tfrac{1}{2}+s_2\epsilon,\tfrac{1}{2}+\tfrac{1}{2}s_{12}\epsilon;1+\tfrac{1}{2}s_{12}\epsilon,1+s_{12}\epsilon;z\right)\,\\
    &\qquad={}_2F_1\!\left(\tfrac{1}{4}+\tfrac{1}{2}s_1\epsilon,\tfrac{1}{4}+\tfrac{1}{2}s_2\epsilon;1+\tfrac{1}{2}s_{12}\epsilon;z\right)^2\,,\\
    _3F_2\!\left(\tfrac{1}{4}+s_1\epsilon,\right.&\!\!\left.\tfrac{3}{4}+s_2\epsilon,\tfrac{2}{4}+\tfrac{1}{2}s_{12}\epsilon;1+\tfrac{1}{2}s_{12}\epsilon,1+s_{12}\epsilon;z\right)\,\\
    &\qquad={}_2F_1\!\left(\tfrac{1}{8}+\tfrac{1}{2}s_1\epsilon,\tfrac{3}{8}+\tfrac{1}{2}s_2\epsilon;1+\tfrac{1}{2}s_{12}\epsilon;z\right)^2\,.
\end{split}\end{equation}


\section{Supplementary materials for additional examples}
\label{app:suppmat}

In this appendix, we give additional material for the examples discussed in~\cref{sec:additional_examples}.

\subsection{Supplementary material ice cone family}
\label{app:icecone}

We list here the first-order differential equations of the eight new functions introduced in the $\epsilon$-factorisation procedure of the four-loop ice cone integral. They satisfy the following relations
\begin{align}
    {G_1^+}'(x) &= \frac{8 (1-8 x) (1+8 x)^3}{3 (1-16 x)^2 (1-4 x)^2 x^2} {\varpi_0^+(x)}^2 \, , \\
     {G_1^-}'(x) &= \frac{(8-x) (8+x)^3}{24 (16-x)^2 (4-x)^2} \varpi_0^-(x)^2 \, , \\
    {G_2^+}'(x) &= \frac{1}{\sqrt{(1-4 x) (1-16 x)}}\frac{G_1^+(x)}{\varpi_0^+(x)} \, , \\
    {G_2^-}'(x) &= \frac{8}{\sqrt{(16-x) (4-x)} x}\frac{G_1^-(x)}{\varpi_0^-(x)} \, , \\
    {G_3}'(x) &= -\frac{p_1(x)}{128 x (1-4 x) (1-16 x) \left(1-x^2\right)^3}\varpi_0^+(x)\varpi_0^-(x)  \\
    &\quad + \frac{63-650 x+532 x^2-5240 x^3+17777 x^4-4510 x^5+128 x^6}{64 (1-x)^2 (1+x)^2 (1-4 x) (1-16 x)}{\varpi_0^+}'(x)\varpi_0^-(x) \nonumber\\
    &\quad + \frac{5 x \left(13-8 x+13 x^2\right)}{128 (1-x) (1+x)} \frac{{\varpi_0^+}'(x)^2\varpi_0^-(x)}{\varpi_0^+(x)} \, , \nonumber\\
    {G_4}'(x) &= \frac{8}{\sqrt{(16-x) (4-x)} x} \frac{G_3(x)}{\varpi_0^-(x)} \, , \\
    {G_5}'(x) &= \frac{63-650 x+532 x^2-5240 x^3+17777 x^4-4510 x^5+128 x^6}{64 \left(1-x^2\right)^2 (1-4 x) (1-16 x)\sqrt{(1-4 x) (1-16 x)}}\varpi_0^-(x)  \\
    &\quad -\frac{1}{\sqrt{(1-4 x) (1-16 x)}}\frac{G_3(x)}{\varpi_0^+(x)} \nonumber \\
    &\quad + \frac{5 x \left(13-8 x+13 x^2\right)}{64 \left(1-x^2\right) \sqrt{(1-4 x) (1-16 x)}} \frac{{\varpi_0^+}'(x)\varpi_0^-(x)}{\varpi_0^+(x)}\, , \nonumber\\
    {G_6}'(x) &= -\frac{4 \left(64-880 x+1767 x^2-880 x^3+64 x^4\right)}{(16-x) (4-x)  (1-4 x) (1-16 x)x}G_3(x) \\
    &\quad +\frac{1}{\sqrt{(1-x) (1-16 x)}}\frac{G_1^+(x)G_5(x)}{\varpi_0^+(x)} \nonumber\\
    &\quad -\frac{8}{\sqrt{(16-x) (4-x)} x}\frac{G_2^-(x)G_3(x)}{\varpi_0^-(x)} \nonumber\\
    &\quad -\frac{5 x \left(13-8 x+13 x^2\right)}{64 (1-x) (1+x) (1-4 x) (1-16 x)}\frac{G_1^+(x)\varpi_0^-(x)}{\varpi_0^+(x)} \nonumber\\
    &\quad +\frac{p_2(x)}{96 (1-16 x)^2 (1-4 x)^2 (1-x)^2  (1+x)^2x}\varpi_0^+(x)\varpi_0^-(x) \nonumber\\
    &\quad +\frac{827-11740 x+24921 x^2-9380 x^3-3008 x^4}{96 (1-x) (1+x) (1-4 x) (1-16 x)}{\varpi_0^+}'(x)\varpi_0^-(x) \nonumber
\end{align}
with the two polynomials
\begin{align}
    p_1(x) &= 63-448 x-2343 x^2+26224 x^3-75991 x^4+48176 x^5-69601 x^6 \\
    &\quad +9248 x^7-128 x^8 \, ,\nonumber\\
    p_2(x) &= 5+1960 x-13630 x^2-105640 x^3+1734457 x^4-5189440 x^5 \\
    &\quad +5811536 x^6-2017280 x^7+69632 x^8 \, . \nonumber
\end{align}

\subsection{Supplementary material gravitational scattering}
\label{app:bh}

Here, we give the first-order differential equations for the seven new functions required in the $\epsilon$-factorisation of the sector containing a Calabi-Yau three-variety 
\begin{align}
     x\,G_1'(x)&= \frac{x^2\, \left(3-x^2\right) \left(1-3 x^2\right)}{\left(1+x^2\right)^3} \varpi_0(x) \, ,\\
     x\,G_2'(x)&= 20 \frac{G_1(x)^2}{ \alpha_1 (x)} 
     + \frac{20 x^2 (1-x^2)^2 \left(1-8 x^2+x^4\right) }{\left(1+x^2\right)^4}\frac{\varpi_0(x)^2}{\alpha_1(x)} \, ,\\
    x\,G_3'(x)&= 
    -\frac{40 \left(1-48 x^2+x^4\right)}{(1-x^2) \left(1+x^2\right) }G_1(x)^2 
    + \frac{1+x^2}{(1-x^2)^3 }\frac{G_2(x)^2\alpha_1(x)^2}{\varpi_0(x)^2} \\
    &\quad + \frac{1280 \left(2-11 x^2+2 x^4\right)}{\left(1+x^2\right)^3 }G_1(x)\varpi_0(x) + \nonumber\\
    &\quad -\frac{4 x^2 \left(41-271 x^2-1412 x^4+2606 x^6-1343 x^8-51 x^{10}-2 x^{12}\right)}{\left(1+x^2\right)^5} \varpi_0(x)^2 \nonumber \\
    &\quad + \frac{2 x^2 (1-x^2)  \left(1+152 x^2-198 x^4+152 x^6+x^8\right)}{\left(1+x^2\right)^3}\varpi_0'(x)^2 \, ,\nonumber\\
    x\,G_4'(x)&= \frac{G_1(x)}{ \alpha_1(x)} \, , \\
    x\,G_5'(x)&= -\frac{1-48 x^2+x^4}{(1-x^2) \left(1+x^2\right)}G_1(x) + \frac{1+x^2}{(1-x^2)^3 }\frac{G_2(x) G_4(x) \alpha_1(x)^2}{\varpi_0(x)^2} \\
            & \quad + \frac{16 \,x^2 \left(2-11 x^2+2 x^4\right) }{\left(1+x^2\right)^3} \varpi_0(x) \, , \nonumber\\
    x\,G_6'(x)&= \frac{1+x^2}{(1-x^2)^3 }\frac{G_2(x)\alpha_1(x)^2}{\varpi_0(x)^2} \, , \\
    x\,G_7'(x)&= \frac{40\, G_1(x)}{\alpha_1(x)} \Big( G_5(x) - G_4(x)G_6(x)  \Big)
    - \frac{G_3(x)}{\alpha_1(x)} \\
      &\quad + \frac{2 (1-x^2) \left(1+152 x^2-198 x^4+152 x^6+x^8\right)}{\left(1+x^2\right)^3}\frac{\varpi_0(x)\varpi_0'(x)}{\alpha_1(x)} \, . \nonumber\\
\end{align}

\bibliographystyle{JHEP} 
\bibliography{CY_eps_form} 

\end{document}